\documentclass[a4paper,10pt]{article}       
\usepackage[utf8]{inputenc}
\usepackage{amsmath}
\pdfoutput=1
\usepackage{amssymb}
\usepackage{array}
\usepackage{bbold}
\usepackage{csquotes}
\usepackage{tcolorbox}
\usepackage{graphicx}
\usepackage{pdfpages}
\usepackage[square,numbers]{natbib}
\usepackage{url}
\usepackage[top=1in, bottom=1.25in, left=1.25in, right=1.25in]{geometry}
\numberwithin{equation}{section}
\usepackage{chngcntr}
\usepackage{booktabs}
\usepackage{mathtools}
\counterwithin{table}{section}
\usepackage{blindtext}

\begin{document}

    \begin{center}
        \vspace*{0.2cm}
        
        \large
        \textbf{Robust reconstruction of fluorescence molecular tomography with an optimized illumination pattern}
        
        \vspace{1.5cm}
        
        \normalsize

		\centerline{\scshape Yan Liu}
		\medskip
		{\footnotesize
		 \centerline{Department of Mathematics, ETH Z\"urich}
		   \centerline{8092 Z\"urich, Switzerland}
		} 

		\medskip

		\centerline{\scshape Wuwei Ren*}
		\medskip
		{\footnotesize
		 \centerline{Institute for Biomedical Engineering, University of Z\"urich and ETH Z\"urich}
		   \centerline{8093 Z\"urich, Switzerland}
		 \centerline{Biomedical Optics Research Laboratory, University Hospital Z\"urich}
		   \centerline{8091 Z\"urich, Switzerland}
		}

		\medskip

		\centerline{\scshape Habib Ammari}
		\medskip
		{\footnotesize
		 \centerline{Department of Mathematics, ETH Z\"urich}
		   \centerline{8092 Z\"urich, Switzerland}
		}

		\bigskip


        \vspace{1cm}

		\footnotetext[2]{The work is supported by the Swiss National Science Foundation and Swiss Innovation Agent BRIDGE proof-of-concept fellowship (No. 178262).} 
		\footnotetext[3]{*Corresponding author: Wuwei Ren.}
    \end{center}
    \begin{abstract} 
\vspace{0.5cm}
 Fluorescence molecular tomography (FMT) is an emerging powerful tool for biomedical research. There are two factors that influence FMT reconstruction most effectively. The first one is the regularization techniques. Traditional methods such as Tikhonov regularization suffer from low resolution and poor signal to noise ratio. Therefore sparse regularization techniques have been introduced to improve the reconstruction quality. The second factor is the illumination pattern. A better illumination pattern ensures the quantity and quality of the information content of the data set thus leading to better reconstructions. In this work, we take advantage of the discrete formulation of the forward problem to give a rigorous definition of an illumination pattern as well as the admissible set of the patterns. We add restrictions in the admissible set as different types of regularizers to a discrepancy functional, generating another inverse problem with the illumination pattern as unknown. Both inverse problems of reconstructing the fluorescence distribution and finding the optimal illumination pattern are solved by fast efficient iterative algorithms. Numerical experiments have shown that with suitable choice of the regularization parameters the two-step approach converges to an optimal illumination pattern quickly and the reconstruction result is improved significantly, regardless of the initial illumination setting.
\end{abstract}


\section{Introduction}

\subsection{Background} 
Fluorescence molecular tomography (FMT) is a medical imaging technique with high sensitivity, noninvasiveness and low cost. The traditional planar imaging strategy suffers from a lack of depth resolution since its dimension and optical properties correspond to a fully diffusive regime. FMT solves such problems by allowing quantitative reconstruction of the three-dimensional fluorescence distribution in the intact organism. Compared to traditional radiative emission tomography such as positron emission tomography (PET) and single photon emission computer tomography (SPECT), it is a relatively cheaper and safer alternative and thus widely used in basic biomedical research as well as drug development \cite{RN9,Rudin,Ntziachristos,RN2}.  

\subsection{Basic principles of FMT}
The major components of FMT include a laser that emits near-infrared light, a highly sensitive camera that captures the outcoming light from the fluorescence probe and a scanner that directs the laser beam onto the object surface.

\begin{figure}[!h]
\begin{center}
 \includegraphics[scale=0.5]{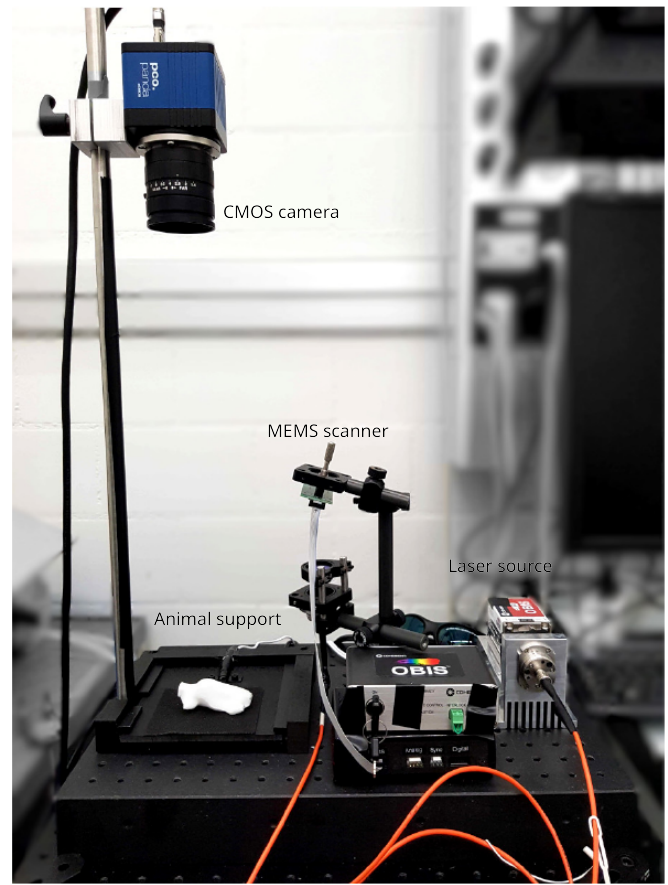}
 \end{center}
 \caption{A snapshot of the latest stand-alone FMT hardware system in the animal imaging center of ETH Zurich. The main components of the hardware system include a sCMOS camera (detector), the animal support (object), the laser source and the MEMS scanner. Photo credit: Andrin G. C. Rickenbacher.}
\end{figure} 
 
A complete FMT experiment is composed of the following processes.
First of all, a fluorescence probe is administrated into the object to be imaged. The object is then illuminated by an external diffuse photon wave source of wavelength $\lambda^e$, giving rise to an excitation field $\phi^e$ in the object. The previously injected fluorescence dye will be excited, absorbs the radiation at wavelength $\lambda^e$ and re-emits as a source at a longer wavelength $\lambda^f$ \cite{Arridge_2009}. Such a physical procedure can be explained as follows. When a photon is absorbed by the probe, it will excite an electron into a higher energy state. Excited electron will soon relax back to ground state, losing some of its energy and sending out another photon with a longer wavelength (Stoke's shift) and less intensity. Measurements taken at both excitation and emission processes are made by a CCD camera such that information from every corner is recorded, which is equivalent to having point detectors densely distributed on the object surface. 

 The data acquisition system of FMT consists of a light source that delivers near-infrared light to the object surface at different points or with certain spatial patterns and a detector system that measures the light transmitted through the tissue. The light source of FMT can be classified into the following three categories: steady-state (or continuous wave) systems for measuring the stationary excitance, time-resolved systems for measuring the temporal distribution of the excitance due to ultrashort input pulse, and frequency-domain systems which uses an intensity  modulated source and measures the phase shift and modulation depth of the detected signal \cite{doi:10.1118/1.598008}. Our FMT model belongs to the category of steady-state systems.     
  
  Particularly in our experiments, we employ a near-infrared light source at a fixed wavelength of 670nm which allows for a drastic increase in the depth of light penetration in the object \cite{Rudin}. In fact, light that does not lie in the near-infrared range (650-900$\mu m$) will suffer from low signal-to-noise ratio due to Rayleigh scattering or photoelectric effect \cite{Ntziachristos}. When electromagnetic waves transmit through biological tissues, different degrees of absorption, reflection and scattering will occur. In order to compute the forward model, the scattering coefficient $\mu_s$ and absorption coefficient $\mu_a$ of the object to be imaged must be given \textit{a priori} or measured by means of other techniques, for instance, diffuse optical tomography. Experimental and theoretical work have demonstrated that under the condition of $\mu_a\ll\mu_s$ which is exactly the case of our application, light propagation in the medium falls into a strongly diffusive regime. Light within the medium essentially behaves like heat \cite{Lyons2019} and the photon transport can be modeled by a diffusion approximation with sufficient accuracy \cite{Groenhuis:83}.  
 
 Finally, the forward problem is formulated into a large linear system by using the finite element method to calculate the light propagation through the object. Together with the measurements collected by a CCD camera, the linear system is properly inversed in order to reconstruct the three-dimensional distribution of the fluorescence dye inside the object. 
 
 \subsection{Current status} 
 The aim of most studies on tomographic imaging in recent years focuses on improving the quality of the reconstruction, especially the spatial resolution, signal to noise ratio (SNR) and the reconstruction speed. Necessary regularization techniques are introduced due to the ill-posedness of the inverse problem. Some effort has been devoted to the development of novel regularization methods such as Tikhonov regularization to replace the traditional Algebraic Reconstruction Technique (ART) which generates diffuse, oversmooth and noisy reconstruction results at a heavy computational cost. Recently, sparse regularization methods become popular because of their advantage in promoting sparsity and thus higher spatial resolution in cases where the target to be imaged is relatively small in size compared to its background.  These sparse regularization methods mainly employ the $l_1$ norm of the image vector combined with other penalty terms to formulate hybrid regularization methods, such as $l_1$ combined with total variation regularization and $l_1$ combined with $l_2$ regularization \cite{Dutta_2012}. Aside from encoding sparsity of the image as the regularizer, other regularization techniques that form regularizers using prior information of the object are also investigated including Laplacian-type regularization that incorporates tissue structural information into reconstruction \cite{Davis_2007}, patch-based anisotropic diffusion regularization that solves the issue of oversmooth and noisy reconstruction \cite{patch}, and maximum-likelihood expectation maximization combined with sparse regularization method that reduces background noise and promotes sparsity at the same time \cite{Zhu:18}. 
 
 Existing work has shown an advantage of sparse regularization methods especially hybrid methods that benefit from the strengths of more than one regularizer compared to the conventional nonsparse methods. Hence, it is necessary to introduce hybrid sparsity-based regularization methods to our discrete FMT inverse model. Considering the complexity of implementation, we choose $l_1$ combined with $l_2$ regularization (also known as elastic net in linear regression) as well as pure $l_1$ regularization. The main advantage of the elastic net is that it promotes sparsity of the solution while preserving its smoothness at the same time, avoiding discontinuities in the reconstructed image. 
 
 Another important factor that affects the reconstruction quality is the data set, which is determined to a large extent by the configuration of detection and illumination. A lot of research groups have investigated designing of the laser/detector pattern in different imaging modalities, such as electrical impedance tomography \cite{1386561} and thermo-acoustic tomography \cite{10.1088/1361-6420/ab0e4d}, etc. Regardless of the modality involved, the common issue to be addressed is the same, namely how to place the source-detector pairs onto the object surface so as to obtain the best reconstruction result. In order to solve this problem, the usual procedure is as follows. First try to find a mathematical description of the geometry of the sensors and the object boundary. Then define the admissible set of possible sensor patterns on the boundary and search for a proper criterion which assesses the quality of a certain sensor pattern. Such a criterion is often defined as a functional of the admissible set and the object boundary. Depending on the mathematical or physical meaning of the defined functional, optimal sensor patterns are obtained by either minimizing or maximizing this functional. Hence, the criterion is also called the optimality condition. 
 
 In most papers, a complex geometric analysis of the boundary of the object as well as the shape of the sensors has to be carried out \cite{10.1088/1361-6420/ab0e4d, eit, 1386561} in order to define the admissible set. And the defined functional is only explicitly solvable in a number of ideal cases that require the object to have some simple geometric structures, for example, a two-dimensional disk. However, our FMT experiment is very different from those settings. First of all, the detectors need not to be optimized. This is because the information of light propagation is acquired by a CCD camera which is able to detect light from every corner of the surface of the object. In other words, we solely focus on optimizing the illumination pattern, namely the laser power and laser positions. Besides, each laser point can be regarded as a small two-dimensional disk projected on the object surface and thus has no thickness, so complex analysis of the geometry of lasers is not needed. Our special problem setting inspires us to define the set of admissible illumination patterns by enforcing the restrictions such as the number of lasers, the power of each laser as well as the position of each laser as constraints directly on the source vector by simply using different norms of the source vector.

 As for the optimality condition, \cite{1386561} employs the A-optimality or D-optimality condition from optimal experiment design which involves the minimization or maximization of a functional of estimation; \cite{fisher} achieves optimal illumination pattern by improving the conditioning of the Fisher information matrix in order to maximize the information content of acquired data; \cite{10.1088/1361-6420/ab0e4d} maximizes the observability functional that stands for the efficiency of a set of sensors to allow for best observation quality of the worst possible case, etc. To benefit most from the discrete formulation of the forward problem, we define our functional as a discrepancy between the measurements and the discrete model in the least square fashion and encode the information in the admissible set as regularizers added to the discrepancy function. Hence, the techniques from existing sparse reconstruction section can be borrowed to solve the optimization of the illumination pattern. Finally, inspired by \cite{10.1088/1361-6420/ab0e4d}, we adopt an iterative approach of using the reconstructed result  to guide the design of the next illumination pattern and using the updated illumination pattern to perform the next step of reconstruction. 
 
 \subsection{Scope of this work}
 Our work focuses on improving the quality of the reconstructed three-dimensional fluorescence distribution from two aspects. The first effort is devoted to reducing the level of diffusion in recovering the fluorescence inclusions through the implementation of novel sparse regularization techniques. The second effort goes to finding the optimal illumination pattern which is achieved by minimizing a discrepancy functional based on the reconstruction result from the previous step. Then we use the updated illumination pattern to run a new FMT experiment and perform the next round of reconstruction. Finally, by combining these two efforts to form a loop of reconstructing the fluorescence distribution and optimizing the illumination pattern, the updated illumination patterns are expected to converge to the optimal illumination pattern regardless of the initial laser setting. The optimal illumination pattern should ensure a better reconstruction result compared to that of a random unoptimized illumination pattern.     

\section{Forward model}
\subsection{Theoretical formulation of FMT}
 Light propagates in tissue with different degrees of absorption, reflection and scattering. The mathematical description of light propagation can be modeled at three different levels: Maxwell's equations at the microscopic scale, radiative transfer equation (RTE) at the mesoscale and the diffusion approximation to the radiative transfer approximation at the macroscale. Here, we model the light propagation as photon transport which is described by the diffusion equation, an asymptotic approximation to the following radiative transfer equation \cite{doi:10.1142/q0067}:      
     
 \begin{equation}\label{RTE}
  \frac{1}{c}\frac{\partial I}{\partial t}+\xi\cdot\nabla I+(\mu_a+\mu_s)I=\mu_s\int_S p(\xi',\xi)I(r,\xi',t)d\xi' 
 \end{equation}
 for $(r,\xi,t)\in\Omega\times S\times\mathbb{R}^+$, where $\mu_a$ and $\mu_s$ are the absorption and scattering coefficients, $c$ is the speed of light in the medium, $S$ is the unit sphere, and $I(r,\xi)$ is the specific intensity defined as the intensity at position $x$ in the direction $\xi$. 
   
The diffusion approximation to RTE is widely used in many applications. It holds when $\mu_s\gg\mu_a$, the point of observation is far away from the medium boundary and the observation time is sufficiently long. If we introduce a rescaling of $\mu_a$, $\mu_s$ and $t$, and  perform the asymptotic expansion for $I(r,\xi,t)$, then \eqref{RTE} yields the following the diffusion equation:
   
 \begin{equation}\label{DE}
  [-\nabla\cdot\kappa(r)\nabla+\mu_a(r)+\frac{1}{c}\frac{\partial}{\partial t}]\phi(r,t)=0
 \end{equation}  
with $\phi(r,t)=\int I(r,\xi,t)d\xi$ being the energy (photon) density, $\kappa(r)=\frac{1}{3(1-g)\mu_t}$ being the diffusion coefficient, $\mu_t=\mu_a+\mu_s$ being the extinction coefficient, and $g=\int_S\xi\cdot\xi'p(\xi\cdot\xi')d\xi'$ being the anisotropy. Here, $g\in(-1,1)$.
    
In our application the medium is a bounded domain, hence, it is necessary to account for boundary layers by adding the following boundary condition to \eqref{DE}:
\begin{equation}\label{bc}
 \phi+l_{\text{ext}}\frac{\partial\phi}{\partial\nu}=q\quad \text{on}\quad \partial\Omega\times\mathbb{R}^+,
\end{equation}
where $q$ is the source term, $\nu$ is the outward normal to $\partial\Omega$, and $l_{\text{ext}}$ is the extrapolation length calculated to be $l_{\text{ext}}=2\zeta(c)\kappa(r)$ in our case. $\zeta$ is the refractive index mismatch at the boundary.

Since we consider our problem in the continuous wave case, we apply Fourier transform to \eqref{DE}, add boundary condition \eqref{bc} and arrive at the following two coupled partial differential equations: 
 
\begin{equation}\label{coupled_PDE1}
-\nabla\cdot\kappa(r)\nabla\phi^e(r;\omega)+[\mu_a(r)+\frac{i\omega}{c}]\phi^e(r;\omega)=0\quad \textrm{in}\;\Omega,
\end{equation}
\begin{equation}\label{bc_1}
\phi^e(r;\omega)+2\zeta(c)\kappa(r)\frac{\partial\phi^e(r;\omega)}{\partial\nu}=q(r;\omega)\quad \textrm{on} \;\partial\Omega,
\end{equation}
\begin{equation}\label{coupled_PDE2}
 -\nabla\cdot\kappa(r)\nabla\phi^f(r;\omega)+[\mu_a(r)+\frac{i\omega}{c}]\phi^f(r;\omega)=q'(r;\omega)\quad \textrm{in} \; \Omega, 
\end{equation}
\begin{equation}\label{bc_2}
 \phi^f(r;\omega)+2\zeta(c)\kappa(r)\frac{\partial\phi^f(r;\omega)}{\partial\nu}=0\quad \textrm{on} \;\partial\Omega. 
\end{equation}  
Here, \eqref{coupled_PDE1}-\eqref{bc_1} describe the propagation of light at the excitation wavelength $\lambda^e$ while \eqref{coupled_PDE2}-\eqref{bc_2} describe propagation of the re-emitted light at the fluorescence wavelength $\lambda^f$. 

The internal source term $q'(r;\omega)$ in \eqref{coupled_PDE2}  represents the portion of the photons absorbed at $\lambda^e$ and re-emitted at $\lambda^f$ during the emission process. To be precise, $q'(r;\omega)$ is a product of the unknown fluorescence concentration $C(r)$ and the photon density $\phi^e(r;\omega)$, that is,

\begin{equation}
 q'(r;\omega)=\eta C(r)\phi^e(r;\omega),
\end{equation}
 where $\eta$ is a constant determined by the fluorescence and the absorption cross section at the excitation wavelength $\lambda^e$.
 
 The detectable flux on the boundary for both excitation and emission processes is the normal current $J_n(\xi;\omega)$ across the boundary:

\begin{equation}\label{flux}
 J_n(\xi;\omega)=-c\kappa(\xi)\frac{\partial\phi(r;\omega)}{\partial\nu}|_{\partial\Omega(\xi)},\quad\xi\in\partial\Omega.
\end{equation}

\subsection{FEM formulation}
Analytical solutions of the two systems of equations (\eqref{coupled_PDE1} and \eqref{bc_1}, \eqref{coupled_PDE2} and \eqref{bc_2}) only exist for simple geometries. For practical applications, a numerical model must be employed in order to incorporate inhomogeneous parameter distributions and complex boundaries. A comprehensive computational model of the diffusion equation has been developed in \cite{doi:10.1118/1.597069,doi:10.1118/1.598008,New1} where a Galerkin finite element method is used to compute the numerical solution of the coupled systems. The details of the FEM formulation is presented as follows.

Since the excitation process and the emission process share the same physical principles with only different source terms, we take the first system \eqref{coupled_PDE1}-\eqref{bc_1} as an example to derive its discrete formulation. \eqref{coupled_PDE2}-\eqref{bc_2} can be dealt with using the same strategy. For convenience we omit the superscript for $\phi^e$.

The Galerkin method proceeds as follows. First write \eqref{coupled_PDE1} in the variational formulation: \eqref{coupled_PDE1} has a solution if there exists a solution of
\begin{equation}\label{weakform}
 \int_\Omega u(r)[-\nabla\cdot\kappa(r)\nabla+[\mu_a(r)+\frac{i\omega}{c}]\phi(r;\omega)d\Omega=0\quad \forall u\in H^1(\Omega),
\end{equation}
where $H^1(\Omega)$ is the set of square-integrable functions with square-integrable weak derivatives in $\Omega$.

Integration by parts yields
\begin{multline}
 \int_\Omega \kappa(r)\nabla u(r)\nabla\phi(r;\omega)d\Omega+\int_{\Omega}\mu_a(r)u(r)\phi(r;\omega)d\Omega+\int_{\Omega}\frac{i\omega}{c}u(r)\phi(r;\omega)\\
 +\int_{\partial\Omega}u(\xi)\kappa(\xi)\nabla\phi(\xi;\omega)\cdot\nu d(\partial\Omega)=0,
\end{multline}
where $\nu$ denotes the outward normal to $\partial\Omega$.

Combining the detectable flux on the boundary \eqref{flux} and boundary condition \eqref{bc_1}, we obtain  
\begin{multline}
 \int_\Omega \kappa(r)\nabla u(r)\nabla\phi(r;\omega)d\Omega+\int_{\Omega}\mu_a(r)u(r)\phi(r;\omega)d\Omega+\frac{i\omega}{c}\int_{\Omega}u(r)\phi(r;\omega)\\
 +\int_\Omega u(r)q(r)d\Omega=\zeta(c)\int_{\partial\Omega}u(\xi)\phi(\xi;\omega) d(\partial\Omega).
\end{multline}

To apply the finite element method, one defines an $N$-dimensional subspace $U^h\subset H^1(\Omega)$ and uses an N-dimensional polynomial basis expansion $\{u_i\}_{i=1}^N\in U^h$ with local support over domain $\Omega$. A piecewise polynomial approximation $\phi^{h}$ of the field $\phi$ can be expressed as 
\begin{equation}
 \phi^{h}(r,\omega)=\sum_{i=1}^N\Phi_i(\omega)u_i(r),
\end{equation}
which is defined by the vector $\Phi\in\mathbb{C}^N$ of nodal coefficients in the excitation process. The problem thus becomes:
find $\phi^{h}(r,\omega)=\sum_{i=1}^N\Phi_i(\omega)u_i(r)\in U^h$ such that 
\begin{multline}\label{fem}
\int_\Omega \kappa(r)\nabla u_j(r)\nabla\phi^h(r;\omega)d\Omega+\int_{\Omega}\mu_a(r)u_j(r)\phi^h(r;\omega)d\Omega+\int_{\Omega}\frac{i\omega}{c}u_j(r)\phi^h(r;\omega)\\
 +\int_\Omega u_i(r)q(r;\omega)d\Omega=\zeta(c)\int_{\partial\Omega}u_i(\xi)u_j(\xi) d(\partial\Omega)
 \end{multline}
holds for all $u_j,j=1,\ldots,N$.

Formulation \eqref{fem} can be expressed by the following linear system:
\begin{equation}
 [K(\kappa^h)+C(\mu_a^h)+\zeta(c) A+i\omega B]\Phi(\omega)=Q(\omega),
\end{equation} 
where $K$, $C$, $A$, and $B$ are sparse symmetric positive definite system matrices and $Q(\omega)$ denotes the nodal values of the source. Individual integrals over an element $\Omega_i^{el}$ are given by 

\begin{displaymath}
 K_{ij}^{(el)}=\sum_{k\in S[\Omega^{(el)}]}\kappa_k\int_{\Omega^{(el)}}u_k(r)\nabla u_j(r)\cdot\nabla u_i(r)\textrm{d}r,
\end{displaymath}
\begin{displaymath}
 C_{ij}^{(el)}=\sum_{k\in S[\Omega^{(el)}]}(\mu_a)_k\int_{\Omega^{(el)}}u_k(r) u_i(r) u_j(r)\textrm{d}r,
\end{displaymath}
\begin{displaymath}
 A_{ij}^{(el)}=\int_{\partial\Omega^{(el)}} u_i(r) u_j(r)\textrm{d}r,
\end{displaymath}
\begin{displaymath}
 B_{ij}^{(el)}=\frac{1}{c}\int_{\Omega^{(el)}} u_i(r) u_j(r)\textrm{d}r,
\end{displaymath}
\begin{displaymath}
 \Phi(\omega)=[\Phi_1(\omega)\;\Phi_2(\omega)\;\ldots\;\Phi_N(\omega)]^T,
\end{displaymath}
and for the external source in the excitation process
\begin{displaymath}
 Q_{ij}(\omega)^{(el)}=\int_{\Omega^{(el)}} u_i(r) q(r;\omega)\textrm{d}r, 
\end{displaymath}
for the internal source in the emission process
\begin{displaymath}
 Q_{ij}'(\omega)^{(el)}=\int_{\Omega^{(el)}} u_i(r) q'(r;\omega)\textrm{d}r. 
\end{displaymath}

Define the stiff matrix $\mathbb{C}^{N\times N} \ni S:=K(\kappa^h)+C(\mu_a^h)+\zeta(c) A+i\omega B$. Then the discrete formulation of \eqref{coupled_PDE1}-\eqref{bc_1} can be simply written as (omitting the frequency $\omega$)
\begin{equation}
 S^e\Phi^e=Q^e.
\end{equation}
Similarly, the system of equations \eqref{coupled_PDE2}-\eqref{bc_2} can be written in the discrete form 
\begin{equation}
 S^f\Phi^f=Q^f, 
\end{equation}
where the superscript $f$ represents the emission process.

In the course of the experiment, the position of the light source is varied in order to generate enough independent observations to allow for reconstruction. 

Suppose there are $L$ illumination points. For the $l$th illumination point $I_r$, denote by $\Phi^e_l,\Phi^f_l\in\mathbb{C}^N$ and $Q^e_l,Q^f_l\in\mathbb{C}^N$ the (discrete) photon densities and the source terms, respectively. We obtain $L$ discrete coupled systems:
\begin{equation}\label{Lcoupledsys1}
 S^e\Phi^e_l=Q^e_l,\quad l=1,\ldots,L,
\end{equation}
\begin{equation}\label{Lcoupledsys2}
 S^f\Phi^f_l=Q^f_l,\quad l=1,\ldots,L.
\end{equation}
Source $Q^f_l$ of the $l$th emission process is given by the following formula:
\begin{equation}\label{sourceE}
Q^f_l=\eta\cdot diag(C)\Phi^e_l,
\end{equation}
where $C\in\mathbb{C}^N$ is the vector of the unknown nodal value of the fluorescence distribution to be recovered. The operation $diag$ turns the vector $C$ into a diagonal matrix by putting all entries of the vector on the diagonal of the square matrix.  

Define the matrices 

\begin{displaymath}
 \Phi^E=[\Phi^e_1\quad\Phi^e_2\quad \ldots \quad \Phi^e_L]\in\mathbb{C}^{N\times L},
\end{displaymath}
\begin{displaymath}
 Q^E=[Q^e_1\quad Q^e_2\quad \ldots \quad Q^e_L]\in\mathbb{C}^{N\times L},
\end{displaymath}
\begin{displaymath}
 \Phi^F=[\Phi^f_1\quad\Phi^f_2\quad \ldots \quad \Phi^f_L]\in\mathbb{C}^{N\times L},
\end{displaymath}
\begin{displaymath}
 Q^F=[Q^f_1\quad Q^f_2\quad \ldots \quad Q^f_L]\in\mathbb{C}^{N\times L}.
\end{displaymath}
The systems \eqref{Lcoupledsys1} and \eqref{Lcoupledsys2} can be written in a compact form
\begin{equation}\label{Ccoupledsys1}
 S^e\Phi^E=Q^E,
\end{equation}
\begin{equation}\label{Ccoupledsys2}
 S^f\Phi^F=Q^F.
\end{equation}
The compactly written systems \eqref{Ccoupledsys1} and \eqref{Ccoupledsys2} complete the FEM formulation of FMT.

\subsection{Measurements}
Early FMT experiments adopted a contact configuration, meaning that the fibers used for photon 
collection need to be in contact with the tissue or require the tissue to conform to simple geometries such as a slab or a cylinder.  
The most common approach to overcome this problem is to insert the subject in some fluid of similar optical properties \cite{PhysRevLett.91.103901}. 
This approach however introduces additional absorption and scattering, significantly degrading the resolution and the signal 
to noise ratio unnecessarily. In the ideal configuration, the detectors should be detached from the object to be imaged, 
offering simple experimental procedures and yielding large numbers of tomographic projections. Hence, a non-contact configuration 
is employed in our detection by placing the detectors away from the object surface. This allows imaging of arbitrary geometries 
in the absence of matching fluid, thus boosting the resolution and quantification accuracy to the maximum. 

\subsubsection{Free-space light propagation model}
A free-space light propagation model to describe the intensity distribution from a diffusive volume of arbitrary
shape $\Omega$ to a set of noncontact detectors is added to \eqref{Ccoupledsys1} and \eqref{Ccoupledsys2}.

Assume that there are $M$ effective pixels on the detector plane. For both the excitation and emission processes during $l$th illumination, $l=1,\ldots, L$, the measured power $P^e_l,P^f_l\in\mathbb{C}^M$ can be expressed by the product of the field $\Phi_l\in\mathbb{C}^N$ and a transportation matrix $\Gamma\in\mathbb{C}^{M\times N}$ of free-space light propagation. $\Gamma$ is assumed identical for both processes:

\begin{equation}\label{measureE}
 P^e_l=\Gamma\Phi^e_l,\quad l=1,\ldots, L,
\end{equation}
\begin{equation}\label{measureF}
 P^f_l=\Gamma\Phi^f_l,\quad l=1,\ldots, L.
\end{equation}

Define the matrices

\begin{displaymath}
 P^E=[P^e_1\quad P^e_2\quad\ldots\quad P^e_L]\in\mathbb{C}^{M\times L},
\end{displaymath}
\begin{displaymath}
 P^F=[P^f_1\quad P^f_2\quad\ldots\quad P^f_L]\in\mathbb{C}^{M\times L},
\end{displaymath}
and recall that 
\begin{displaymath}
 \Phi^E=[\Phi^e_1\quad \Phi^e_2\quad\ldots\quad \Phi^e_L]\in\mathbb{C}^{N\times L}.
\end{displaymath}

The system \eqref{measureE} and \eqref{measureF} can be written compactly as
\begin{equation}\label{CmeasureE}
 P^E=\Gamma\Phi^E,
\end{equation}
\begin{equation}\label{CmeasureF}
 P^F=\Gamma\Phi^F.
\end{equation}
    
The calculation of the transportation matrix $\Gamma$ can be found in \cite{stift}. 

\subsubsection{Linear representation of a FMT experiment}
\eqref{Ccoupledsys1}, \eqref{Ccoupledsys2}, \eqref{CmeasureE} and \eqref{CmeasureF} give a complete description of the 
whole procedure of FMT in discrete form. The formulation of the forward problem can thus be derived.

During a complete FMT experiment, after one illumination measurements are made for both the excitation and emission processes. 
In order to cancel out location-dependent gaining factors, we calculate the measurement vector $Y$ using a normalization technique. 

For the $l$th illumination, define the measurement $Y_{l,k}$ at the $k$th effective pixel by the ratio of the measured power 
$P^f_l(k)$ ($k$th entry of the vector $P^f_l$) of the emission process at $k$th pixel and the measured power $P^e_l(k)$ of the excitation process at the $k$th pixel:

\begin{equation}\label{Y_lk}
 Y_{l,k}:=\frac{P^f_l(k)}{P^e_l(k)}.
\end{equation}

Using \eqref{measureE}, \eqref{measureF}, \eqref{Lcoupledsys1} and \eqref{sourceE}, $Y_{l,k}$ can be written as

\begin{align}
 Y_{l,k}&=\frac{\Gamma^T_k\Phi^f_l}{\Gamma^T_k\Phi^e_l}=\frac{\Gamma^T_k(S^f)^{-1}Q^f_l}{\Gamma_k^T\Phi^e_l}=\frac{\eta\Gamma_k^T(S^f)^{-1}diag(C)\Phi^e_l}{\Gamma_k^T\Phi^e_l}\nonumber\\
 &=\frac{\eta}{\Gamma^T_k\Phi^e_l}\begin{bmatrix}\Gamma^T_k(S^f)^{-1}(1)\Phi^e_l(1)\\\Gamma^T_k(S^f)^{-1}(2)\Phi^e_l(2)\\\vdots\\\Gamma^T_k(S^f)^{-1}(N)\Phi^e_l(N)\end{bmatrix}^T
 \begin{bmatrix}C(1)\\C(2)\\\vdots\\C(N)\nonumber\end{bmatrix}. 
\end{align}

Here, $\Gamma_k$ denotes the $k$th column vector of $\Gamma$, $(S^f)^{-1}(i), i=1,\ldots,N$ is the $i$th column vector of the inverse of the matrix $S^f$ during the emission process, $C$ is the unknown vector of fluorescence distribution and $C(i), i=1,\ldots, N$ denotes the $i$th entry of $C$. 

Define the row vector $$W_{l,k}:=\frac{\eta}{\Gamma^T_k\Phi^e_l}\begin{bmatrix}\Gamma^T_k(S^f)^{-1}(1)\Phi^e_l(1)\\\Gamma^T_k(S^f)^{-1}(2)\Phi^e_l(2)\\\vdots\\\Gamma^T_k(S^f)^{-1}(N)\Phi^e_l(N)\end{bmatrix}^T\in\mathbb{C}^N.$$ Then a single measurement point $Y_{l,k}$ at the $k$th pixel during $l$th illumination can be expressed by the inner product of two vectors $W_{l,k}$ and $C$. 

Stack all $M$ row vectors $W_{l,k},\;k=1,\ldots,M$ during $l$th illumination to get the $l$th weighting matrix $W_l^T=
\begin{bmatrix}
 W_{l,1}^T, W_{l,2}^T, \ldots, W_{l,M}^T
\end{bmatrix}^T\in\mathbb{C}^{M\times N}$, and define the $l$th measurement vector $Y_l=
\begin{bmatrix}
 Y_{l,1},Y_{l,2},\ldots,Y_{l,M}
\end{bmatrix}^T
\in\mathbb{C}^M$, it is easy to see that 
\begin{displaymath}
Y_l=W_lC,
\end{displaymath}
a linear representation of the $l$th FMT experiment.

In order to obtain sufficient measurement data, put together all $L$ measurement vectors by stacking all column vectors $Y_l,\;l=1,\ldots,L$ into a long column vector $Y=
\begin{bmatrix}
 Y_1^T, Y_2^T, \ldots,Y_L^T 
\end{bmatrix}^T
\in\mathbb{C}^{LM}$
to obtain the final measurement vector $Y$. Perform the same technique for all $L$ weighting matrices $W_l,l=1,\ldots,L$ to get the final weighting matrix $W=
\begin{bmatrix}
 W_1^T, W_2^T, \ldots, W_L^T
\end{bmatrix}^T
\in\mathbb{C}^{LM\times N}$, it is readily seen that 
\begin{equation}\label{forward}
 Y=WC, 
\end{equation}
a linear system representing the complete FMT experiment with $L$ illumination processes.

\subsection{Designing optimal illumination pattern}

\subsubsection{Design matrix $V$}

To be precise, designing an optimal illumination pattern refers to optimizing the spatial positions of the lasers used in the experiment as well as the power of each laser. Both the location information and the laser power are encoded in the group of external source vectors $\{Q^e_l\}_{l=1}^L$ defined in \eqref{Lcoupledsys1}. In order to find out how the external source impacts the reconstruction result, we rewrite the measurement $Y_{l,k}$ from \eqref{Y_lk} as a function of the external source vector $Q^e_l$.

Let $(S^e)^{-1}$ be the inverse of the matrix $S^e$. According to \eqref{Lcoupledsys1}, $\Phi^e_l=(S^e)^{-1}Q^e_l$, hence,
\begin{align}
 Y_{l,k}&=\frac{\Gamma^T_k\Phi^f_l}{\Gamma^T_k\Phi^e_l}=\frac{\Gamma^T_k(S^f)^{-1}Q^f_l}{\Gamma_k^T\Phi^e_l}=\frac{\eta\Gamma_k^T(S^f)^{-1}diag(C)\Phi^e_l}{\Gamma_k^T\Phi^e_l}\nonumber\\
 &=\frac{\eta\Gamma^T_k}{\Gamma^T_k\Phi^e_l}(S^{f})^{-1}diag(C)(S^e)^{-1}Q^e_l \\\nonumber
 &=\underbrace{\frac{\eta\Gamma^T_k}{\Gamma^T_k\Phi^e_l}(S^{f})^{-1}diag(C)(S^e)^{-1}}_{\text{$=V_{l,k}$}}Q^e_l.\nonumber
\end{align}

Define the row vector $V_{l,k}:=\frac{\eta\Gamma^T_k}{\Gamma^T_k\Phi^e_l}(S^{f})^{-1}diag(C)(S^e)^{-1}$. Then the measurement point $Y_{l,k}$ can be written as the inner product of two vectors $V_{l,k}$ and $Q^e_l$.
Using the same strategy as the last section by stacking all $M$ row vectors $V_{l,k}$ to get the $l$th design matrix 
$V_l=\begin{bmatrix}V_{l,1}, V_{l,2},\ldots,V_{l,M}\end{bmatrix}^T\in\mathbb{C}^{M\times N}$ associated with the $l$th illumination $I_l$, it is easy to see that 
\begin{displaymath}
 Y_l=V_lQ^e_l.
\end{displaymath}
 Once again assemble all $L$ measurement vectors into $Y=\begin{bmatrix}Y_1^T,Y_2^T,\ldots,Y_L^T\end{bmatrix}^T\in\mathbb{C}^{LM}$and $L$ design matrices into $V=\begin{bmatrix}V_1^T,V_2^T,\ldots,V_L^T\end{bmatrix}^T\in\mathbb{C}^{LM\times N}$, we have 
 \begin{equation}\label{illu}
  Y=VQ^e,
\end{equation}
a linear system that reveals the relation between the external source $Q^e$ and the measurement $Y$. Note that $V$ is a function of the fluorescence distribution vector $C$, which indicates that designing a new illumination pattern requires the information of the reconstructed fluorescence distribution as a prior.  \eqref{forward} and \eqref{illu} constitute a complete forward model for FMT with a designed illumination pattern.   

\subsubsection{Definition of the illumination pattern $\Sigma$}
The illumination pattern physically refers to the spatial distribution of the lasers (illumination points) on the object surface. Moreover, it also encodes the information of the power strength of each laser. Mathematically, the illumination pattern vector $\Sigma$ is defined by means of the set of external source vectors $\{Q^e_l\}_{l=1}^{L}$, $L$ is the number of illumination points used in an FMT experiment. Recall that the illumination process in an FMT experiment proceeds as follows: the first laser beam shoots at location $\xi_1\in\partial\Omega$ on the surface of $\Omega$. Light propagates through the tissue exciting the fluorescence inclusion and then reaches the detector plane, measured power $P^e_1$ is recorded. Re-emitted light with shifted wavelength $\lambda^f$ due to the relaxation of the excited fluorephore reaches the detector plane and measured power $P^f_1$ is recorded. Then the second laser beam shoots at location $\xi_2\in\partial\Omega$ and measured power $P^e_2$ and $P^f_2$ are recorded and so on, until the $L$th laser beam. 

To improve the quality of the reconstruction result, a straightforward strategy is to use a large number of illumination points to increase the amount of detected information. However, this strategy is not always numerically tractable due to the size of the data set. Moreover, it leads to a significant increase in the acquisition time \cite{Ducros_2012}. To remain compatible with in vivo measurements, we need to limit the number of illumination points. In other words, the number of illumination points $L$ is expected to be relatively small and should be under control.

Notice that during each illumination, the source vector $Q^e_l$ contains exactly one
\footnote{This is based on the assumption that the laser power perfectly concentrates on a node $\xi_l$ of the finite element mesh. In a real FMT experiment, the laser power decays around its center obeying some certain rule. In most cases, Gaussian distribution is accurate enough to describe such a phenomenon. Hence, we can amend our statement to be more rigorous: $Q^e_l$ has only a small number of nonzero entries.} 
nonzero entry at the corresponding location $\xi_l\in\partial\Omega$. Hence, the vectors $\{Q^e_l\}_{l=1}^L$ are all sparse. Adding up all $L$ vectors we define the illumination pattern $\Sigma:=\sum_{i=1}^LQ^e_l$ that is an $N$-dimensional column vector with exactly $L$ nonzero entries. 

  Under the FEM framework, $\Omega$ is discretized into a set $\Omega^{el}$ of $N$ elements. All the elements on the surface of $\Omega$ constitute the surface element set $\partial\Omega^{el}$ and the rest of the elements belong to the inner element set $\Omega^{el}_{in}=\Omega^{el}\backslash\partial\Omega^{el}$. Regard $\Sigma$ as a function mapping an element $\Omega^{el}_i\in\Omega^{el}$ to $\mathbb{R}^N_{\ge0}$. Here we only concern the location information carried by an element $\Omega^{el}_i$. It is easy to see that $\Sigma$ maps the set $\Omega^{el}_{in}$ to $\{0\}$ because only the surface of the target can be illuminated. For the surface set $\partial\Omega^{el}$, only $L$ elements have nonzero function values since there are $L$ illumination points. Additionally, we demand that the number of illumination points to be small, hence, $L\ll N$. Meanwhile, the laser power should be under a maximal value due to safety reasons. In other words, $||\Sigma||_{\infty}\leq p_{\max}$ with $p_{\max}$ the maximally allowed value of the laser power.

In summary, the set $\Delta$ of admissible illumination patterns is defined as follows:

\begin{equation}
 \Delta=\left\{\Sigma: \Omega^{el}\rightarrow\mathbb{R}_{\geq0}^N\ s.t.\  \Sigma(\Omega^{el}_{in})=0,\ ||\Sigma||_0\le L,\  ||\Sigma||_{\infty}\leq p_{\max}\right\},
\end{equation}
where $||\cdot||_0$ denotes the $l_0$ norm that counts the nonzero entries of a vector, $||\cdot||_{\infty}$ denotes the supremum norm that returns the maximal amplitude of a vector.
Note that we impose the inequality constraint $||\Sigma||_0\le L$ instead of the equality constraint $||\Sigma||_0=L$ to allow for more freedom in controlling the number of illumination points.

\section{Inverse model: Sparse reconstruction and illumination pattern optimization}
This section dedicates to solving two inverse problems. The first one is reconstructing the fluorescence distribution with sparse regularizers. The second one is finding the optimal illumination pattern given a reconstructed fluorescence distribution. Both problems involve an objective function to be minimized. We first define the objective functions and then design suitable iterative algorithms to solve these optimization problems. In the end, we explain in detail our two-step approach of reconstruction with an optimized illumination pattern. 

\subsection{Recover fluorescence distribution $C$}
In this section we formulate the inverse problem for recovering the fluorescence distribution, discuss the choice of suitable sparse regularizers, and then present the algorithm we use for this inverse problem. 
\subsubsection{Formulation of the inverse problem} 
 Given an illumination pattern $\Sigma$, we aim to recover the fluorescence distribution in the object $\Omega$ so as to image the location and shape of the target of interest. This corresponds to reconstructing the unknown vector $C$ in \eqref{forward}:
 \begin{displaymath}
  Y=WC.
 \end{displaymath}
 Such a problem is usually ill-posed. Instead of direct inversion, we take an optimal design approach which formulates the inversion of \eqref{forward} as an equivalent optimization problem:
 
 \begin{equation}\label{step1}
  \arg\min_{C\in\mathbb{R}_{\geq0}^N} \left\{J(C;\Sigma)=A(C;\Sigma)+\mathcal{R}(C)\right\},
 \end{equation}
where the objective function $J$ is defined by a data-fitting term $A(C;\Sigma)$ plus a regularization term $\mathcal{R}(C)$. $A(C;\Sigma)$ is defined in the least-squared sense:

\begin{equation}\label{admiset}
 A(C;\Sigma)=\frac{1}{2}||Y-W(\Sigma)C||^2_2,
\end{equation}
where $\Sigma$ denotes the illumination pattern defined in \eqref{admiset}.

\subsubsection{Basic choice of $\mathcal{R}(C)$: Lasso}

In most imaging problems we are considering, the target is confined in a small region. In other words, the fluorescence distribution 
$C$ is a sparse vector. We can thus employ $l_1$ regularization to promote sparsity and define:
\begin{displaymath}
 \mathcal{R}(C)=\lambda||C||_{1},
\end{displaymath}
where $||C||_1=\sum_{i=1}^N|C(i)|$ is the $l_1$ norm of $C$, $\lambda>0$ is the regularization weight.
\eqref{step1} thus becomes the following Lasso estimator:
\begin{equation}\label{purel1}
 \arg\min_{C\in\mathbb{R}_{\geq0}^N}\left\{J(C;\Sigma)=\frac{1}{2}||Y-W(\Sigma)C||^2_2+\lambda||C||_{1}\right\}.      
\end{equation}

\subsubsection{Advanced choice of $\mathcal{R}(C)$: Elastic net}
Numerical experiments show that while pure $l_1$ regularization captures the sparsity of the solution, it tends to bring discontinuities 
into the reconstructed image. In order to improve the smoothness of the solution, we replace the pure $l_1$ penalty
with the elastic net penalty which makes a compromise between the $l_2$ and the $l_1$ penalties:
\begin{displaymath}
 \mathcal{R}_e(C)=\lambda(\alpha||C||_{1}+\frac{1-\alpha}{2}||C||_2^2),\quad\text{where }\alpha\in[0,1] \text{ is a constant.}
\end{displaymath}
Such a combination of norms takes 
advantage of both regularizers and finds a balance between sparsity and the integrity of the image.

The optimization problem thus writes as follows: 
\begin{equation}\label{l1l2}
 \arg\min_{C\in\mathbb{R}_{\geq0}^N}\left\{J(C;\Sigma)=\frac{1}{2}||Y-W(\Sigma)C||^2_2+\lambda(\alpha||C||_{1}+\frac{1-\alpha}{2}||C||_2^2)\right\},\quad \alpha\in[0,1].
\end{equation}
Note that when $\alpha=1,$ \eqref{l1l2} becomes pure $l_1$ regularization, and when $\alpha=0,$ \eqref{l1l2} is Tikhonov regularization.

In summary, the problem of reconstructing the fluorescence distribution $C$ is formulated as the following minimization problem:
\begin{equation}\label{J}
 \arg\min_{C\in\mathbb{R}_{\geq0}^N}\left\{J(C;\Sigma)=\frac{1}{2}||Y-W(\Sigma)C||^2_2+\mathcal{R}(C)\right\}, 
\end{equation}
where $\mathcal{R}(C)$ can be chosen between $\lambda||C||_1$ and $\lambda(\alpha||C||_{1}+\frac{1-\alpha}{2}||C||_2^2)$.  
 \subsubsection{Algorithm for the inverse problem: proximal gradient descent}
 
This section focuses on designing a fast and accurate iterative algorithm to solve \eqref{J} based on gradient descent methods. Notice that both the Lasso and the elastic net regularizers are convex nondifferentiable due to the $l_1$-norm, so it is necessary to seek a variation of the usual gradient descent method in order to handle the nondifferentiality. 
        
Recall that the gradient descent for a convex differentiable function $f(x)$ usually takes the form
\begin{equation}\label{gd} 
 x_{t+1}= x_t-s_t\nabla f(x_t),\qquad t=0,1,2\ldots,
\end{equation}
where $s_t>0$ is the step size.  
  
\eqref{gd} has the following alternative representation:
\begin{equation}
 x_{t+1}=\underset{x\in\mathbb{R}^N}{\text{argmin}}\left\{f(x_t)+\langle\nabla f(x_t),x-x_t\rangle+\frac{1}{2s_t} ||x-x_t||_2^2\right\},
\end{equation}   
where $\langle\cdot,\cdot\rangle$ is the euclidean scalar product.

Notice that the objective function $J$ in \eqref{J} is nondifferentiable but $J$ can be decomposed into a sum of a convex differentiable part \eqref{admiset} and a convex nondifferentiable part $\mathcal{R}$. Hence, the usual gradient step \eqref{gd} needs to be generalized:    
\begin{equation}\label{ggd}
 x_{t+1}=\underset{x\in\mathbb{R}^N}{\text{argmin}}\left\{g(x_t)+\langle\nabla g(x_t),x-x_t\rangle+\frac{1}{2s_t} ||x-x_t||_2^2+h(x)\right\},
\end{equation}
where $f$ has the aforementioned type of decomposition: $f=g+h$. \eqref{ggd} approximates the differentiable part $g$ and retains the nondifferentiable component $h$.    

Define the proximal map $\text{prox}_h$ of the convex function $h$ to be
\begin{equation}
 \text{prox}_h(z):=\underset{y\in\mathbb{R}^N}{\text{argmin}}\left\{\frac{1}{2}||z-y||^2_2+h(y)\right\}.
\end{equation}
Then \eqref{ggd} has the following equivalent representation
\begin{equation}\label{prox}
 x_{t+1}=\text{prox}_{s_th}(x_t-s_t\nabla g(x_t)).  
\end{equation}   

The update \eqref{prox} has an explicit formula when $h$ is the lasso or the elastic net penalty. We simply take a gradient step and then perform elementwise soft-thresholding. To be precise, when $\mathcal{R}(C)=\lambda(\alpha||C||_{1}+\frac{1-\alpha}{2}||C||_2^2)$,    
\begin{equation*}
 \textrm{prox}_{s_t\mathcal{R}}(z)=\frac{1}{1+(1-\alpha)\lambda s_t}\tau_{\alpha\lambda s_t}(z),
\end{equation*}
where $\tau_{\theta}:\mathbb{R}^N\rightarrow\mathbb{R}^N$ is the soft-thresholding shrinkage operator defined by 
\begin{equation}
 \tau_{\theta}(x)_i=(|x_i|-\theta)_+\textrm{sign}(x_i),\quad, i=1\cdots,N
\end{equation}
with $\theta\in\mathbb{R}$ a given threshold. $(x)_+:=\max\{x,0\}$ is the rectifier function.

 Following the idea of proximal gradient descent plus an accelerating process \cite{doi:10.1137/080716542}, we implement the following algorithm for the reconstruction procedure. Due to sparsity, the initial guess of $C$ is set to be the zero vector.
\begin{tcolorbox}
\textbf{FISTA with constant step size}\\    
 Input $\lambda, \varepsilon$\\
 Compute $L=\textrm{eig}_{\max}(W^TW);$\\
 Set $s_t=1/L, C_1=x_0=0, p=1;$\\
 Step $k(k\ge1)$
 \begin{align}
  &z_k=C_k-tW^T(Y-WC_k);\\
  &x_k=\textrm{prox}_{s_t\mathcal{R}}(z_k);\\
  &p_{k+1}=\frac{1+\sqrt{1+4p_k^2}}{2};\\
  &C_{k+1}=(x_k+\frac{p_k-1}{p_k+1}(x_k-x_{k-1}))_+;\\
  &\text{check stopping criterion}.\nonumber
 \end{align}
\end{tcolorbox}
Here, $L$ is the largest eigenvalue of $W^TW$. 

\subsection{Design illumination pattern $\Sigma$}
This section first introduces the optimality condition and then formulates the minimization problem by adding the constraints from the admissible set $\Delta$ as different regularizers one by one. In the end, the algorithm to solve this constrained minimization problem is explained.
\subsubsection{Optimality condition}
Each entry of the vector $\Sigma$ encodes the location as well as the laser power information at the corresponding node on the object surface. Given a reconstructed fluorescence distribution vector $C$, we calculate the design matrix $V(C)$ defined in \eqref{illu}. The design of the optimal illumination pattern is achieved by solving the following linear system:  
\begin{equation}\label{pattern_linear}     
 Y=V(C)\Sigma,      
\end{equation}
subject to the restriction $\Sigma\in\Delta$. In other words, we want to solve the following optimization problem:
\begin{equation}\label{l0}
 \arg\min_{\Sigma} \frac{1}{2}||Y-V(C)\Sigma||^2_2,\quad \text{s.t. }\left\{
 \begin{array}{l}
 ||\Sigma||_0\le L,\\
 0\le\Sigma_j\le p_{\max},\\
 \Sigma(\Omega_{in})=0.\\
 \end{array}\right.
\end{equation}

\subsubsection{Solving the constrained minimization problem \eqref{l0}}
We try to solve this multi-constrained problem
by adding one constraint at a time to \eqref{pattern_linear}. 
First add $||\Sigma||_0\le L$ and we arrive at
\begin{equation}\label{add1}
\arg\min_{\Sigma} \frac{1}{2}||Y-V(C)\Sigma||^2_2,\quad \text{s.t. }
||\Sigma||_0\le L.
\end{equation}
 Note that \eqref{add1} is nonconvex due to the 0-norm and thus computationally intractable.
To make it solvable, consider its $l_1$ relaxation formulated 
as the following basis-pursuit program:
\begin{equation}\label{relax_add1}
 \arg\min_{\Sigma} \frac{1}{2}||Y-V(C)\Sigma||^2_2,\quad \text{s.t. }
 ||\Sigma||_1\le R. 
\end{equation}
Problem \eqref{relax_add1} can be written in the equivalent Lagrangian form:
\begin{equation}\label{equiv_relaex_add1}
 \arg\min_{\Sigma}\left\{ \frac{1}{2}||Y-V(C)\Sigma||^2_2+\mu||\Sigma||_1\right\},
\end{equation}
where $\mu>0$ is the regularization weight.  

Solving the basis-pursuit program \eqref{relax_add1} is only equivalent to solving the original problem \eqref{add1} under specific conditions, see chapter 10 of \cite{Hastie:2015:SLS:2834535} for details. In general, the solutions of \eqref{relax_add1} and \eqref{add1} do not coincide. This is because $l_0$ norm and $l_1$ norm act differently as a regularizer: $l_1$ norm        
penalizes large coefficients more heavily than smaller coefficients in terms of magnitude. 
In order to rectify the influence of the $l_1$ norm, 
we can adopt a reweighted $l_1$ minimization approach to make $l_1$ norm behave more like $l_0$ norm \cite{4193451}.

The idea is as follows: 
replace $||\Sigma||_1$ with $||H\Sigma||_1$ in \eqref{equiv_relaex_add1} to get
\begin{equation}\label{reweighted}
 \arg\min_{\Sigma}\left\{ \frac{1}{2}||Y-V(C)\Sigma||^2_2+\mu||H\Sigma||_1\right\},
\end{equation}
where H is a $N\times N$ diagonal 
matrix with nonnegative diagonal entries $h_1,\ldots,h_N$ defined inversely proportional to $\Sigma_1,\ldots,\Sigma_N$, respectively.
An iterative procedure that alternates between recovering $\Sigma$ and redefining $H$ can be used to solve \eqref{reweighted}.
 
\begin{tcolorbox}    
\textbf{Reweighted $l_1-$minimization}\\
Input $\mu,\epsilon$\\
Set $h_i^1=1,i=1,\ldots,N$\\
Step $k(k\ge1)$
\begin{align}
  &H^k=\text{diag}\left\{h_1^k,\ldots,h_N^k\right\};\nonumber\\
  &\Sigma^k=\arg\min_{\Sigma}\left\{ \frac{1}{2}||Y-V(C)\Sigma||^2_2+\mu||H^k\Sigma||_1\right\};\label{l1}\\
  &h^{k+1}_i=\frac{1}{|\Sigma^k_i|+\epsilon},i=1,\ldots,N;\nonumber\\
  &\text{check stopping criterion.}\nonumber
  \end{align}        
\end{tcolorbox}      
$\epsilon$ is a parameter that should be set slightly smaller than the expected entries of $\Sigma$ to avoid numerical overflow.
 
Now we only need to solve \eqref{l1} plus the remaining constraints, i.e.,
\begin{equation}\label{final_l1}
 \arg\min_{\Sigma}\left\{f(\Sigma)=\frac{1}{2}||Y-V(C)\Sigma||^2_2+\mu||H\Sigma||_1\right\} \text{ s.t. }
 \left\{\begin{array}{l}
  0\le\Sigma\le p_{\max},\\
  \Sigma(\Omega_{in})=0.\\
 \end{array}\right. 
\end{equation}
  
Notice that the objective function $f(\Sigma)$ in \eqref{final_l1} has an additive
decomposition:      
\begin{equation}\label{separability} 
 f(\Sigma)=g(\Sigma)+\sum_{j=1}^Nd_j(\Sigma_j),    
\end{equation}
where $g(\Sigma)=\frac{1}{2}||Y-V(C)\Sigma||^2_2$ is convex differentiable,  
and the univariate functions $d_j(\Sigma(j))=\mu h_j|\Sigma(j)|$ are convex.
  
Coordinate descent algorithm is perfectly tailored for such a separable structure as the separability shown in \eqref{separability} guarantees convergence to a global minimizer.    
Meanwhile, coordinate descent also makes it easy to set upper and lower 
bounds on $\Sigma(j)$ by simply computing the coordinate update and setting the 
$\Sigma(j)$ that violates the bounds to the closest boundary. To this end, the second constraint is successfully added. 
   
The last constraint can be added by applying the Lagrange multiplier to 
the linear system $Y=V\Sigma$ to arrive at the following extended linear system  
\begin{equation}\label{extended}
 \tilde{Y}=\tilde{V}\Sigma,
\end{equation}
where
\begin{displaymath}
 \tilde{V}=\begin{bmatrix}
            V\\
            S_p
           \end{bmatrix}
 \text{ and }\tilde{Y}=
 \begin{bmatrix}
  Y\\
  O
 \end{bmatrix}.  
\end{displaymath}
Here, $S_p\in\mathbb{R}^{N\times N}$ is a diagonal matrix with only a few entries of value 1 on the diagonal and all the remaining entries 0. These nonzero entries correspond to all the internal nodes of the mesh. $O\in\mathbb{R}^N$ is a zero vector. Such an extended linear system will force all the entries in $\Sigma$ that correspond to the internal nodes to have value 0. Hence, the last constraint is satisfied.

In summary, we are dealing with \eqref{extended} and trying to solve 
the following minimization problem:
\begin{equation}\label{final_extended}
 \underset{0\le\Sigma\le p_{\max}}{\arg\min}\left\{\tilde{f}=\frac{1}{2}||\tilde{Y}-\tilde{V}(C)\Sigma||^2_2+\mu||H\Sigma||_1\right\},
\end{equation} by using a coordinate descent method. In the next section, 
a detailed coordinate descent algorithm for solving \eqref{final_extended}
is presented.

\subsubsection{Coordinate descent as the main solver}
Coordinate descent is an iterative algorithm that updates from the $t$th step 
$\Sigma^t$ to next step $\Sigma^{t+1}$ by choosing one coordinate to update
and then performing a univariate minimization over this coordinate \cite{Hastie:2015:SLS:2834535}.
More precisely, if the coordinate $j$ is chosen at iteration $t$, the 
update is given by
\begin{align}
 &\Sigma^{t+1}_j=\arg\min_{\Sigma_j}f(\Sigma^t_1,\ldots,\Sigma^t_{j-1},\Sigma_j,\Sigma^t_{j+1},\ldots,\Sigma^t_N),\nonumber\\ 
 &\Sigma^{t+1}_i=\Sigma^{t}_i,\text{ for }i\ne j.\nonumber 
 \end{align}
As we cycle through all the coordinates a complete update $\Sigma^{t+1}$ is obtained. The algorithm is hence usually referred to as Cyclic Coordinate Descent. An explicit solution for \eqref{final_extended} can be derived using the soft-thresholding operator
$\tau$ introduced in previous sections .

\begin{tcolorbox}
 \textbf{Cyclic Coordinate Descent algorithm for \eqref{final_extended}}\\
 Input $\mu$, $\Sigma^1$ (original laser setting), $p$ (number of rows of $\tilde{V}$)\\
 Step $t(t\ge1)$\\
 For $j=1:N$\\
 \begin{align}
  &r_i^{(j)}=\tilde{Y}_i-\sum_{k\ne j}\tilde{Y}_{ik}\Sigma^t_j;\quad i=1,\ldots,p\nonumber\\
  &\Sigma^{t+1}_j=\frac{\tau_{\mu h_j}(\sum_{i=1}^pr_i^{(j)}\tilde{V}_{ij})}{\sum_{i=1}^p\tilde{V}_{ij}^2};\nonumber\\
   &\Sigma^{t+1}_j=\min(\max(0,\Sigma^{t+1}_j),p_{\max});\nonumber   
 \end{align}
 end\\   
 Check stopping condition.
\end{tcolorbox}

\subsection{Combining fluorescence reconstruction and illumination pattern optimization}
It is easy to notice that the problem of recovering $C$ and of determining $\Sigma$ are mutually dependent, each 
takes the output of the other as an input. This inspires us to design the following two-step approach in order to obtain the best possible reconstruction quality.

\begin{tcolorbox}
\begin{itemize}
 \item[1.] Input an initial illumination pattern $\Sigma_0$;
 \item[2.] Use $\Sigma^k(k\ge0)$ to calculate the weighting matrix $W_k$, solve \eqref{J} to obtain the fluorescence distribution $C_k$;
 \item[3.] Use $C_k$ to calculate the extended design matrix $\tilde{V_k}(C)$, solve \eqref{l0} to obtain the updated illumination pattern $\Sigma^{k+1}$;
 \item[4.] Check if $\Sigma^{k+1}$ varies much from is $\Sigma^{k}$, otherwise, return to step 2.       
 \end{itemize}
\end{tcolorbox}

By iterating back and forth between step 2 and step 3, we are updating the illumination pattern for the 
next reconstruction based on the reconstructed fluorescence distribution from the previous step. In the end, this two-step approach is expected to produce a better reconstruction result compared to that of a fixed and unoptimized illumination pattern.  
 
\section{Numerical experiments}

The two-step approach from the last section contains a fluorescence distribution reconstruction step and an illumination pattern optimization step. A combination of these two steps constitutes a complete loop for updating the illumination pattern and is referred to as one round. Several rounds are performed in order to improve the final reconstruction quality. We validate our two-step approach by performing numerical experiments on a virtual cubic phantom under the reflection mode, i.e. the laser array and detector array are restricted to the same face of the cubic phantom as opposed to the transmission mode where lasers and detectors are placed on the opposite faces of the phantom. 
\subsection{Experiment setup}
\begin{table}[h!]
  \begin{tabular}{cc}
   \includegraphics[width=0.45\textwidth]{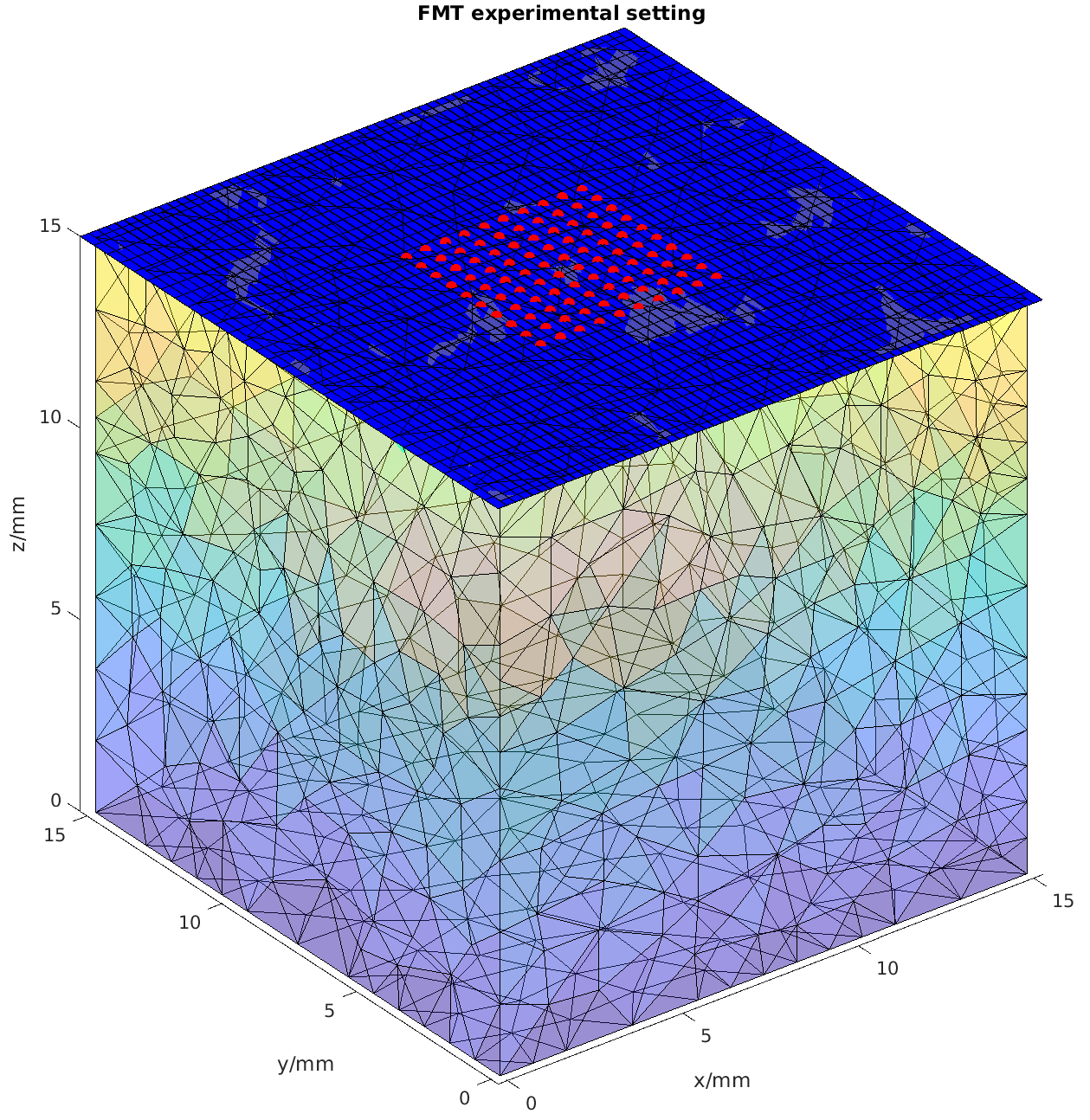}&
   \includegraphics[width=0.4\textwidth]{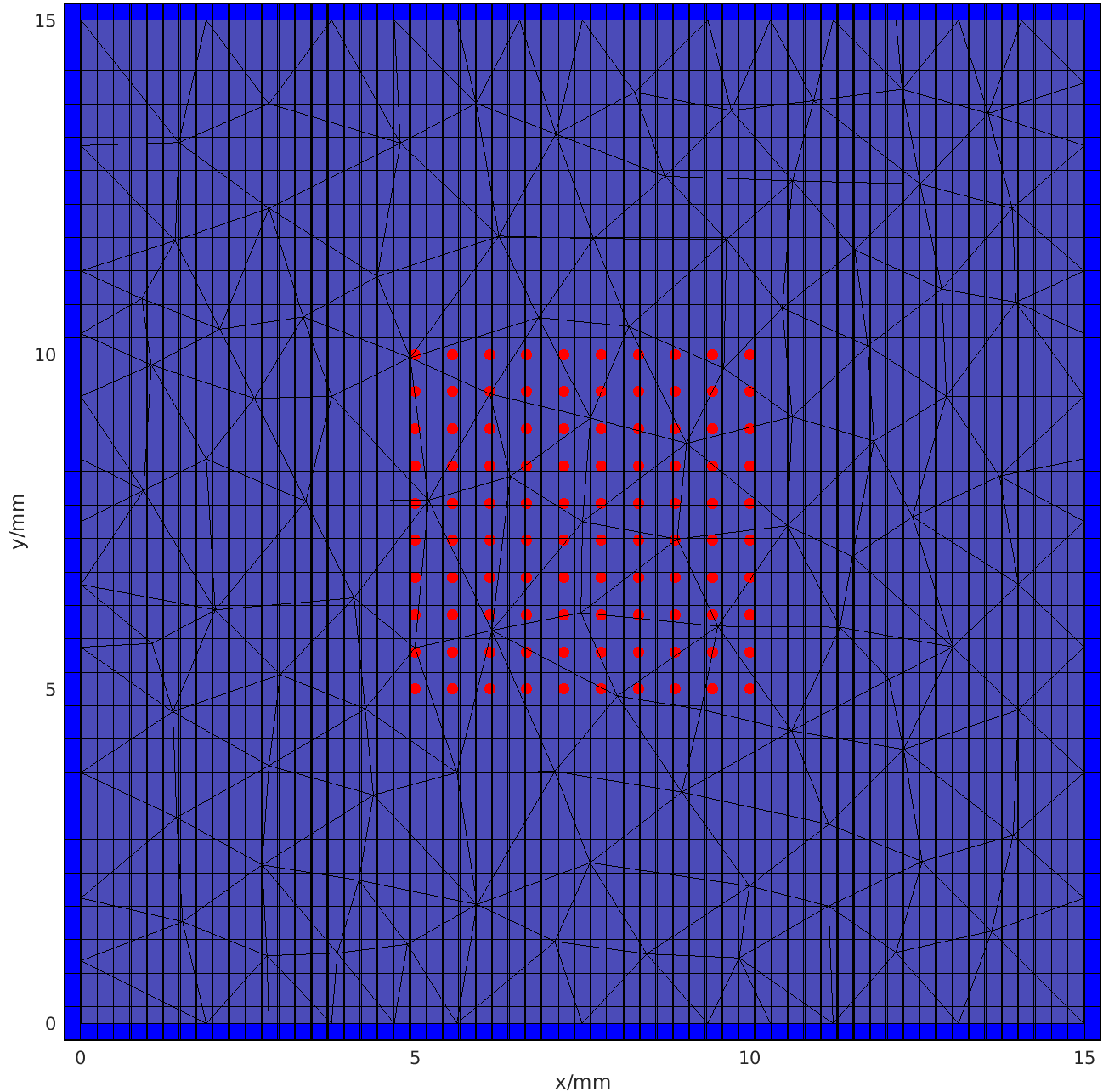}
  \end{tabular}
\caption{Left: 15$\times$15$\times$15 $mm^3$ phantom with absorption coefficient $\mu_a=0.01$ and scattering coefficient $\mu_s=1$. Blue region represents the virtual detector plane that is made up of a 60 by 30 detector array. Each detector has size 1$\times$1 $mm^2$ and an efficiency of 100\%. Red points represent the 10 by 10 laser array. Each laser point has power 1 $W/mm^2$ and wavelength 670 $nm$. The setup of the phantom is performed by the TOAST++ software \cite{New1}. Right: birdview of the laser and detector array from the top surface of the phantom. Laser array is placed at the center of the detector plane.}
\label{phantom}
 \end{table}

 \begin{table}[h!]
  \begin{tabular}{ccc}
   \includegraphics[width=0.3\textwidth]{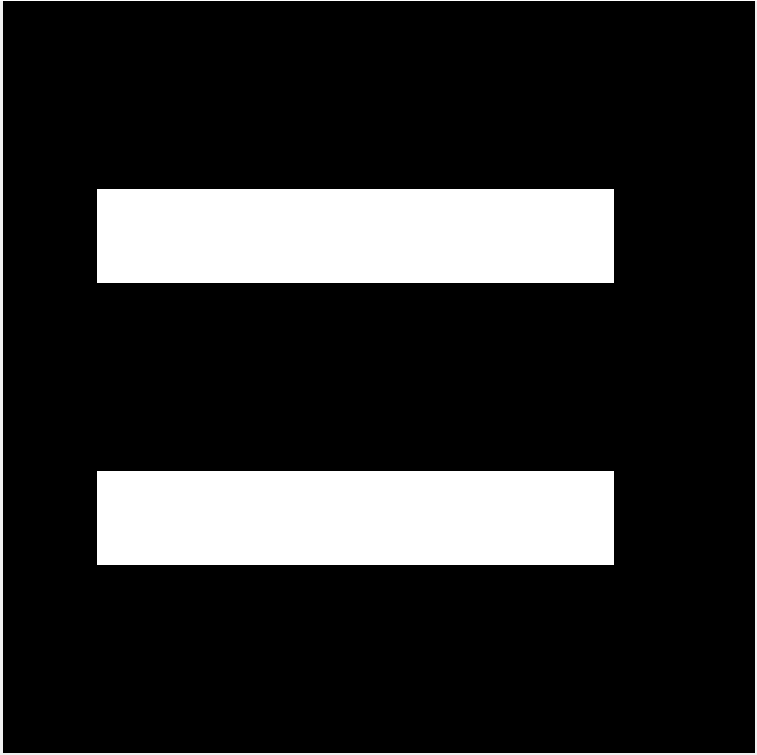}&
   \includegraphics[width=0.3\textwidth]{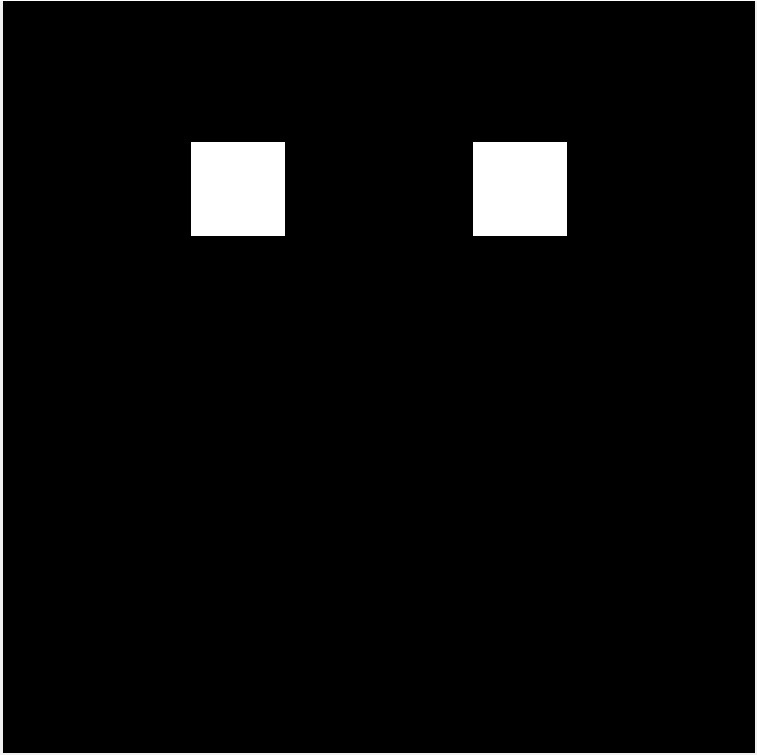}&
   \includegraphics[width=0.3\textwidth]{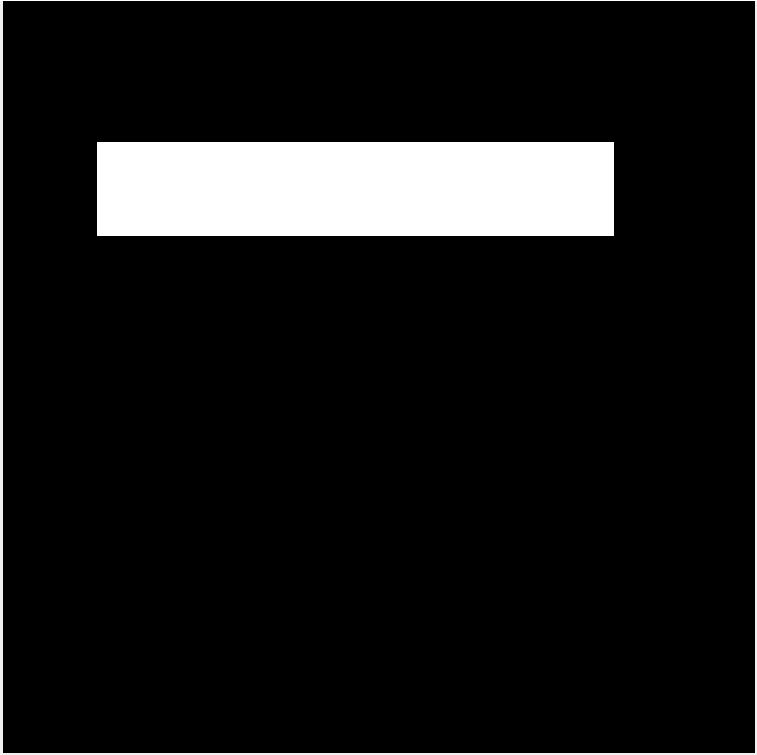}
  \end{tabular}
  \caption{Fluorescence ground truth setting sliced at reference point $(13,9,6)$ viewed from three directions: top, left and front of the phantom (from left to right). The fluorescence ground truth is composed of two identical bars of size 1$\times$1$\times$10 $mm^3$ and fluorescence intensity 100 $a.u.$. The two bars are both embedded at a depth of 3 $mm$ from the top surface of the phantom and $6mm$ away from each other.}
  \label{fluo_gt}
 \end{table}

During the reconstruction step, $l_1$ regularization together with FISTA is used to recover the fluorescence distribution vector $C$. Regularization weight $\lambda$ during the reconstruction is empirically chosen to be $10^{-4}$ and kept the same throughout all rounds. During the illumination pattern optimization step, the regularization weight $\mu$ is chosen to be $1.5\times10^{-8}$ and kept the same throughout all rounds of the experiment.   

\subsection{Result: updated illumination patterns at each round}
\begin{table}[h!]
\begin{center}
  \begin{tabular}{cc}
   \includegraphics[width=0.37\textwidth]{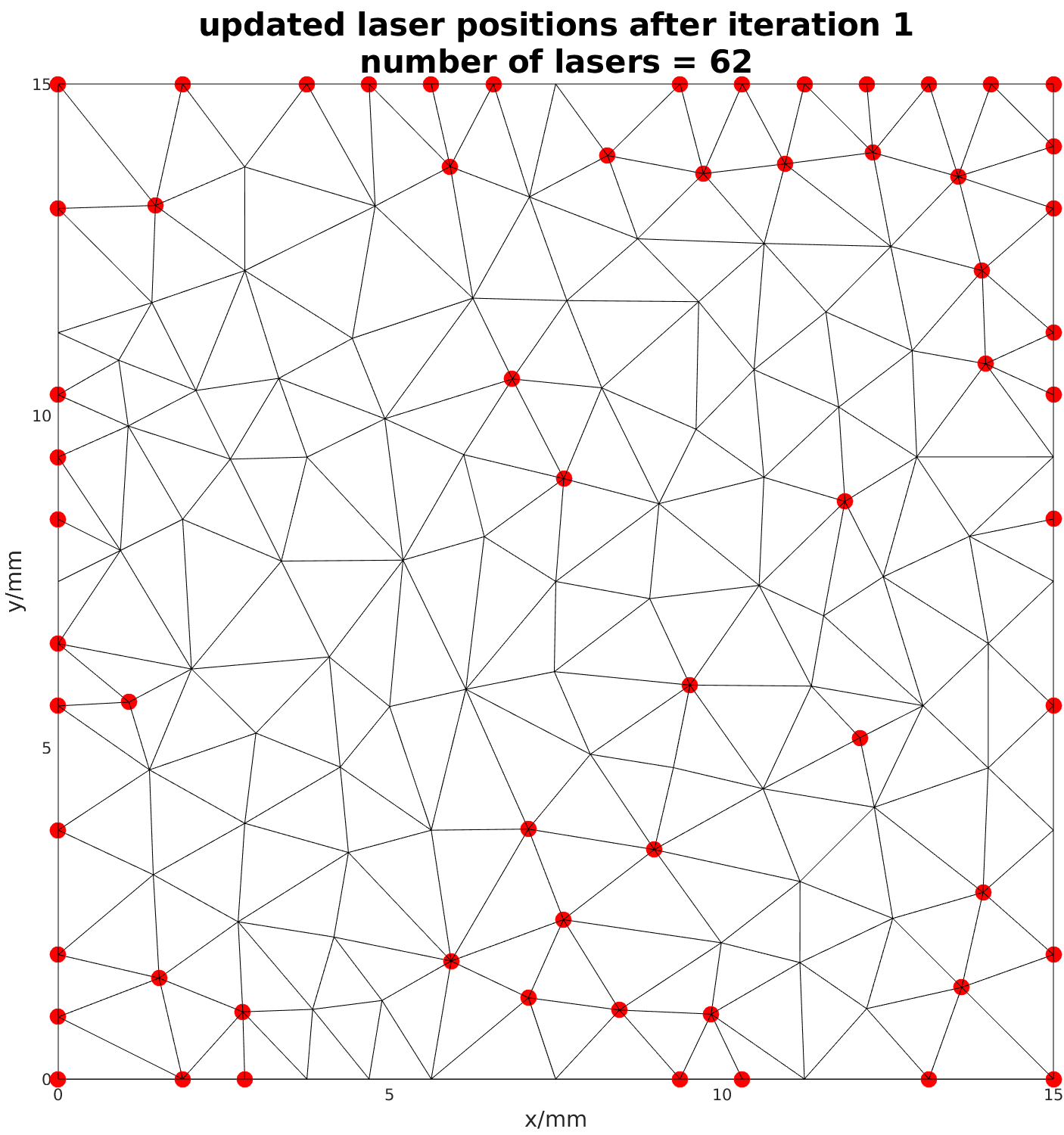}&
   \includegraphics[width=0.37\textwidth]{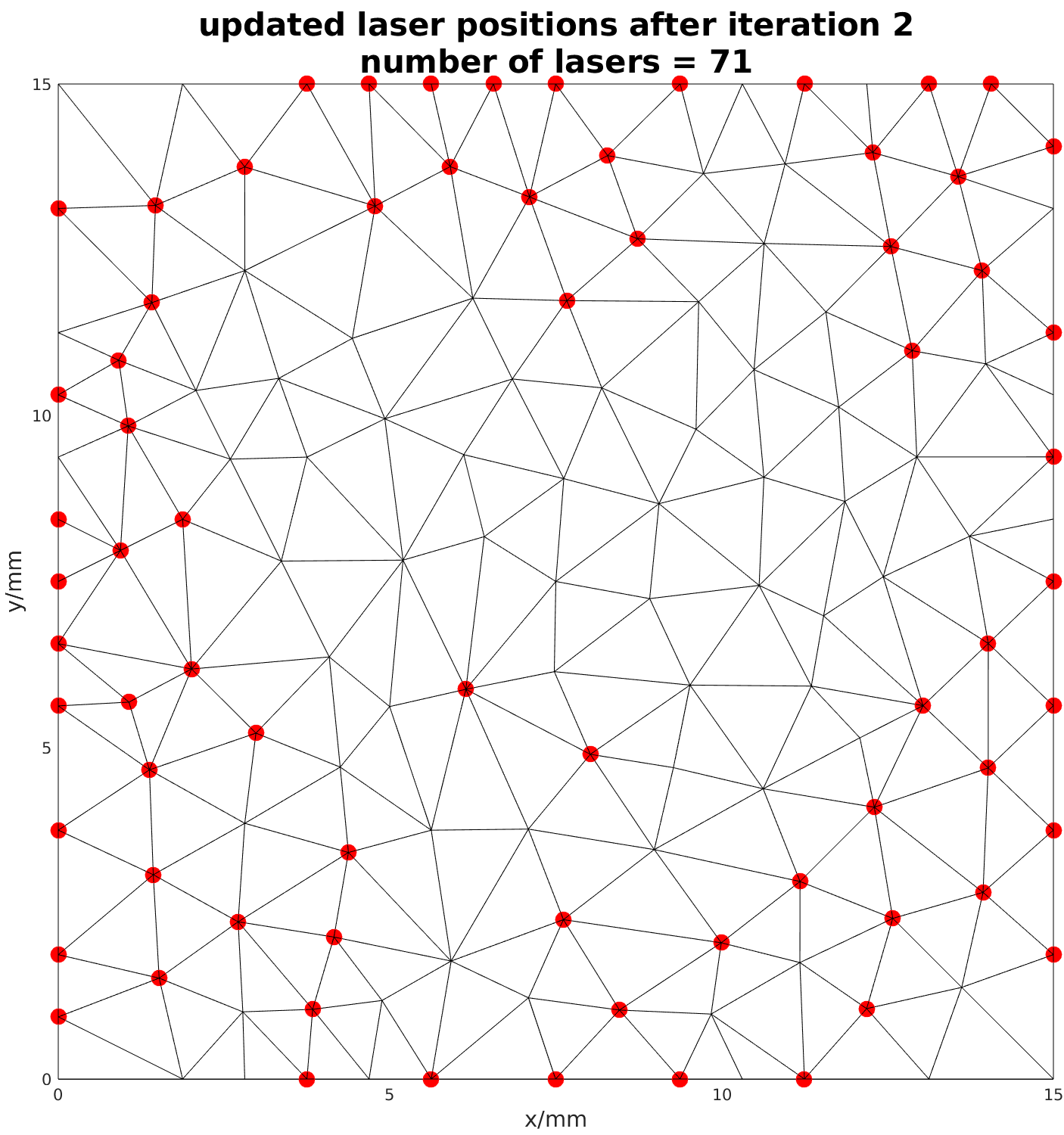}\\
   \includegraphics[width=0.37\textwidth]{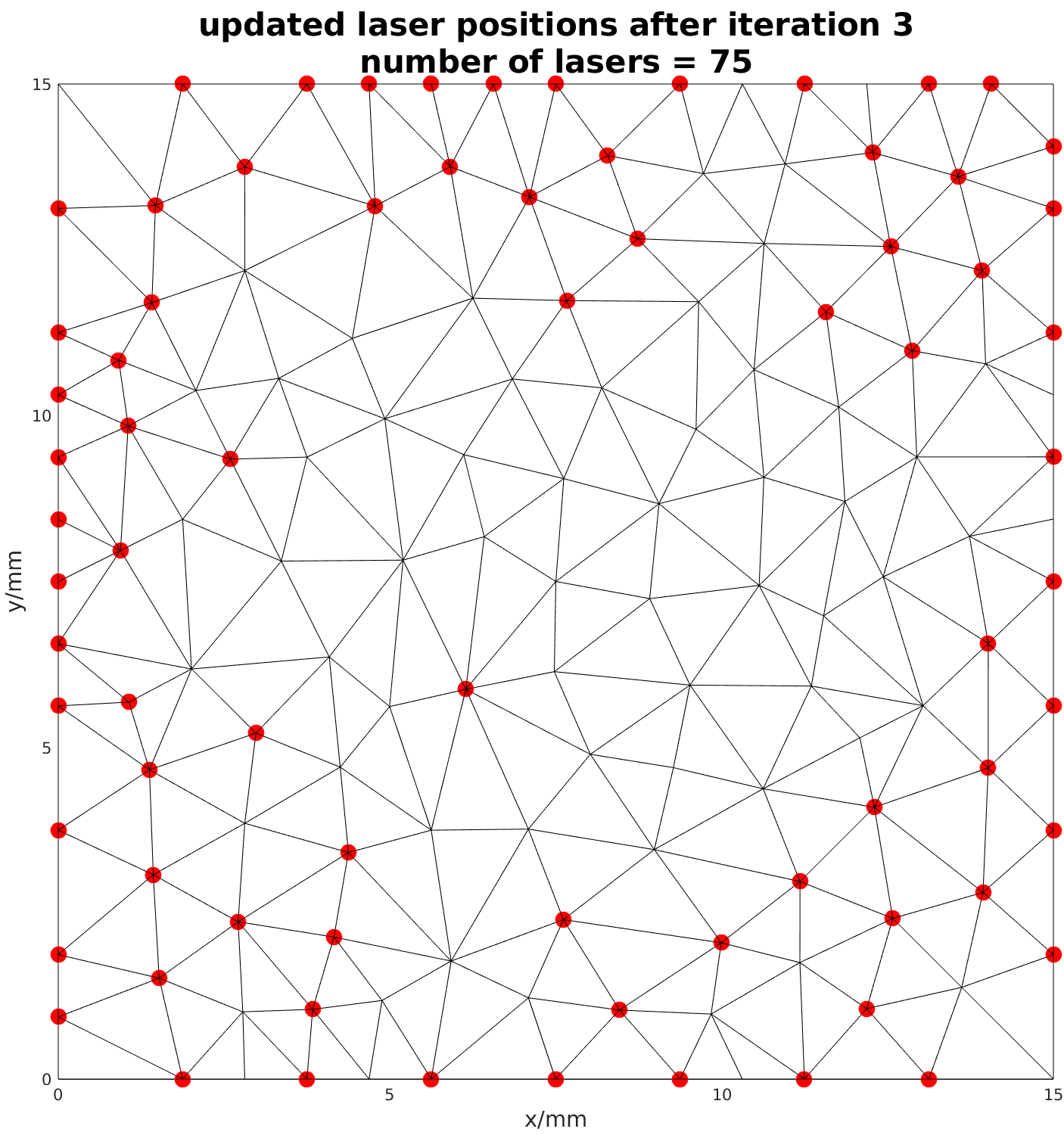}&
   \includegraphics[width=0.37\textwidth]{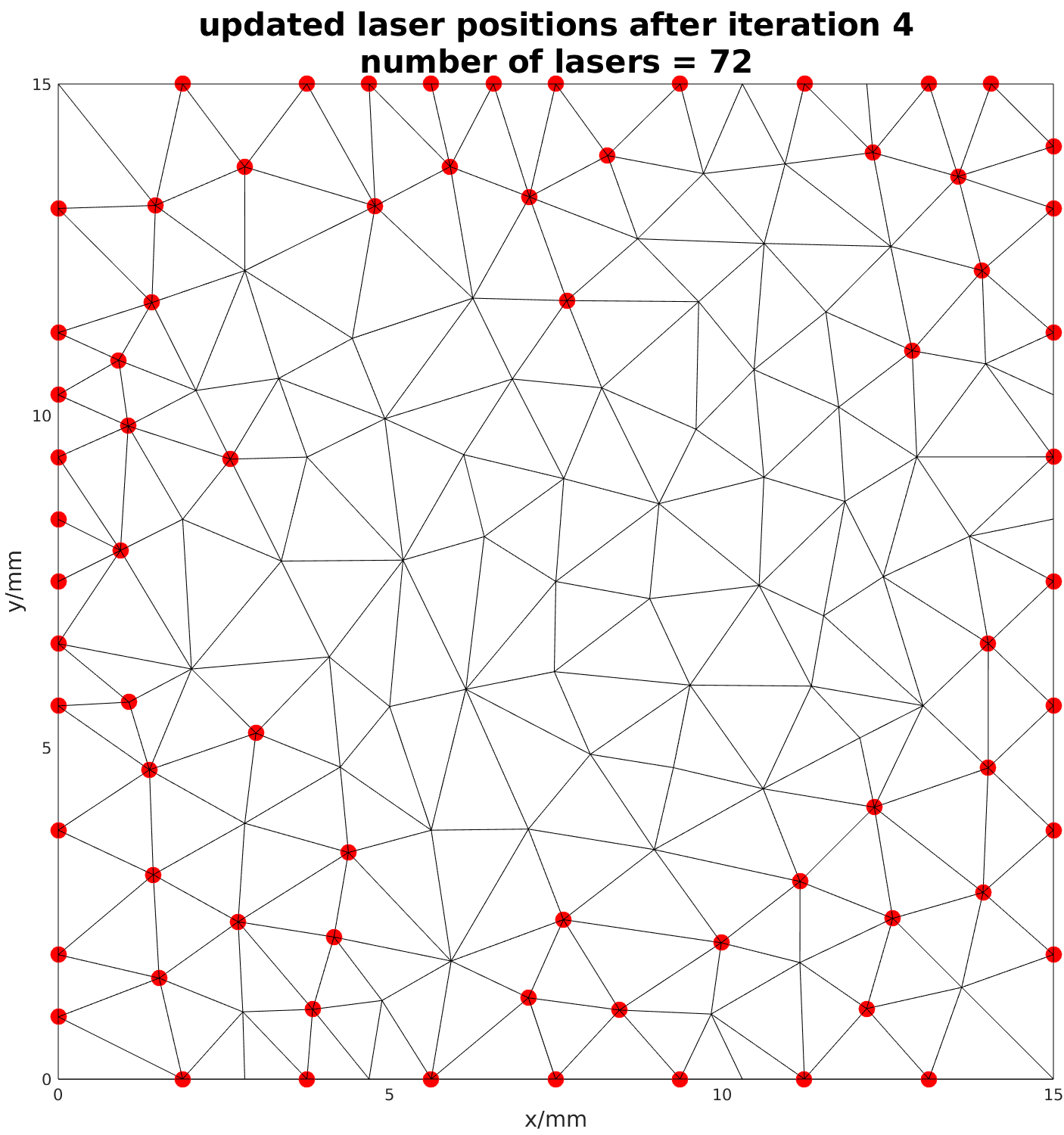}\\
   \includegraphics[width=0.37\textwidth]{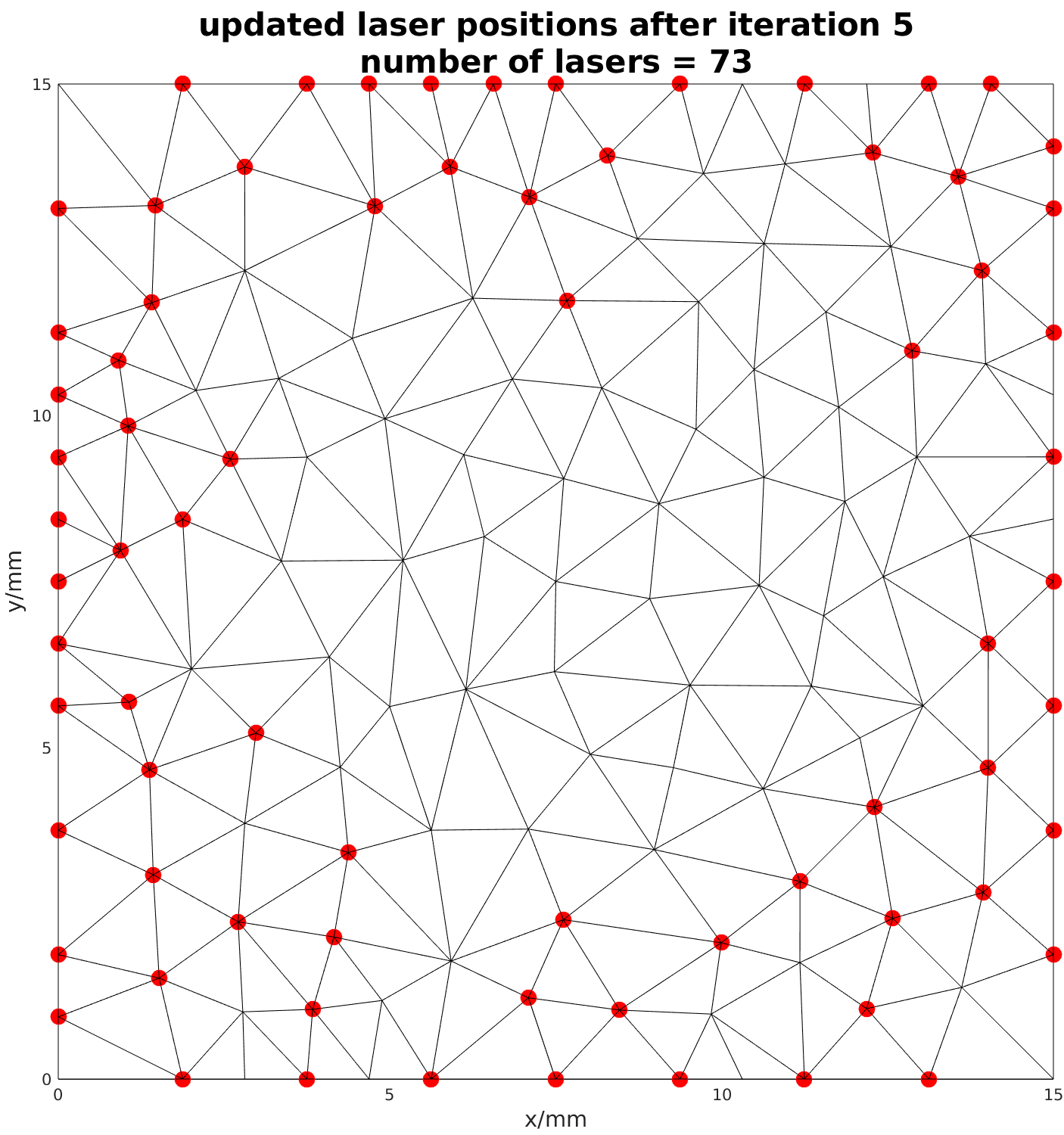}&
   \includegraphics[width=0.37\textwidth]{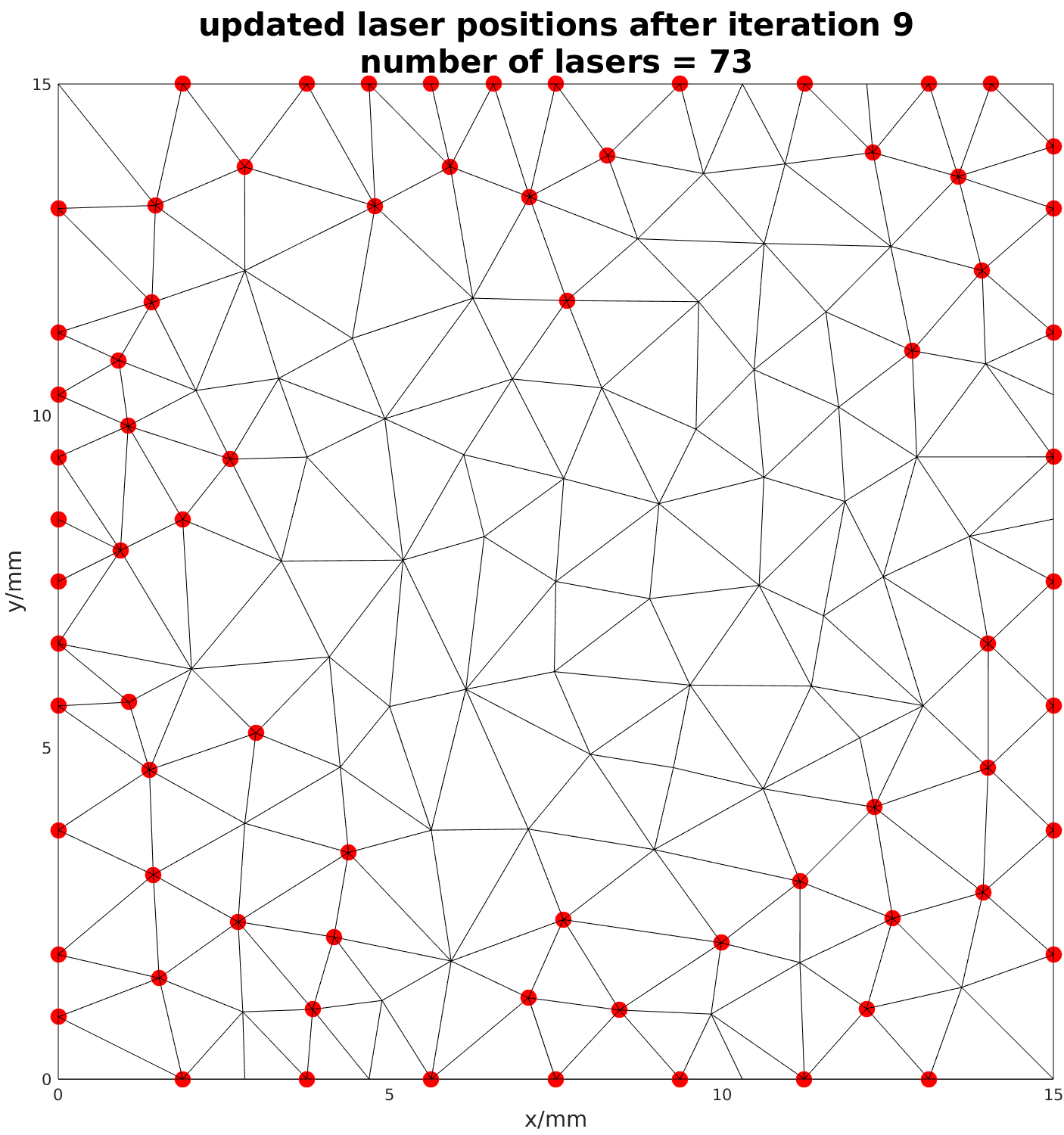}
  \end{tabular}
  \end{center}
\caption{Updated spatial distribution of lasers after round 1, 2, 3, 4, 5 and 9. Red points represent the lasers. Laser positions are attached to nodes of the finite element mesh on the top surface of the phantom. The number of lasers (illumination points) are indicated above each figure.}
\label{illupattern}
 \end{table}
 
 \begin{table}[h!]
 \setlength\tabcolsep{2pt}
  \begin{center}
   \begin{tabular}{cc}
    \includegraphics[width=0.5\textwidth]{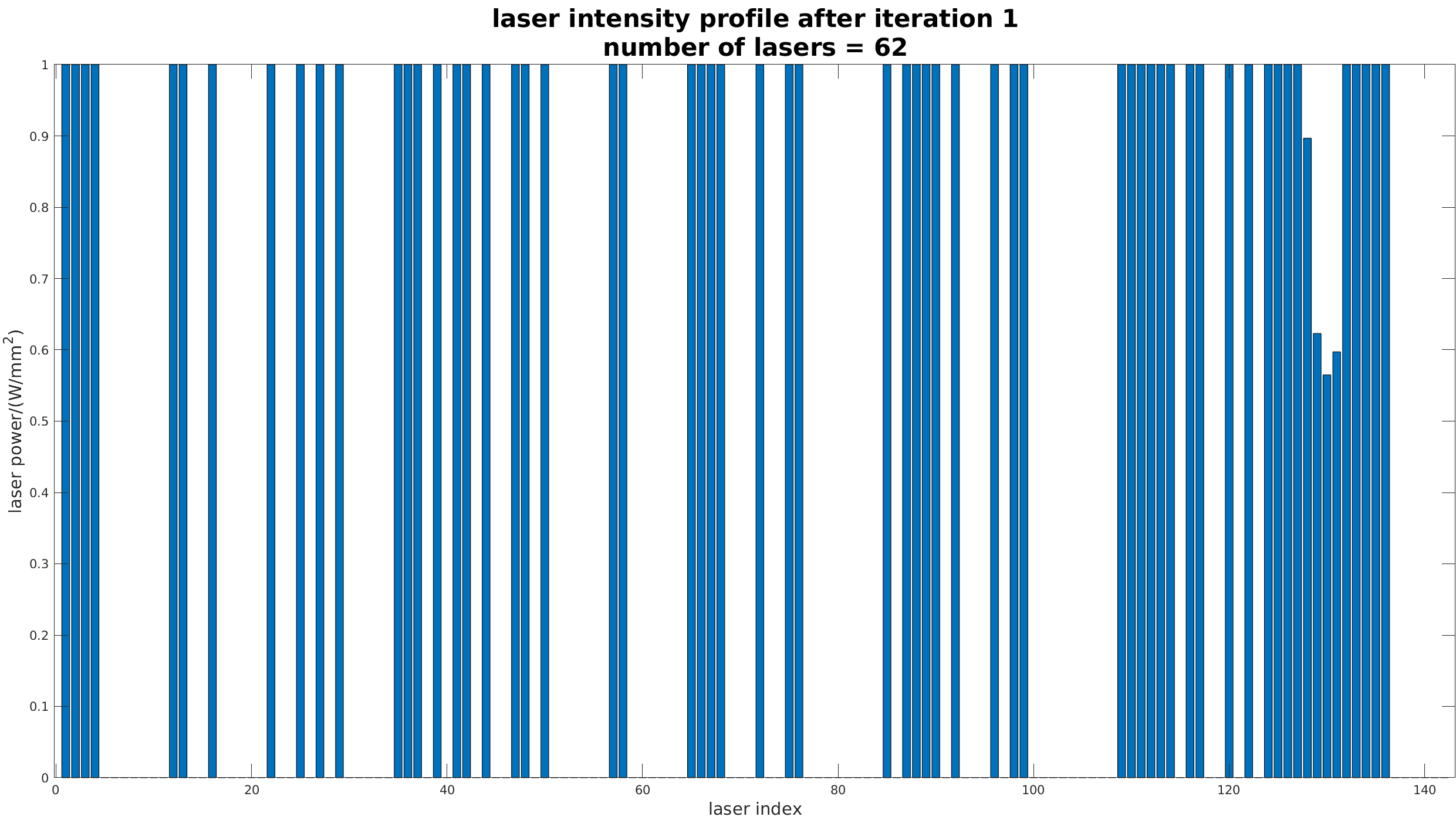}&
    \includegraphics[width=0.5\textwidth]{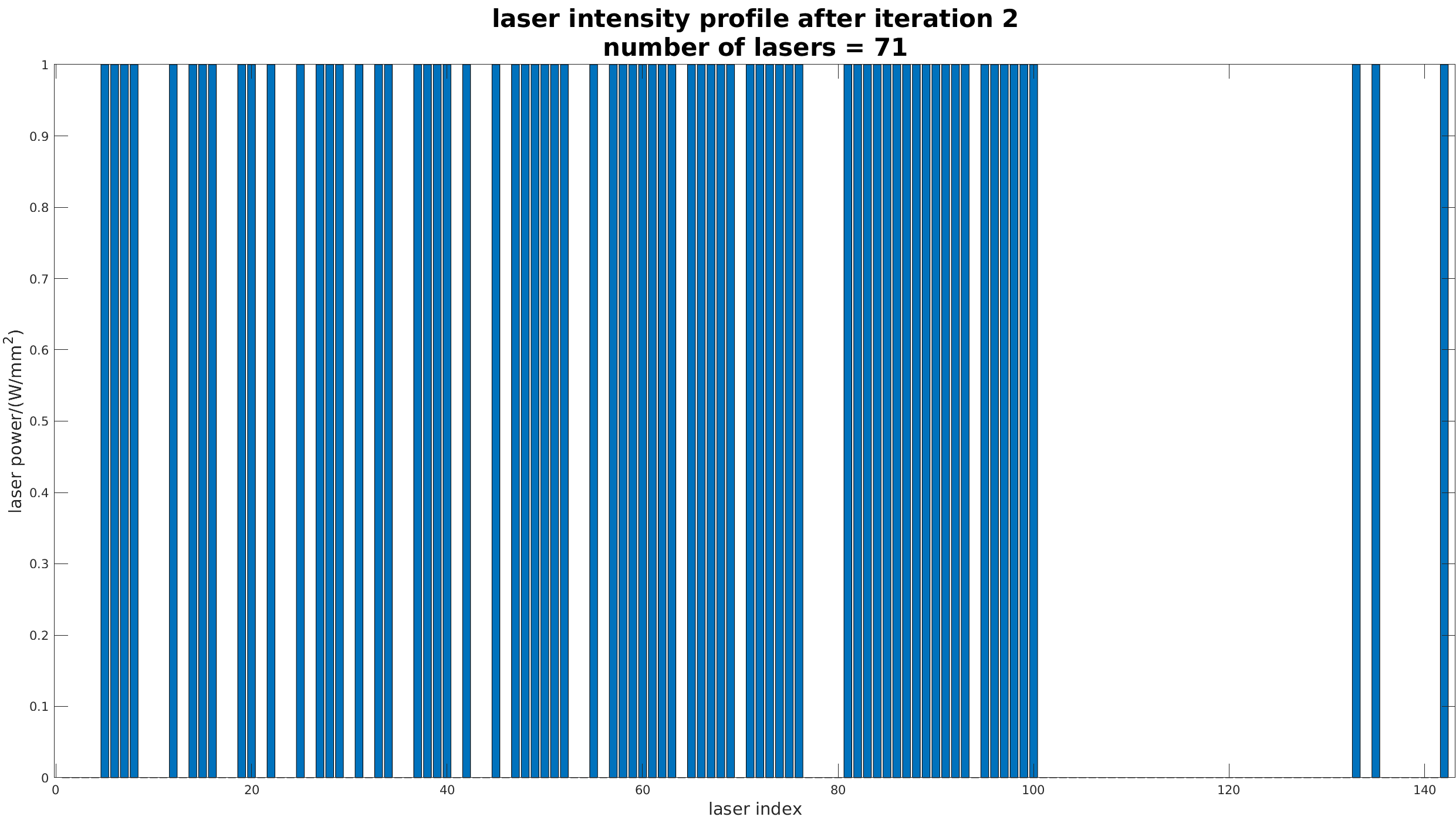}\\
    \includegraphics[width=0.5\textwidth]{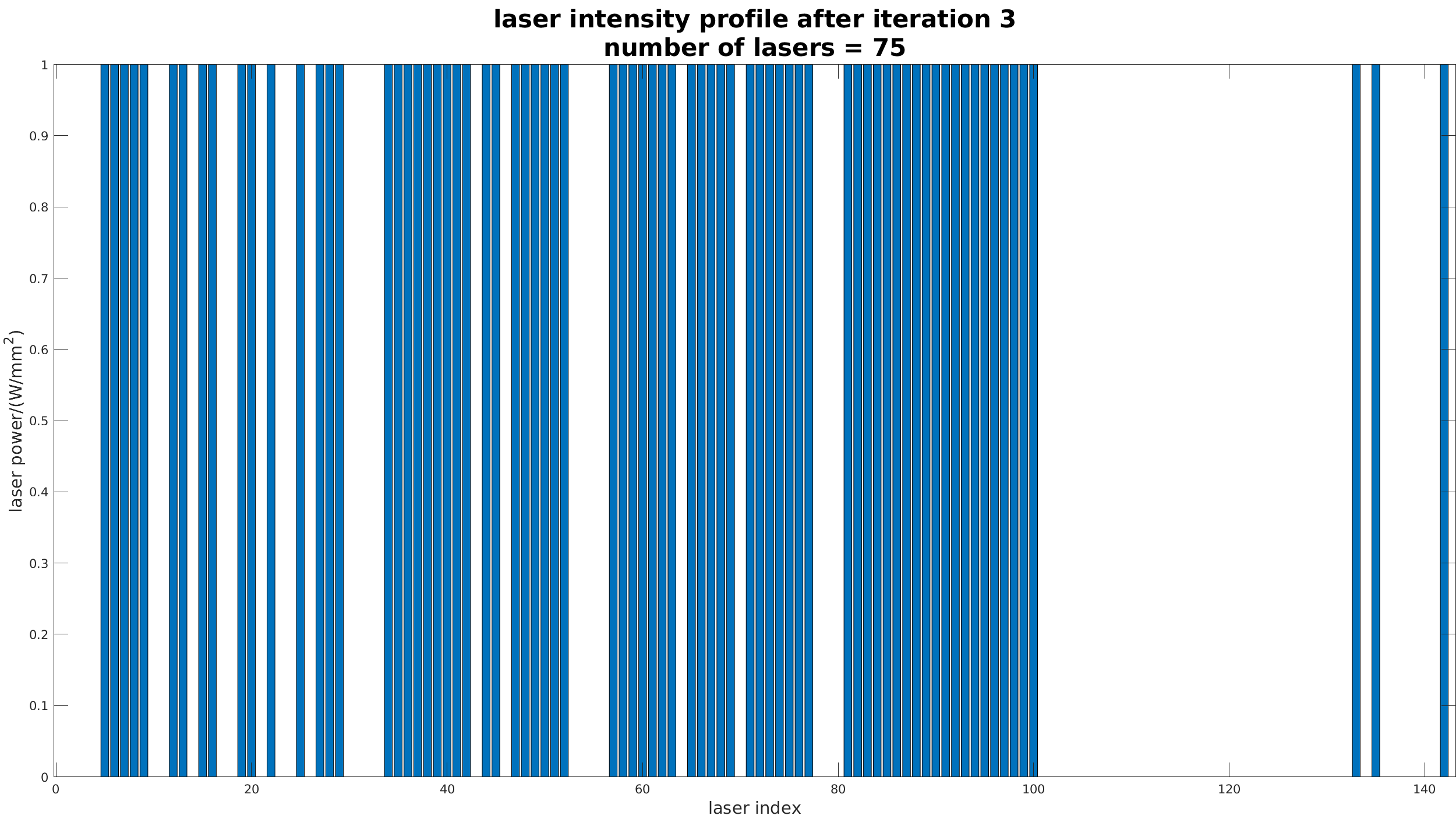}&
    \includegraphics[width=0.5\textwidth]{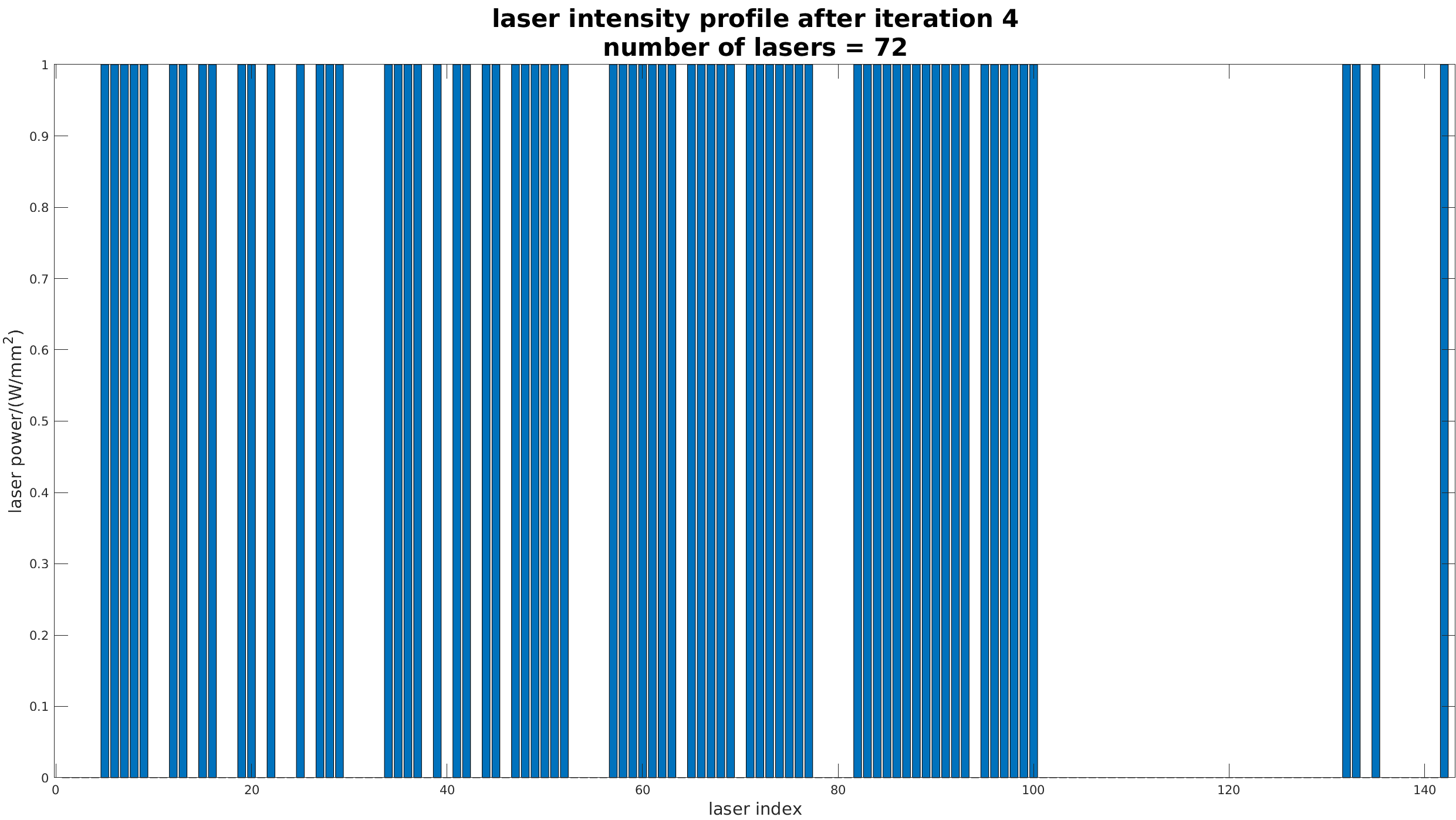}\\
    \includegraphics[width=0.5\textwidth]{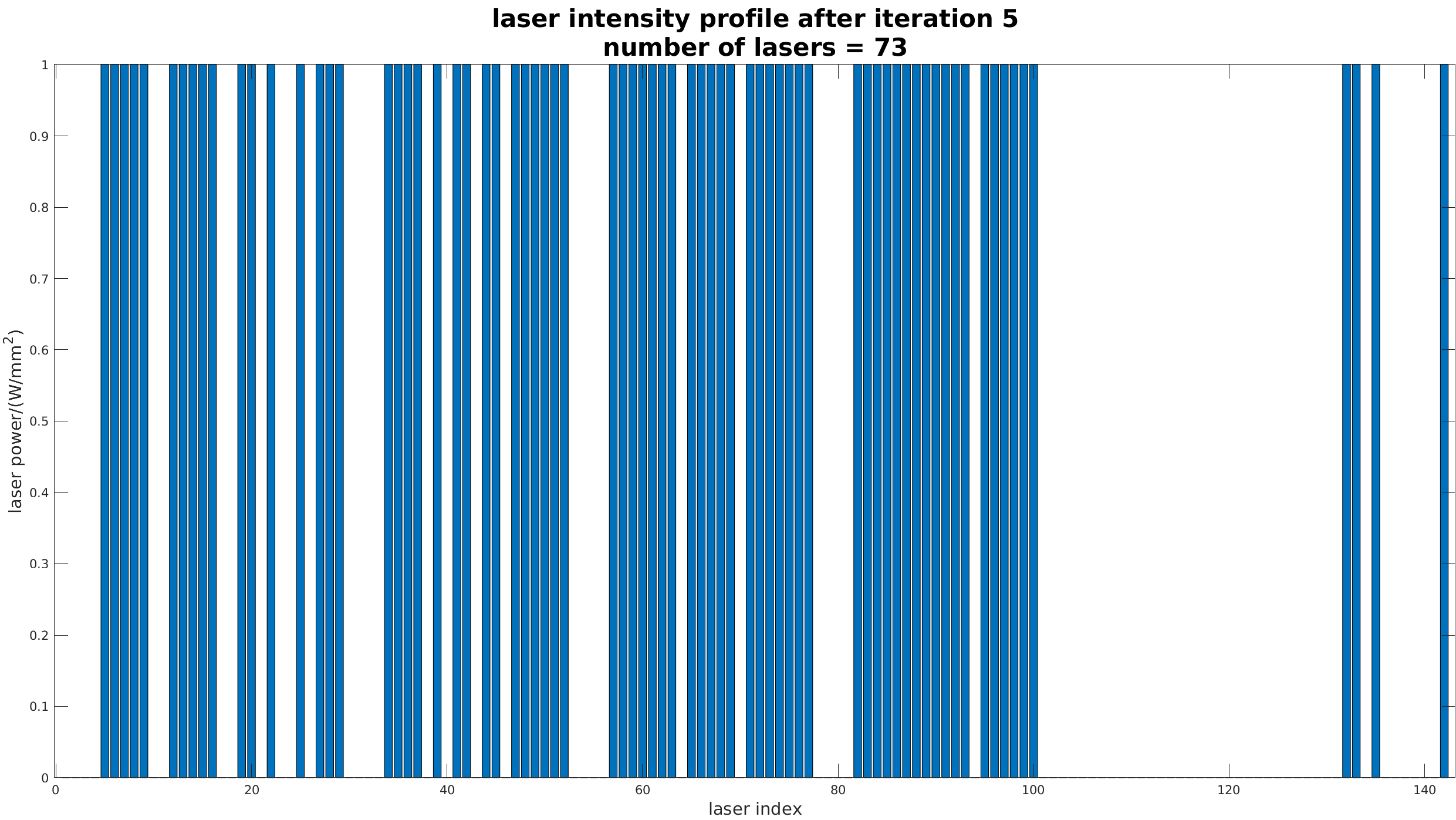}&
    \includegraphics[width=0.5\textwidth]{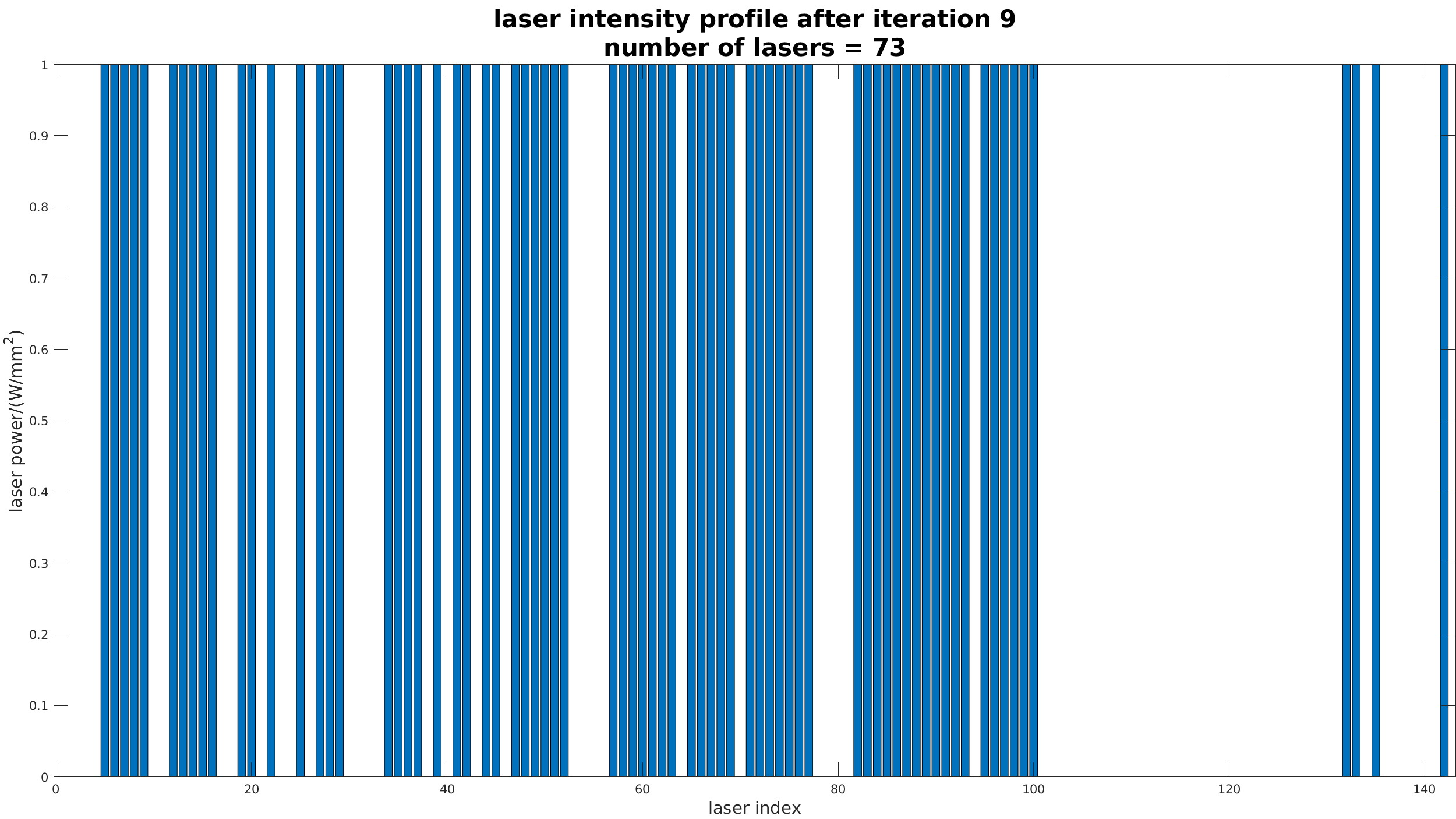}
   \end{tabular}
  \end{center}
\caption{Laser intensity profiles of each updated illumination pattern. First column from top to bottom: laser profile for 1st, 3rd, and 5th updated illumination patterns. Second column from top to bottom: laser profile for 2nd, 4rd, and 9th updated illumination pattern. On each panel the $x-$axis indicates the node number of lasers and the $y-$axis indicates the intensity ($W/mm^2$) of each laser. }
 \label{laserprofile}
 \end{table}

 It can be seen in Table \ref{illupattern} that with the value of $\mu=1.5\times 10^{-8}$, the number of lasers is restricted to approximately 70. After the first round of iteration, the illumination pattern is already very different from the initial spatial layout of the lasers, instead, it is shaping into the optimal illumination pattern as is shown in the following updated illumination patterns. It seems that most lasers are taking their positions near the edge of the top surface and only a few of them are intruding the center of the top surface. After the second round of iteration, both the number and the spatial distribution of the lasers already tend to stabilize with only a few laser points' position changed. After the 9th rounds of iteration the illumination pattern is still very stable. Besides, the laser intensity profiles demonstrate that after the second round of iteration, all lasers have reached maximal possible intensity 1$W/mm^2$, in other words, the laser intensity has a homogeneous distribution. One plausible explanation is that with the limited number of lasers available, all the lasers reach maximal value of intensity at their optimal position so that more useful information can be obtained. In some sense, the intensity of the lasers may not be the dominating factor compared to the spatial layout of lasers in terms of improving the reconstruction result.  
 \subsection{Result: reconstructed fluorescence distribution at each round} 
\begin{table}[h!]
\begin{center}
 \begin{tabular}{cccc}
   Illumination 0&\includegraphics[width=0.18\textwidth]{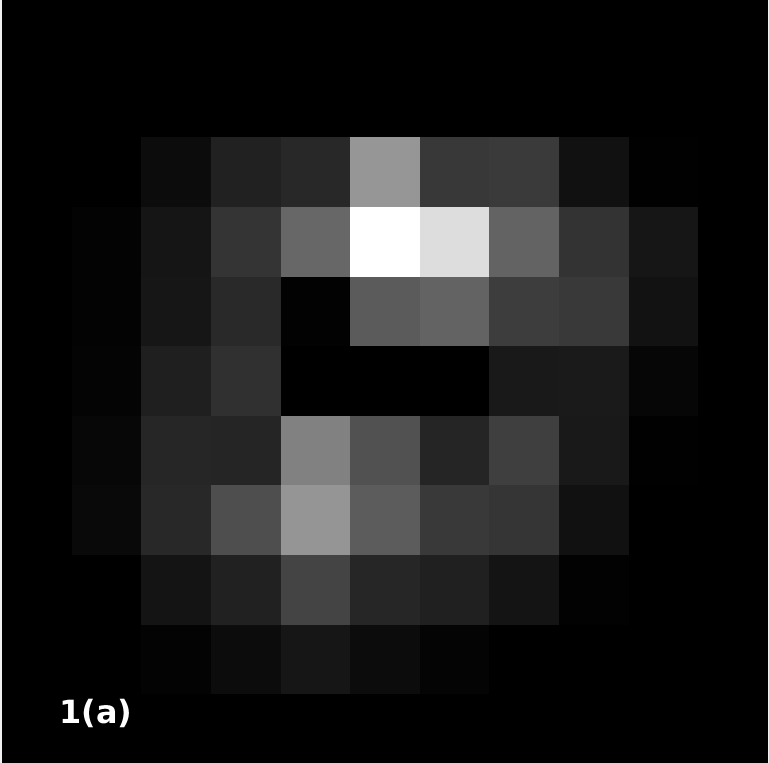}&
   \includegraphics[width=0.18\textwidth]{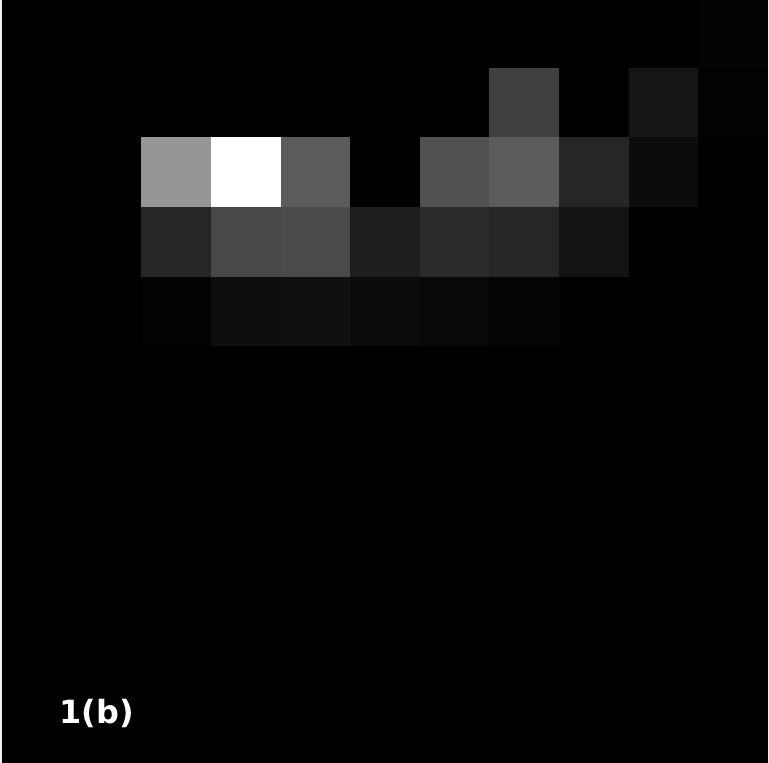}&
   \includegraphics[width=0.18\textwidth]{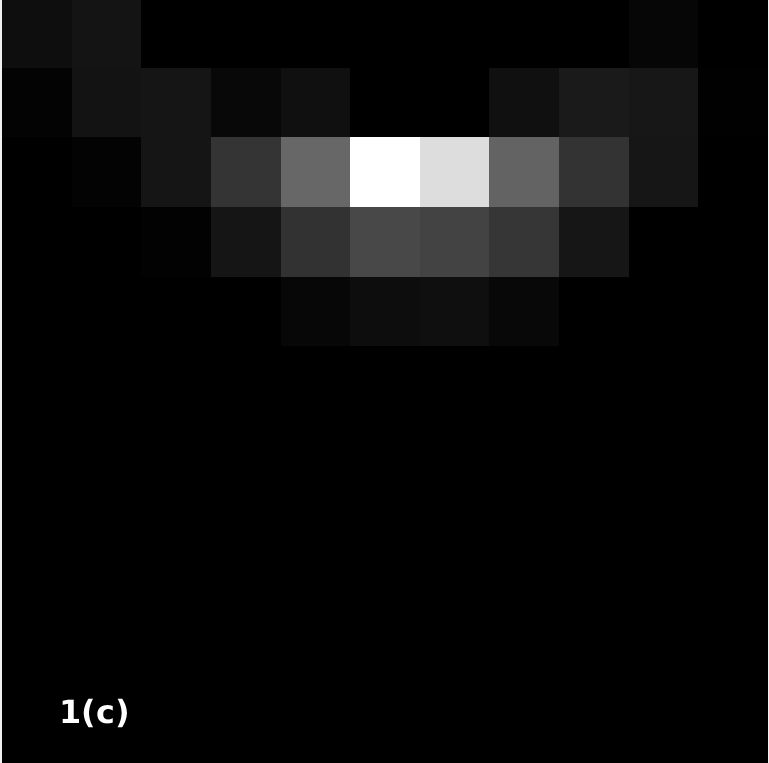}\\
   Illumination 1&\includegraphics[width=0.18\textwidth]{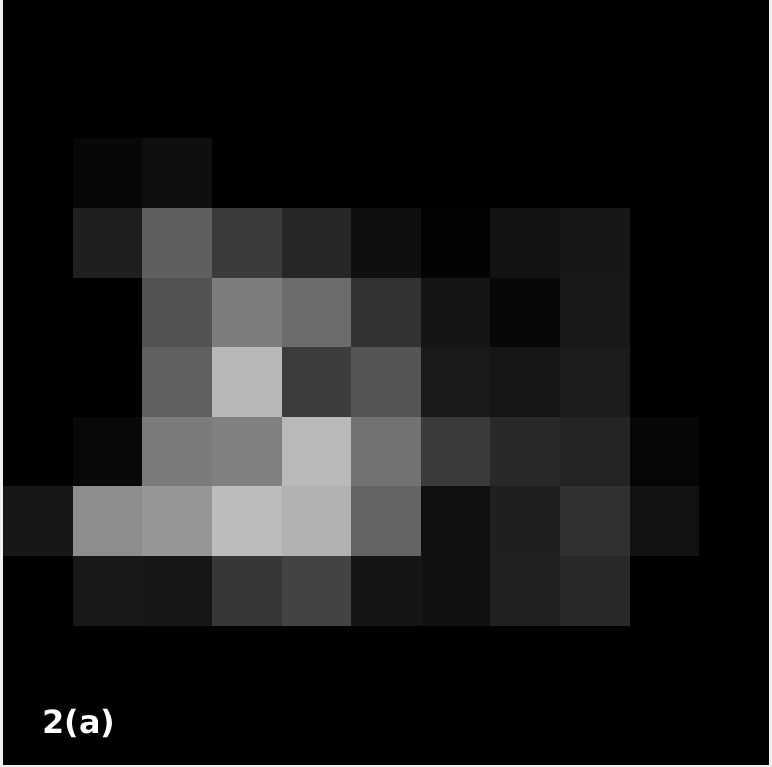}& 
   \includegraphics[width=0.18\textwidth]{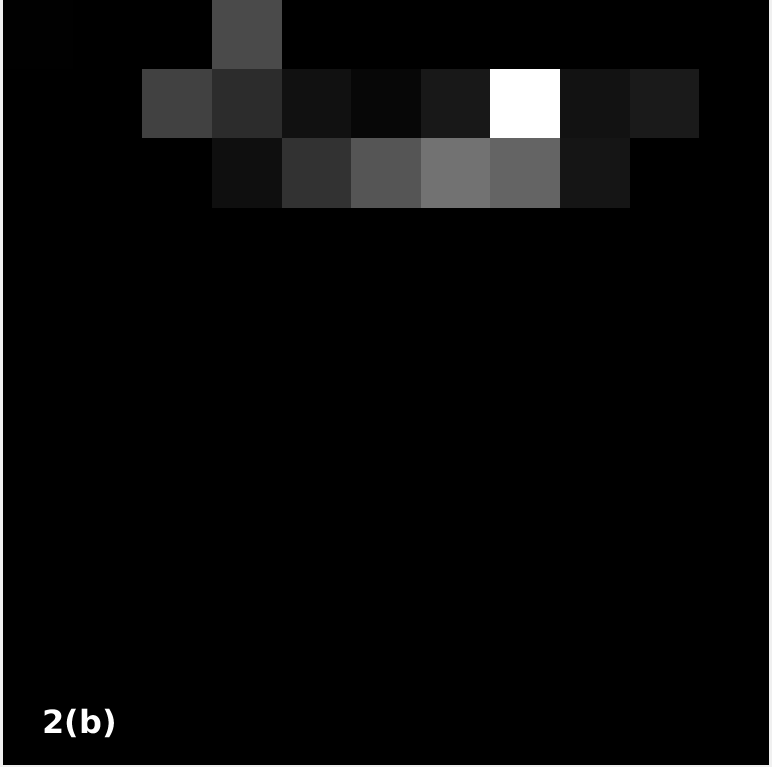}& 
   \includegraphics[width=0.18\textwidth]{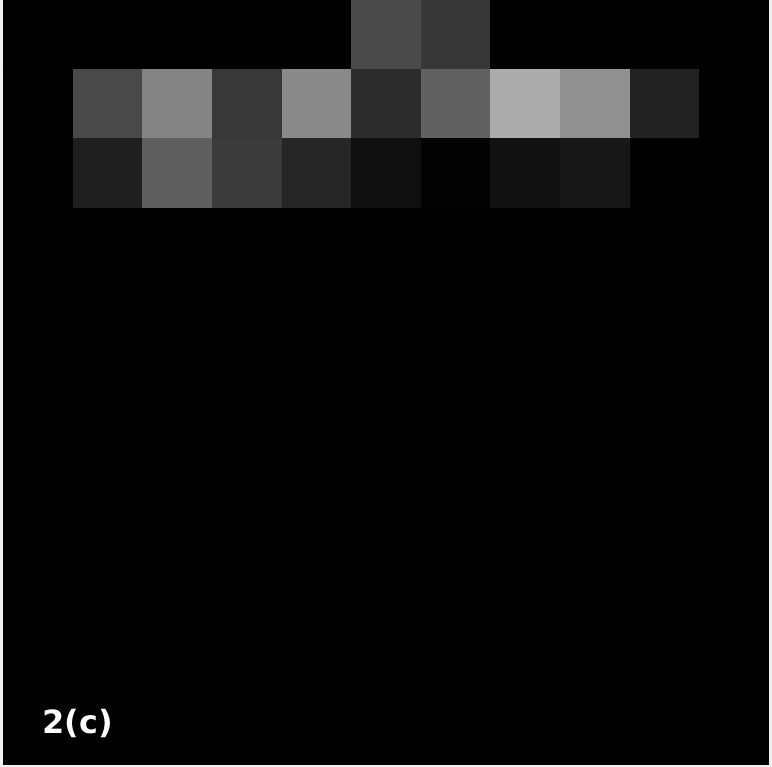}\\
   Illumination 2&\includegraphics[width=0.18\textwidth]{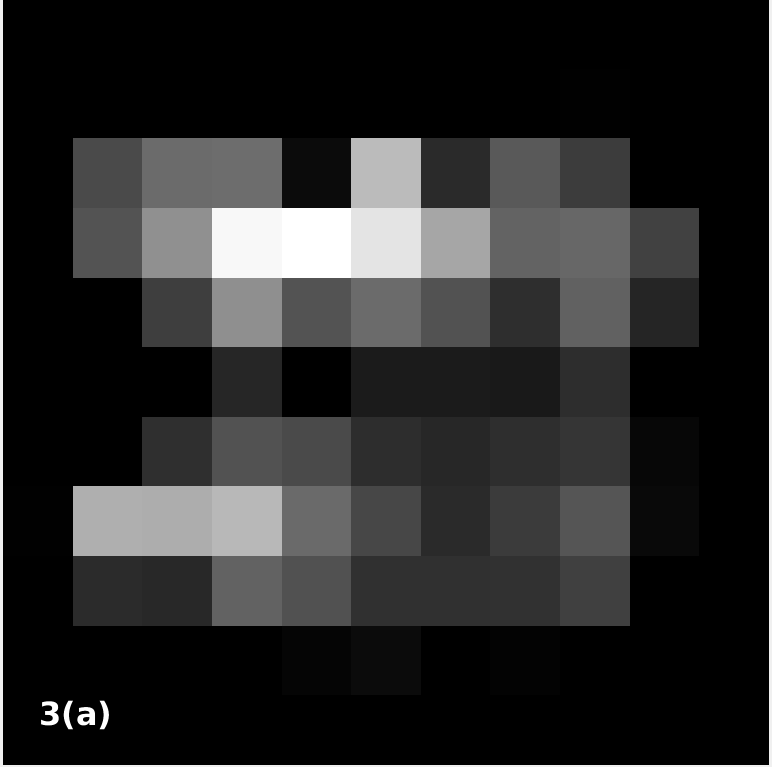}&
   \includegraphics[width=0.18\textwidth]{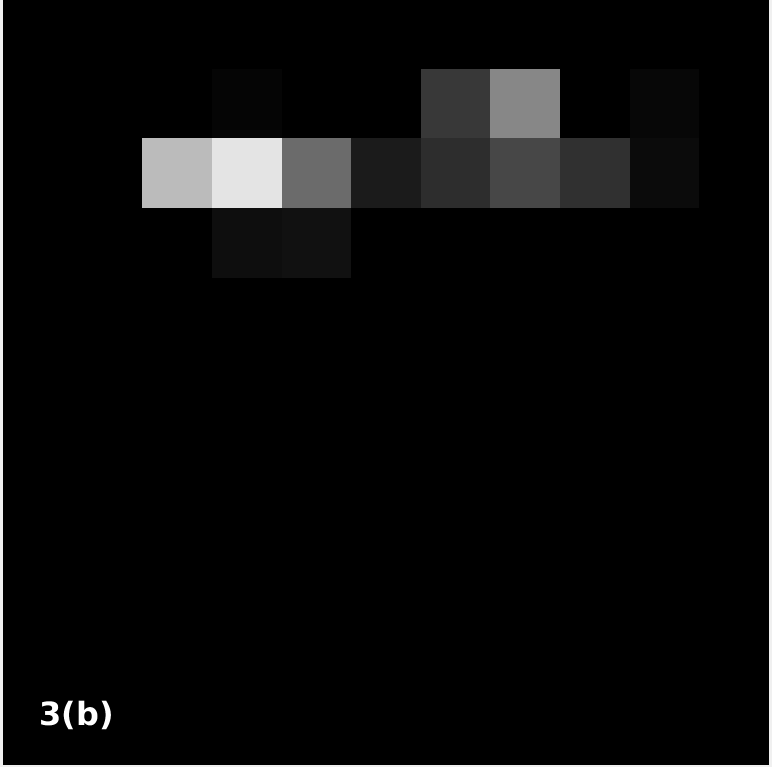}&
   \includegraphics[width=0.18\textwidth]{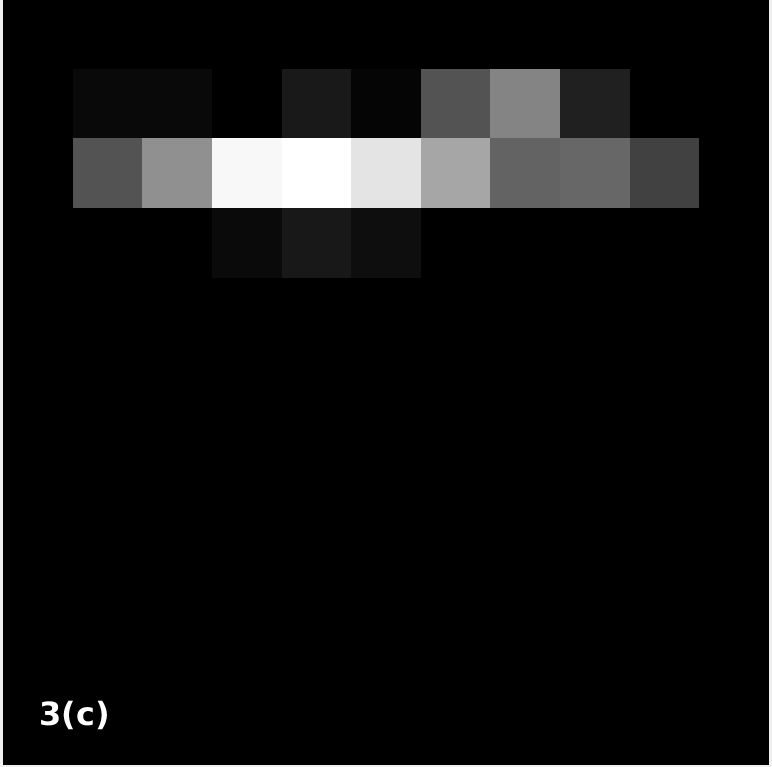}\\
   Illumination 3&\includegraphics[width=0.18\textwidth]{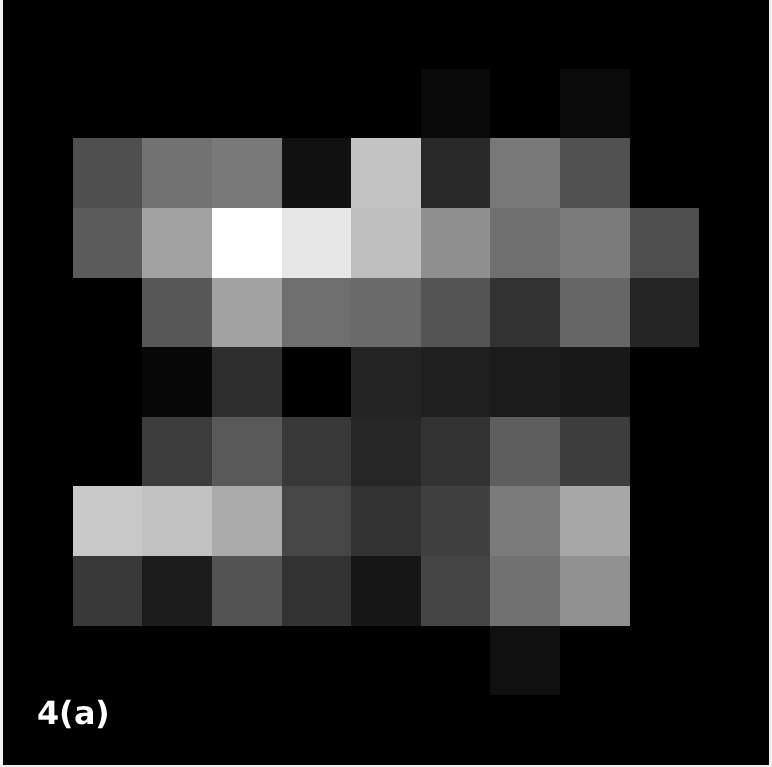}&
   \includegraphics[width=0.18\textwidth]{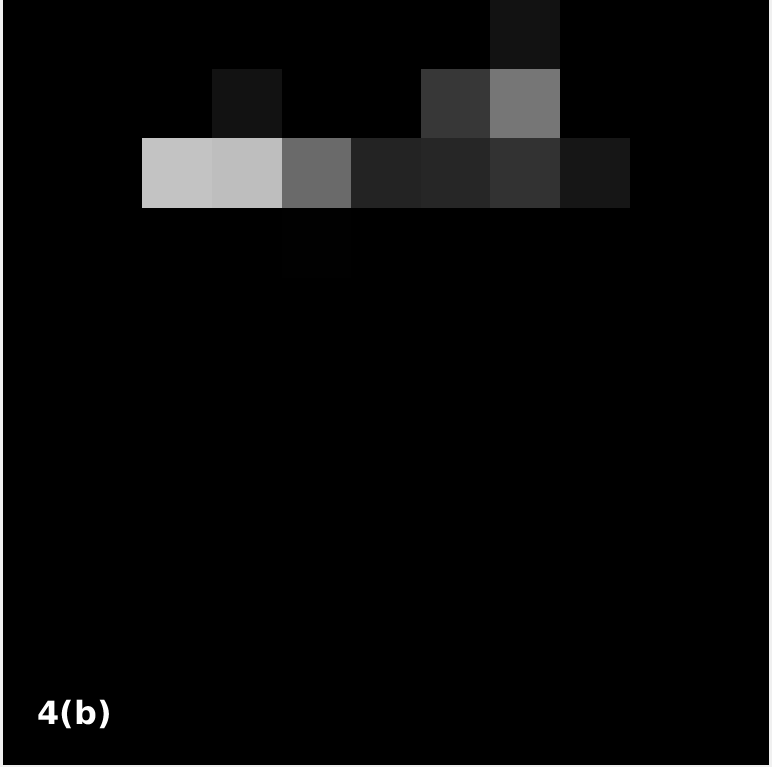}&
   \includegraphics[width=0.18\textwidth]{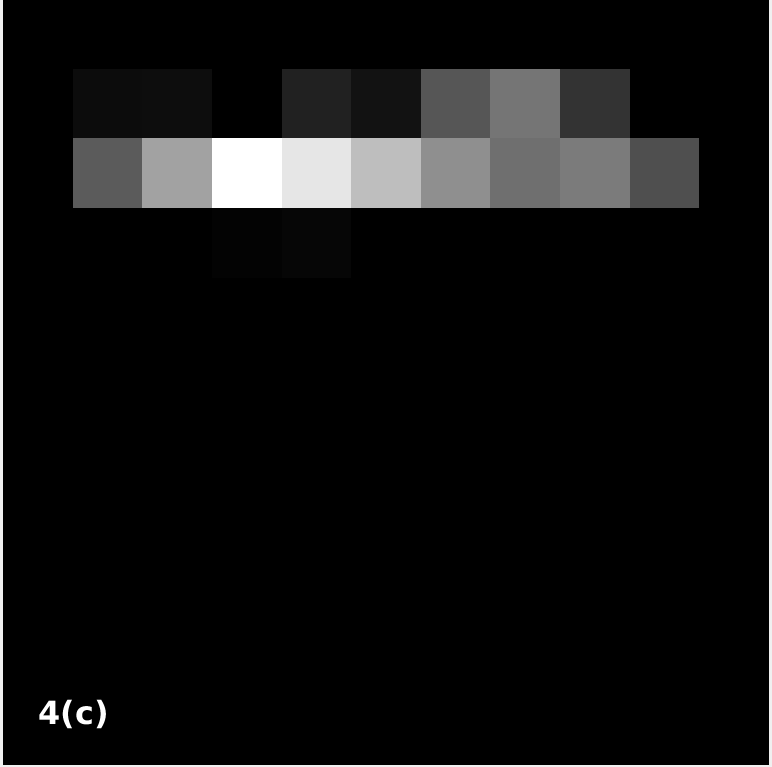}\\
   Illumination 4&\includegraphics[width=0.18\textwidth]{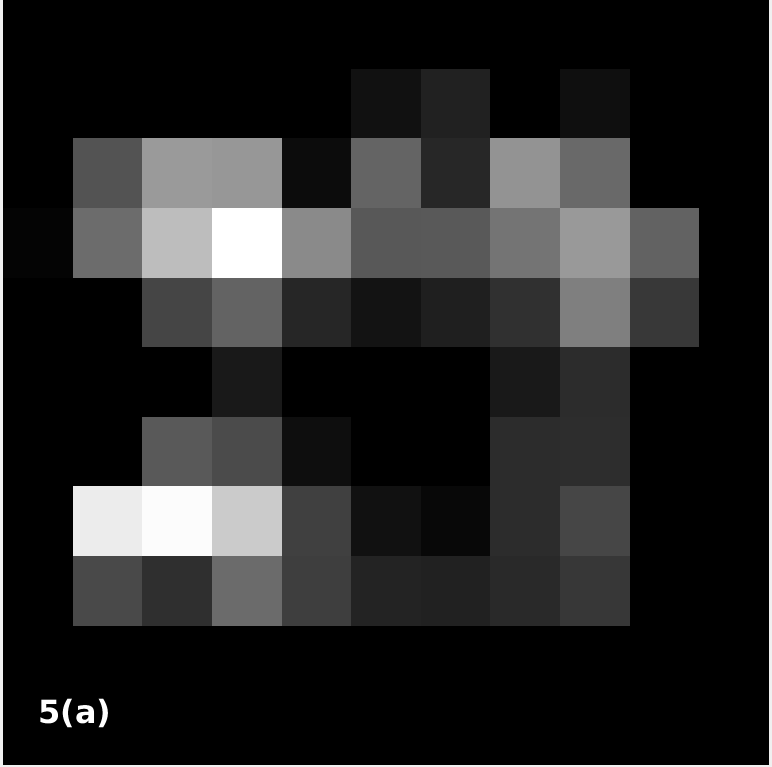}& 
   \includegraphics[width=0.18\textwidth]{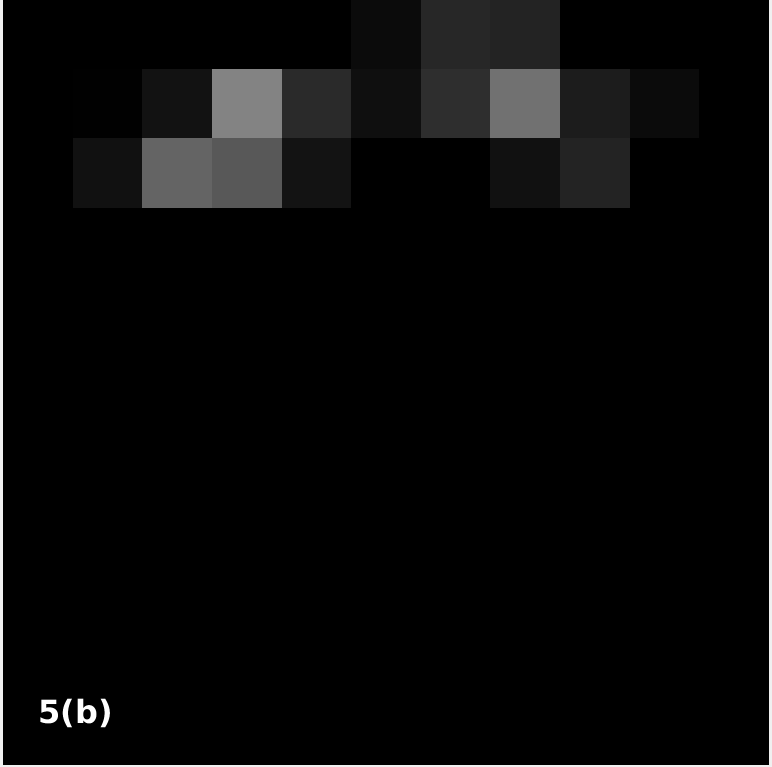}& 
   \includegraphics[width=0.18\textwidth]{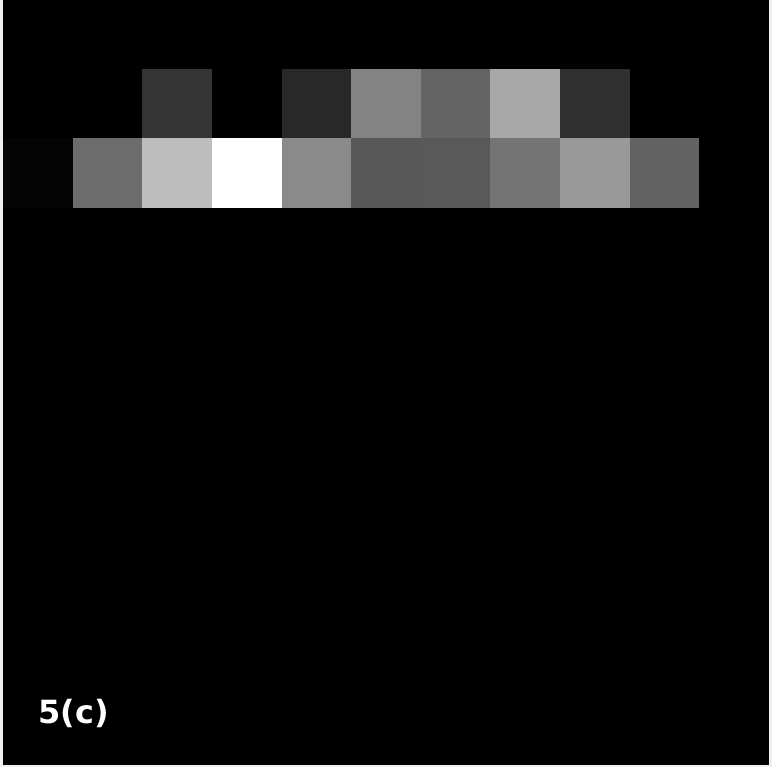}\\
   Illumination 5&\includegraphics[width=0.18\textwidth]{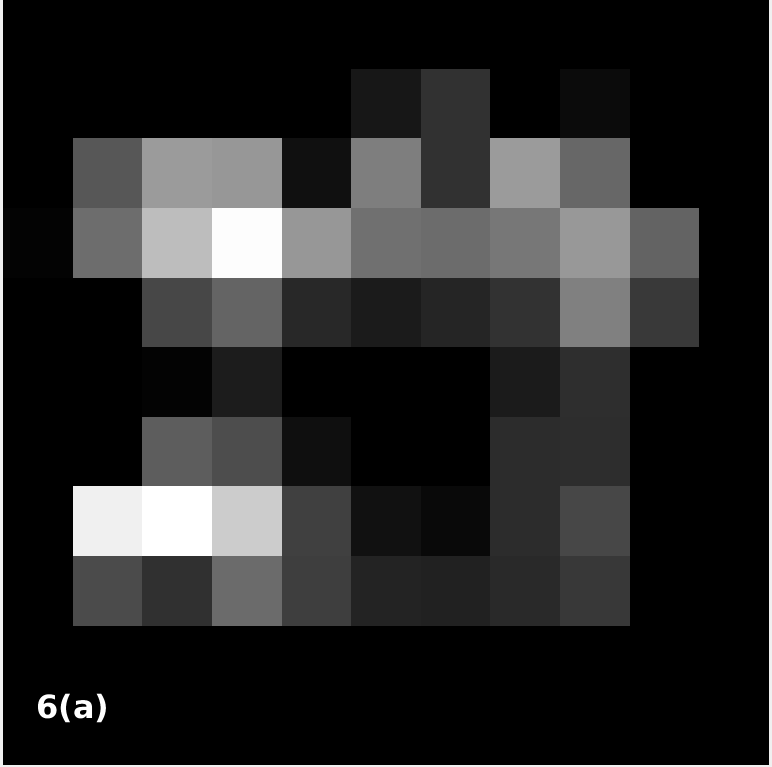}& 
   \includegraphics[width=0.18\textwidth]{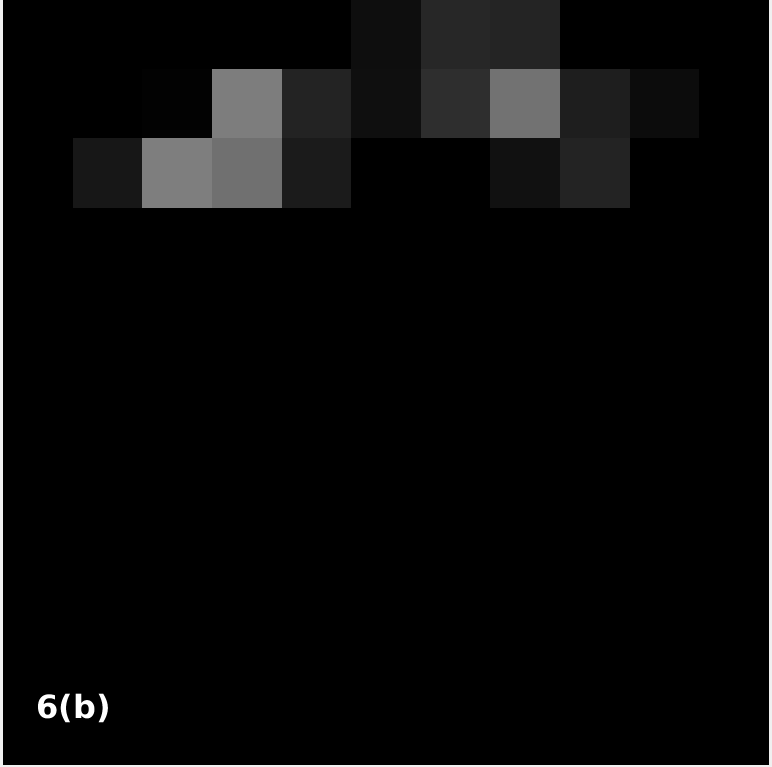}& 
   \includegraphics[width=0.18\textwidth]{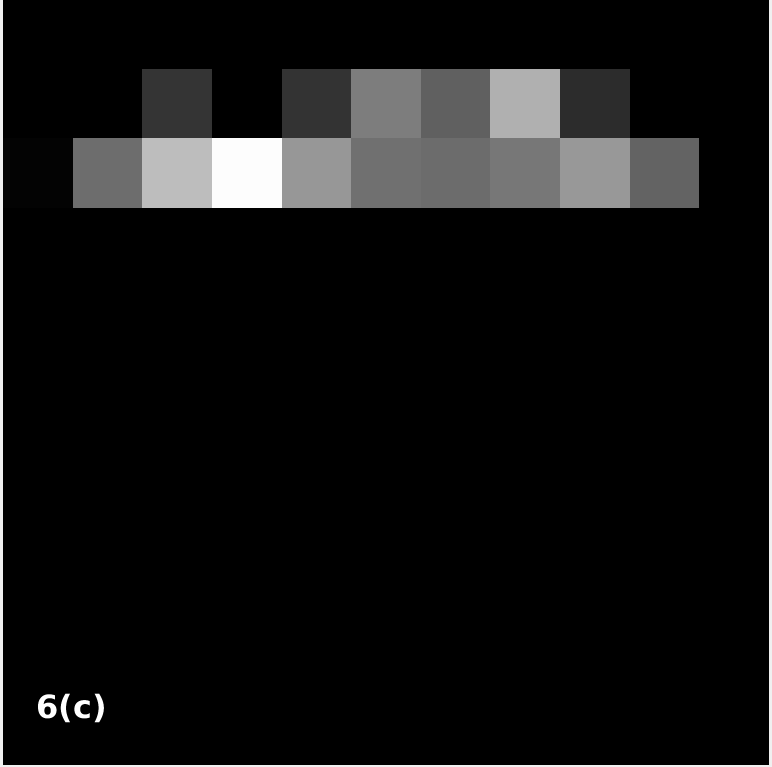}\\
   Illumination 9&\includegraphics[width=0.18\textwidth]{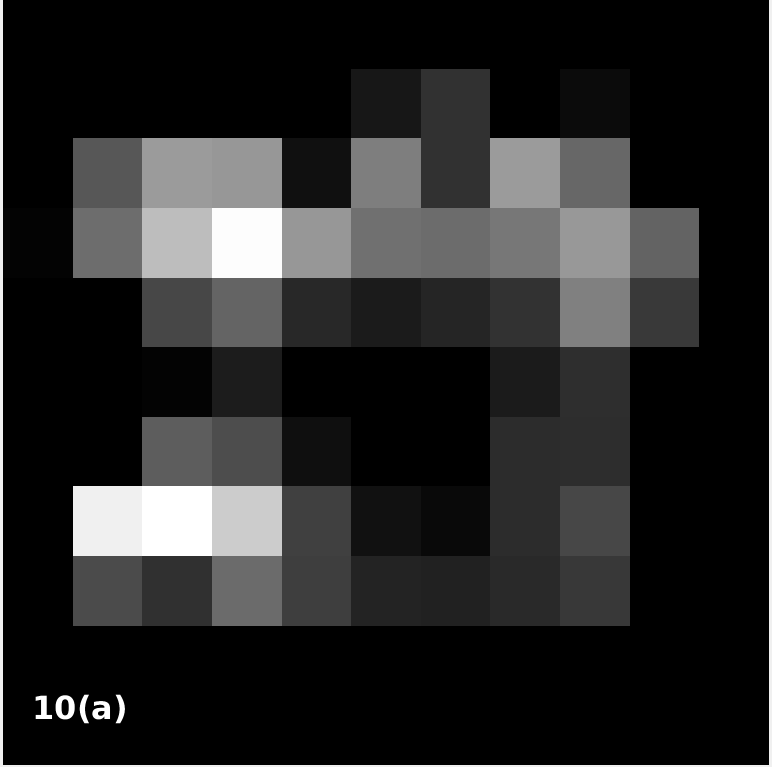}& 
   \includegraphics[width=0.18\textwidth]{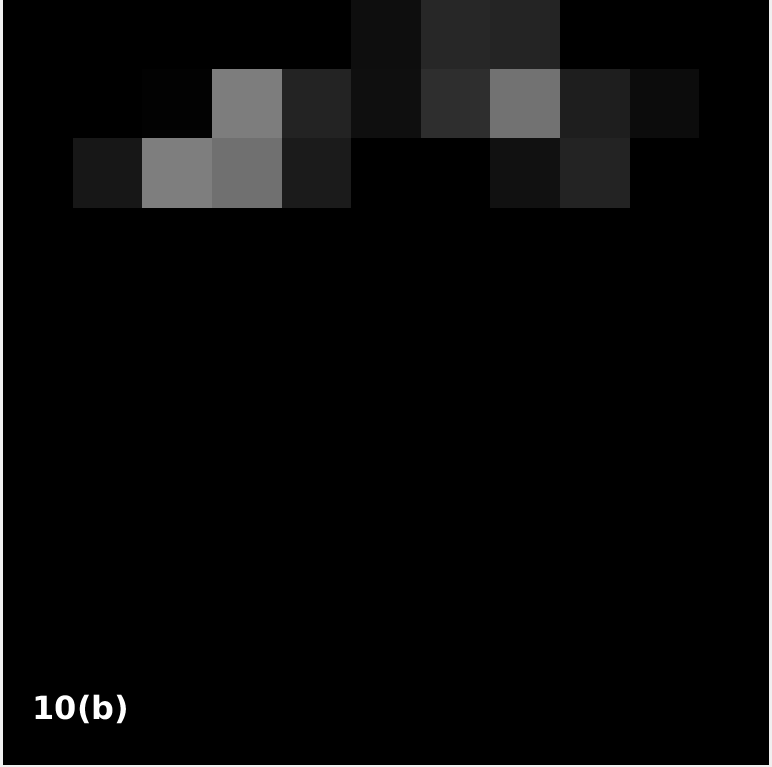}& 
   \includegraphics[width=0.18\textwidth]{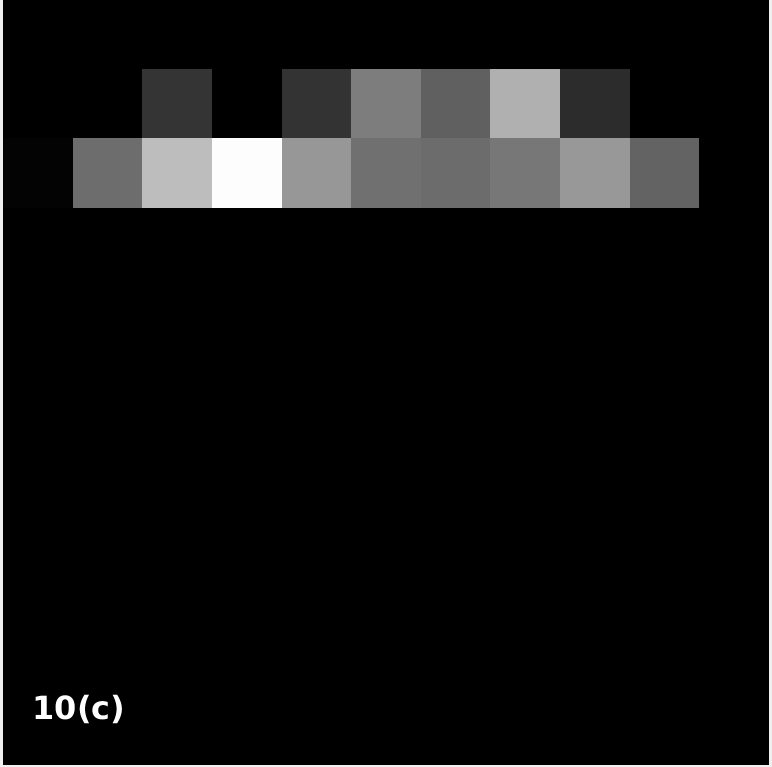}
   \end{tabular}
 \end{center}
 \caption{Reconstruction results sliced at the same reference point $(13,9,6)$ as the ground truth in Table \ref{fluo_gt}. The first row (Illumination 0) contains reconstruction results of the initial laser setting viewed from the top (a), left (b) and front (c) of the phantom. Row 2 to row 7 are reconstruction results from the 1st, 2nd, 3rd, 4th, 5th and 9th updated illumination pattern.}
 \label{reconimage}
\end{table}

It is easy to see that the reconstructed fluorescence distribution (first row) based on the initial laser setting loses information of the object's edges, especially in 1(a). The reason for this might be that in the initial laser setting all illumination points are densely distributed at the center of the top surface so that information about the edges can not be effectively captured. After the second round of iteration, with the illumination points moving towards the edges of the top surface, more information is obtained and the reconstruction result gets improved. This is seen from 2(b) and 2(c) where images are brighter than those in the first row. However, 2(a) is still problematic. Fortunately, as we run more rounds of updates, the quality of the reconstruction results drastically exceeds that of the initial laser setting. The shape, location and the concentration of the fluorescence distribution are well recovered. What's more, the reconstruction results are quite stable up to 10th round of iteration. This is in accordance with the updated illumination patterns in Table \ref{illupattern} where the illumination patterns tend to stabilize after 9 rounds of updates.
  
  \subsection{Error analysis}
  This section carries out quantitative error analysis of the reconstruction results to evaluate the quality of the reconstructed three-dimensional images. We first define the four metrics used for quantifying the error and then present the error analysis results for the experiment.
  \subsubsection{Metrics for error analysis}
  We employ four most common metrics for image processing: mean square error (MSE), Dice similarity (Dice), volume ratio (VR) \cite{metrics} and signal to noise ratio (SNR) \cite{GonzalezRafaelC2008Dip}. Suppose image vector $x$ is the reconstructed image  and $x^*$ is the ground truth of the same dimension as $x$. In case $x$ and $x^*$ are two-dimensional or three-dimension image matrices, serialize the image matrix into a vector first. Denote $N$ the total number of entries of $x$. Define the region of interest (ROI) the set of entries that have intensity greater than one third of the maximal intensity of the image $x$. In other words,
  \begin{displaymath}
 \textrm{ROI}(x)=\{x_i:x_i>\frac{1}{3}\max\{x\}, i=1,\ldots, N\}.
\end{displaymath}
The four metrics are defined as follows:
\begin{itemize}
 \item The mean square error (MSE) describes the general deviation of the reconstruction result from the ground truth:
 \begin{equation*}
  \textrm{MSE}=\frac{1}{N}\sum_{i=1}^N(x_i-x^*_i)^2.
 \end{equation*}
 The smaller the MSE, the better the reconstruction result.
 \item Dice similarity (Dice) between the reconstructed ROI and the true ROI measures the shape and location accuracy of the reconstructed image:
 \begin{equation*}
  \textrm{Dice}=\frac{2|\textrm{ROI}(x)\cap\textrm{ROI}(x^*)|}{|\textrm{ROI}(x)|+|\textrm{ROI}(x^*)|}.
 \end{equation*}
 Here, $|\cdot|$  denotes the cardinality of a set. Note that Dice ranges from 0 to 1.
 Hence, If Dice is close to 1, the reconstructed image is well overlapping with the ground truth; otherwise if Dice is close to 0, the reconstructed image is not recovering the true image at the correct location or shape.
 \item Volume Ratio (VR) measures the ratio between cardinality of the true region of interest (ROI) and that of the reconstructed ROI:
 \begin{displaymath}
  \textrm{VR}=\frac{|\textrm{ROI}(x)|}{|\textrm{ROI}(x^*)|}.
 \end{displaymath}
In ideal conditions, VR should be 1. If VR is smaller than 1, it indicates that the reconstructed image is oversparsified; if VR is greater than 1, the reconstruction is diffuse.
 \item Signal to Noise Ratio (SNR) expressed in decibels (dB) measures how well the reconstructed image is distinguished from the background noise:
 \begin{equation*}
  \text{SNR}(x)=10\log_{10}\left\{\frac{\sum_{i=1}^N(x^*_i)^2}{\sum_{i=1}^N(x_i-x^*)^2}\right\}.
 \end{equation*}
 The higher the SNR the better image we have produced.
\end{itemize}

\subsubsection{Reconstruction error at each round}
Using the metrics defined in the last section, the error of the reconstruction results from the 7 rounds of iterations are displayed in Table \ref{errorana}.
  \begin{table}[h!]
  \begin{center}
   \begin{tabular}{cc}
    \includegraphics[width=0.49\textwidth]{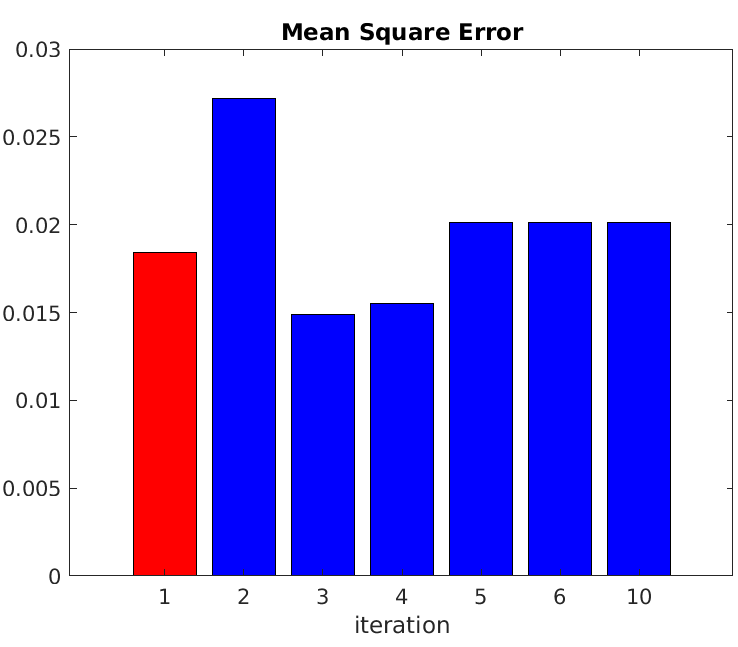}&
    \includegraphics[width=0.48\textwidth]{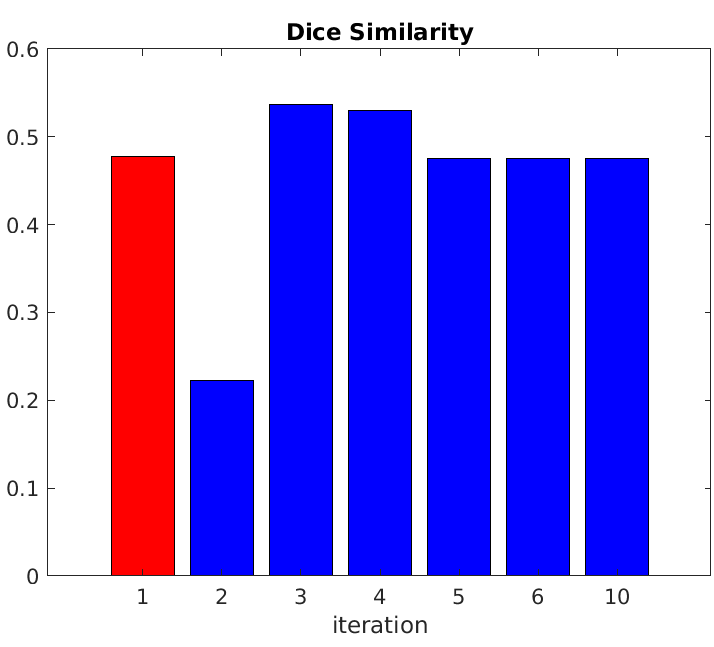}\\
    \includegraphics[width=0.48\textwidth]{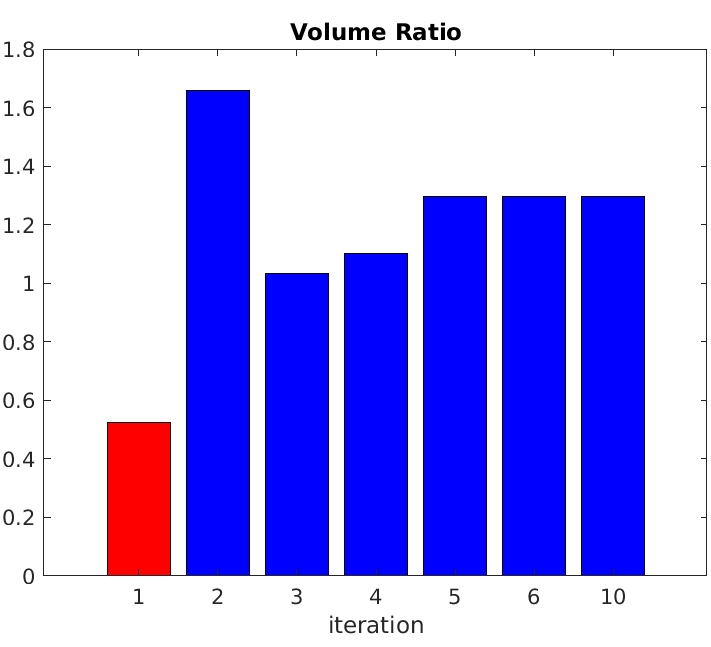}&
    \includegraphics[width=0.48\textwidth]{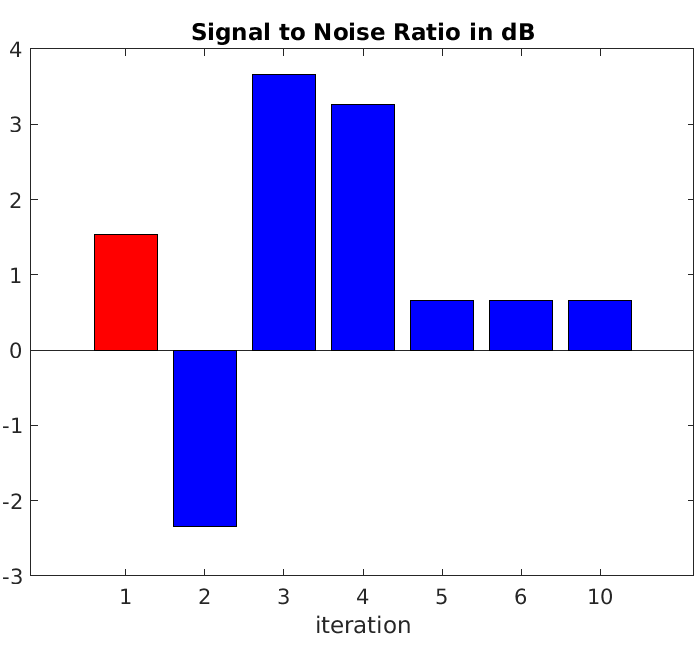}
   \end{tabular}
   \end{center}
   \caption{Mean square error, Dice similarity, volume ratio and signal to noise ratio of the reconstruction results at round 1-6 and round 10. The error of the reconstruction result based on the initial laser setting, i.e., the first round of iteration is indicated by the red bars. The blue bars represent the error of following rounds of iterations that are based on updated illumination patterns.}
   \label{errorana}
  \end{table}
  
According to the definition of the four metrics, the reconstruction result at the third round has the best quality since it has the lowest MSE, highest Dice similarity and SNR, and an almost perfect volume ratio ($\approx$ 1). It is safe to say that the reconstruction result is indeed improved after merely two rounds of iterations.  

It is interesting to notice that the reconstruction result produced by the first updated illumination pattern is actually worse than that produced by the initial laser setting with a fairly low Dice similarity and a negative SNR. Negative SNR means that the signal is drowning in the noise. However, after updating the illumination pattern again based on this bad reconstruction result, the produced reconstruction result enjoys a big improvement in quality. In the following fourth and fifth rounds, the reconstruction quality is slightly declining. Hence, the best reconstructed result is achieved after two rounds of experiments. As we increase the number of iteration rounds the quality of the reconstruction result becomes stable.

\subsection{Uniqueness of the optimal illumination pattern}
Now that we have achieved the optimal illumination pattern for our experiment, it is natural to ask if the optimal illumination pattern depends on the initial illumination pattern. To be precise, using the same phantom as in Table \ref{phantom}, the same fluorescence ground truth as in Table \ref{fluo_gt} and the same regularization weight $\mu=1.5\times10^{-8}$, does the optimal illumination pattern change as the initial illumination pattern varies? If the optimal illumination pattern does not depend on the initial laser setting, it is  unique. We perform another set of experiments starting from a different initial laser pattern to find out if the updated illumination patterns converge to the optimal one from the previous experiment. 
\subsubsection{A different initial laser setting}
Without loss of generality, the initial illumination pattern is still the 10 by 10 laser array but with the initial intensity of each laser changed to half the intensity of the previous experiment namely $0.5W/mm^2$. In addition, the center of the laser array is moved to the bottom left corner of the top surface of the phantom: 
\begin{table}[!h]
\begin{center}
 \begin{tabular}{c}
  \includegraphics[width=0.45\textwidth]{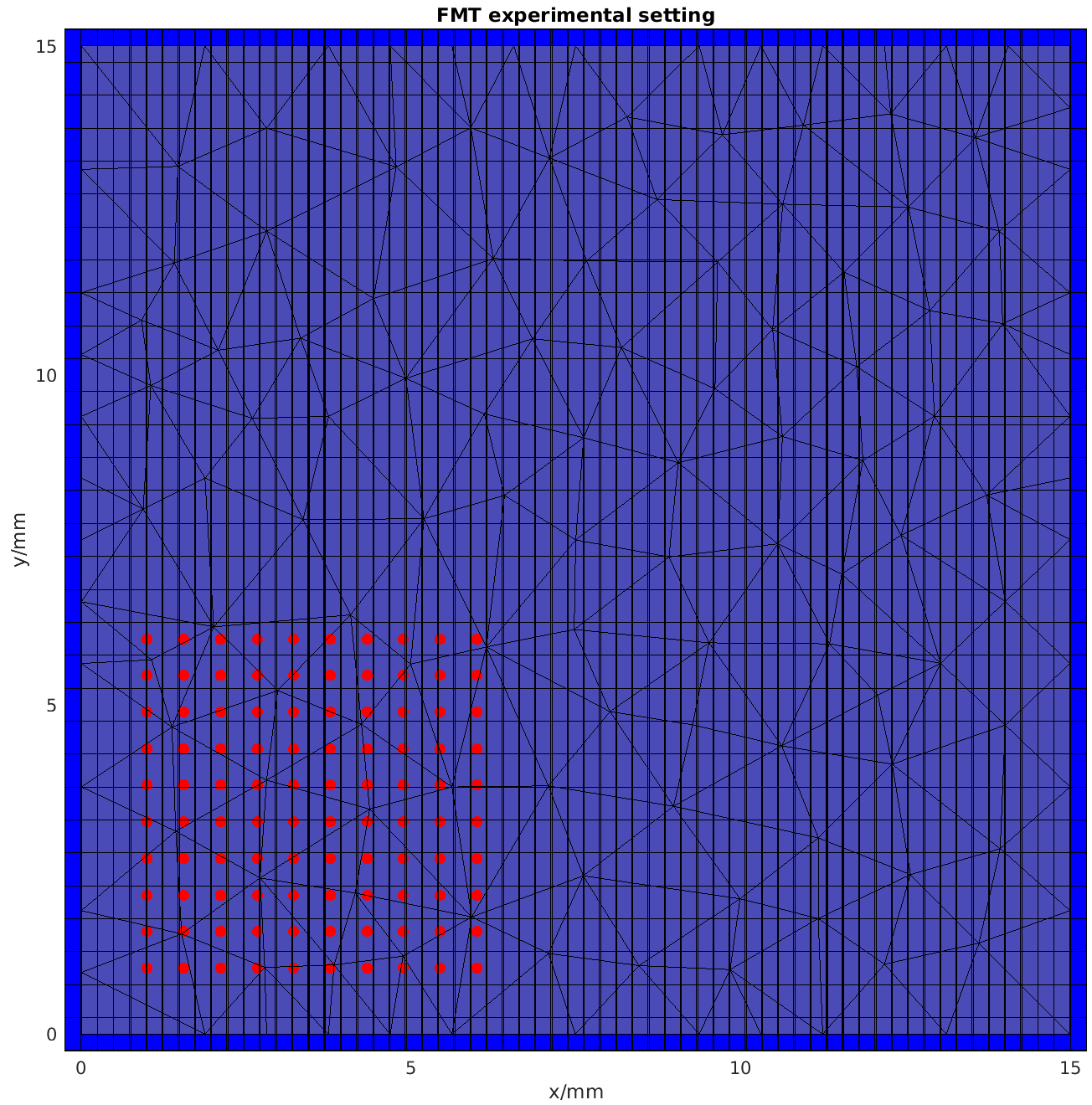}
  \end{tabular}
  \end{center}
\caption{A different initial illumination pattern viewed from the top surface of the same phantom. Except for the initial laser intensities and the location of the laser array, everything else remains the same as the previous experiment. The detector array is the same 60 by 30 array represented by the blue area; the laser array is the 10 by 10 point array represented by the red points.}
\label{cornerfmt}
\end{table}

\subsubsection{Results and error analysis for the new laser setting}
\begin{table}[h!]
\begin{center}
 \begin{tabular}{cc}
   \includegraphics[width=0.4\textwidth]{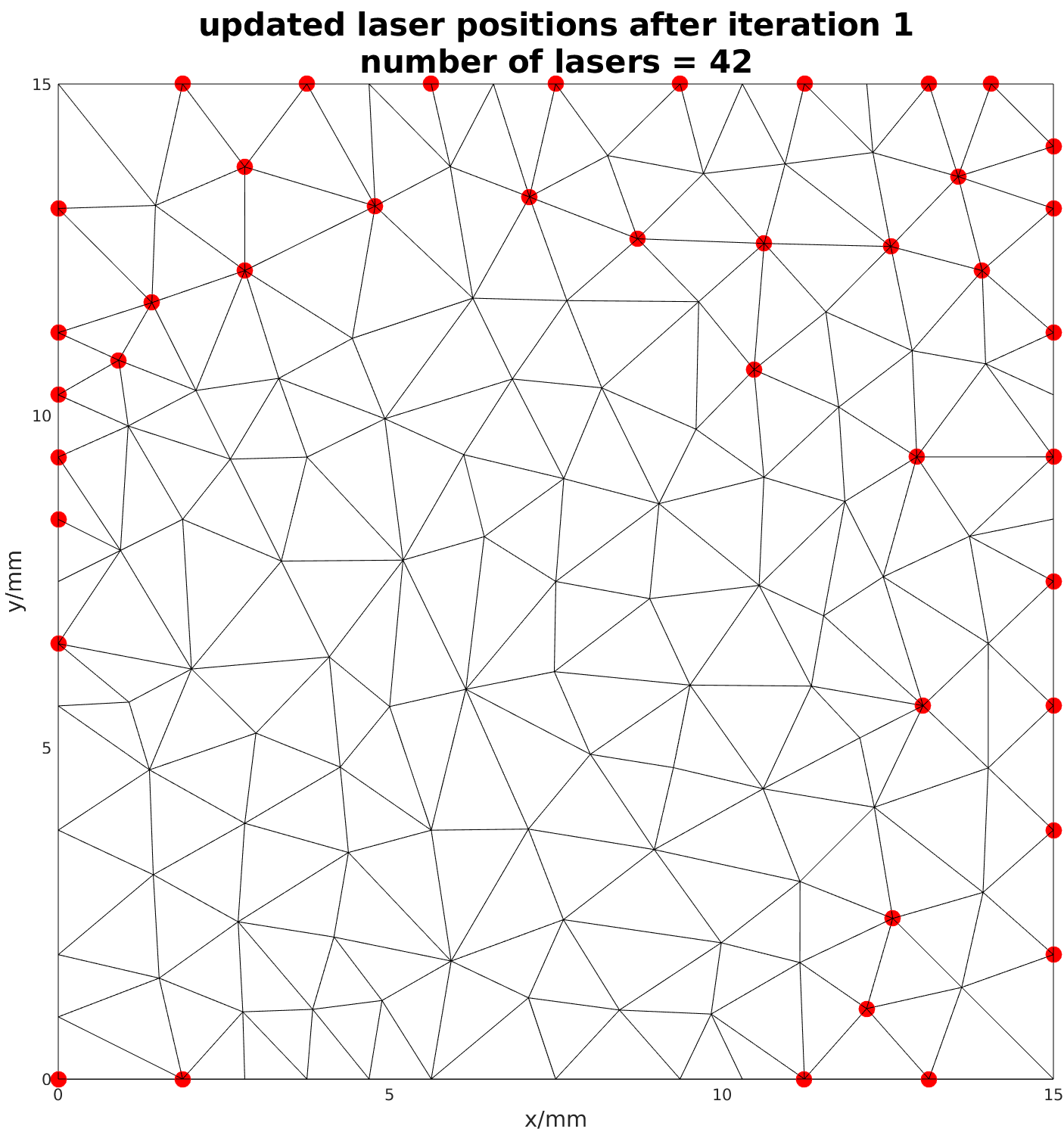}&
   \includegraphics[width=0.4\textwidth]{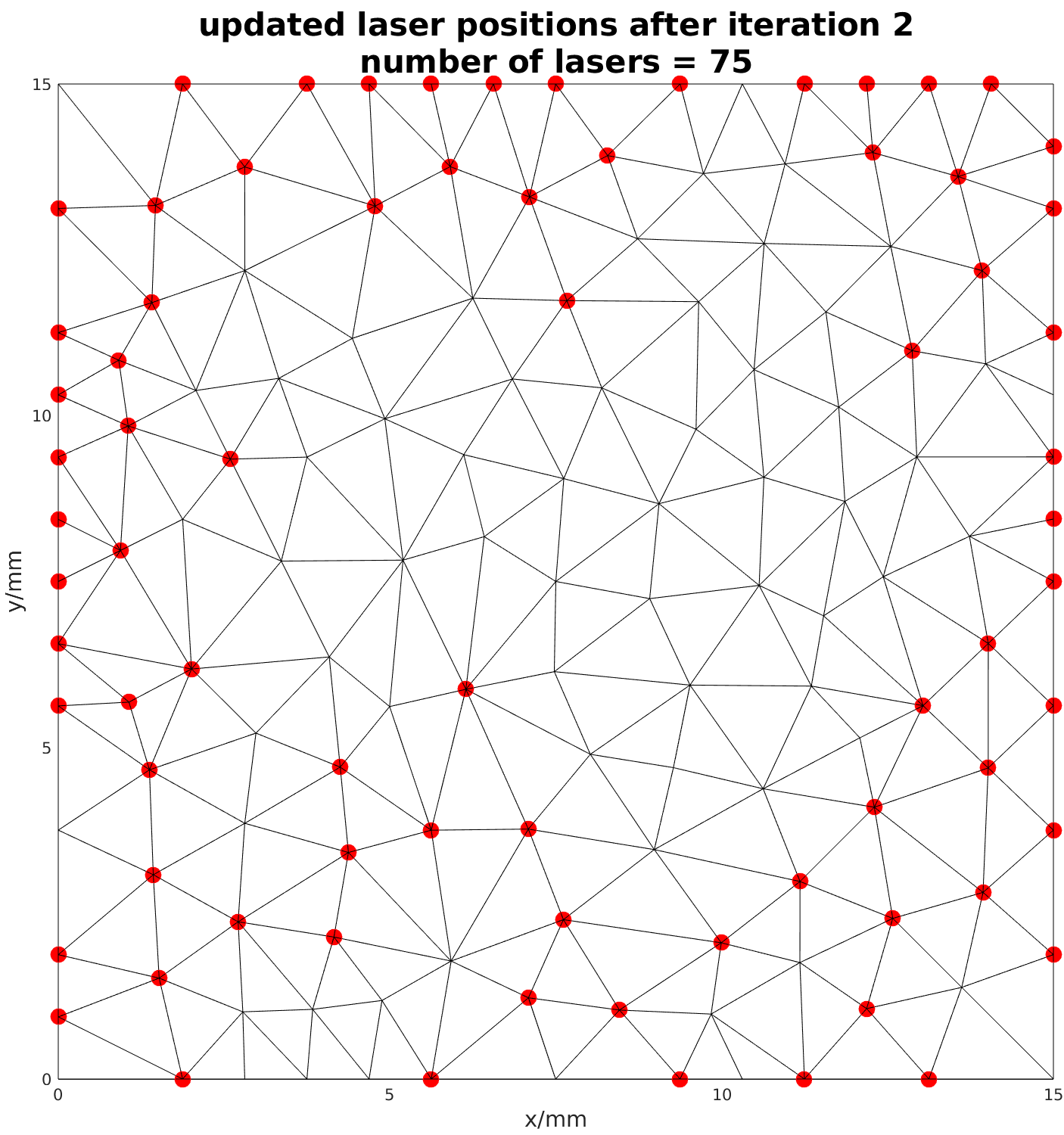}\\
   \includegraphics[width=0.4\textwidth]{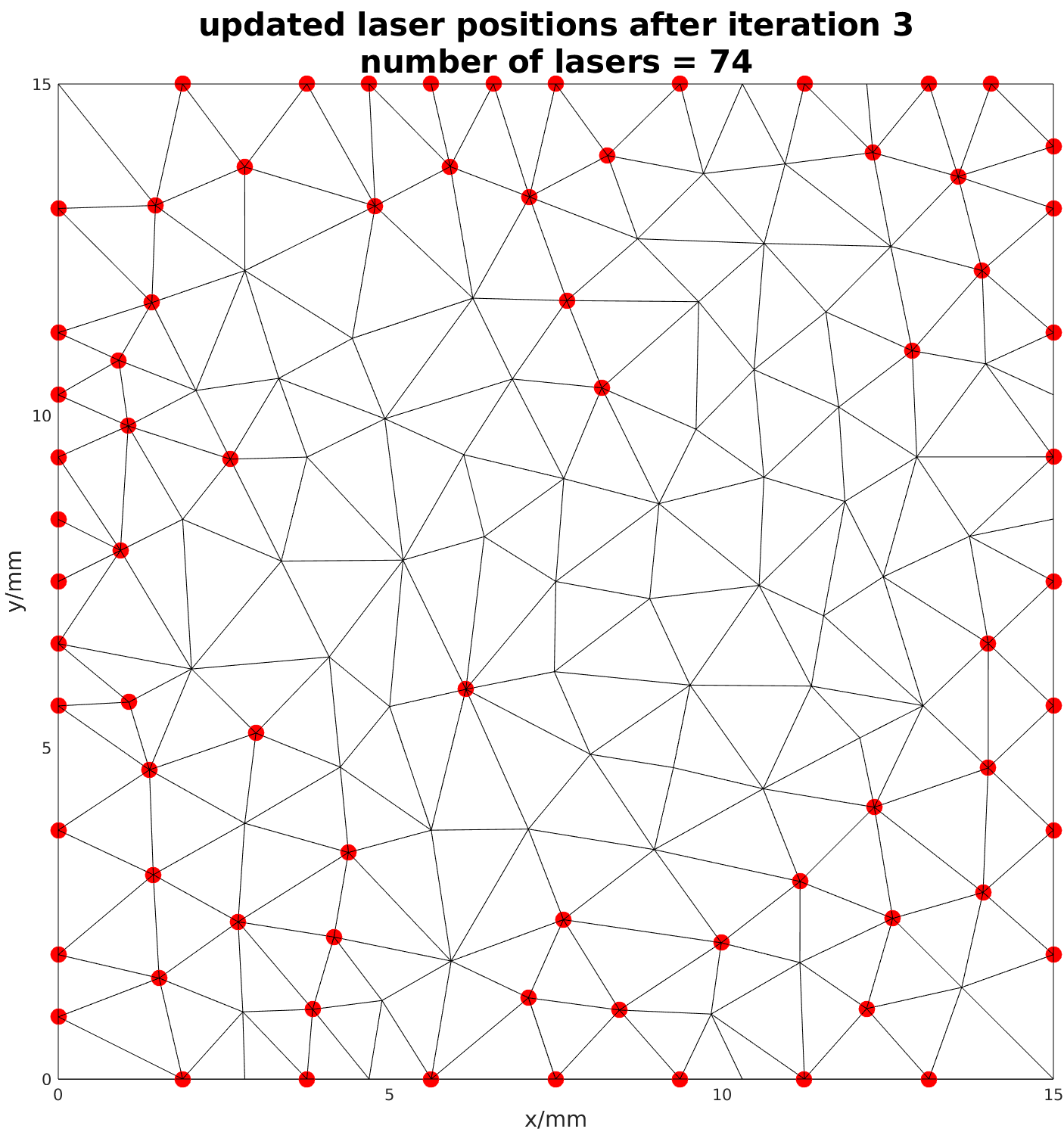}&
   \includegraphics[width=0.4\textwidth]{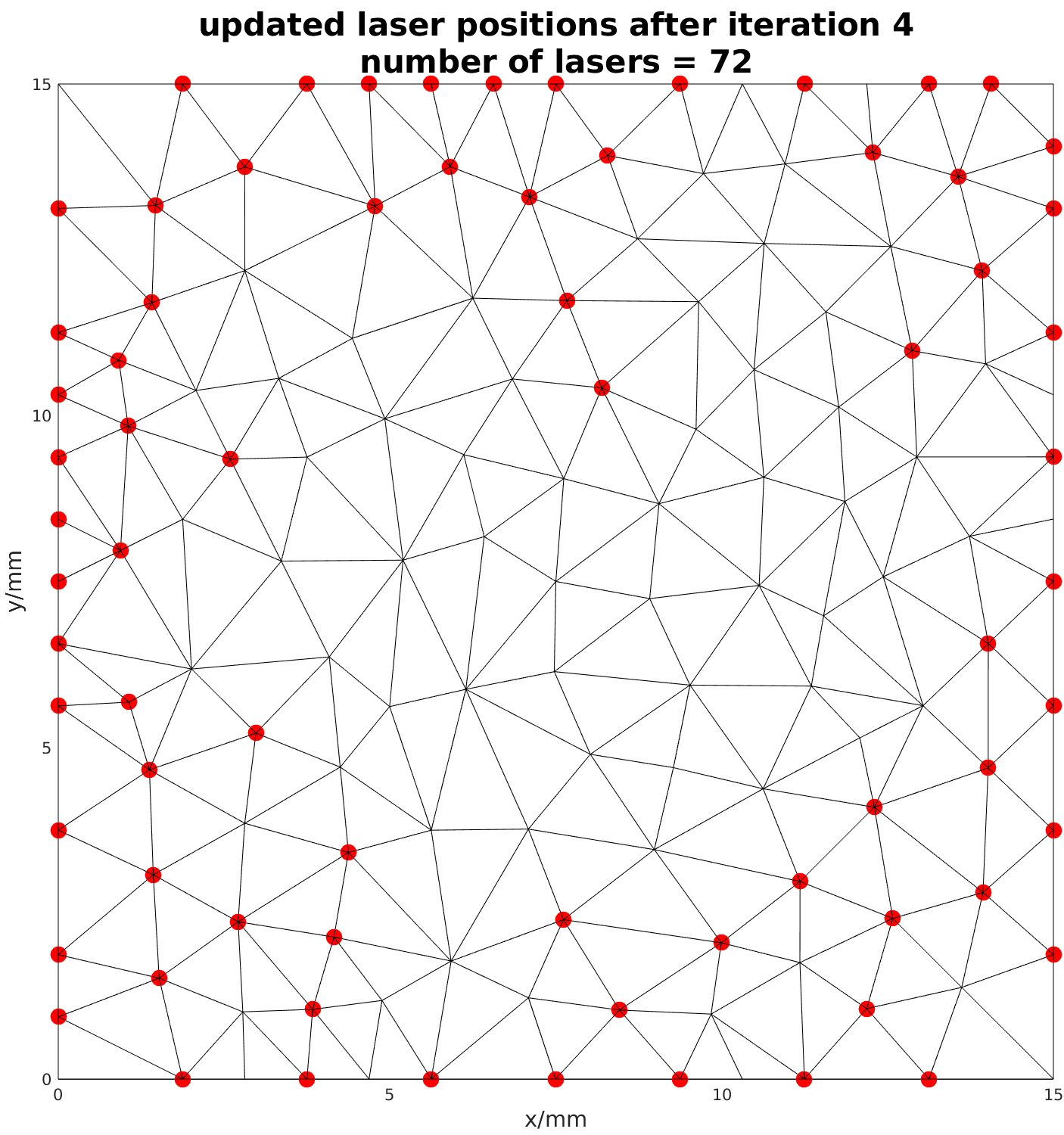}\\
   \includegraphics[width=0.4\textwidth]{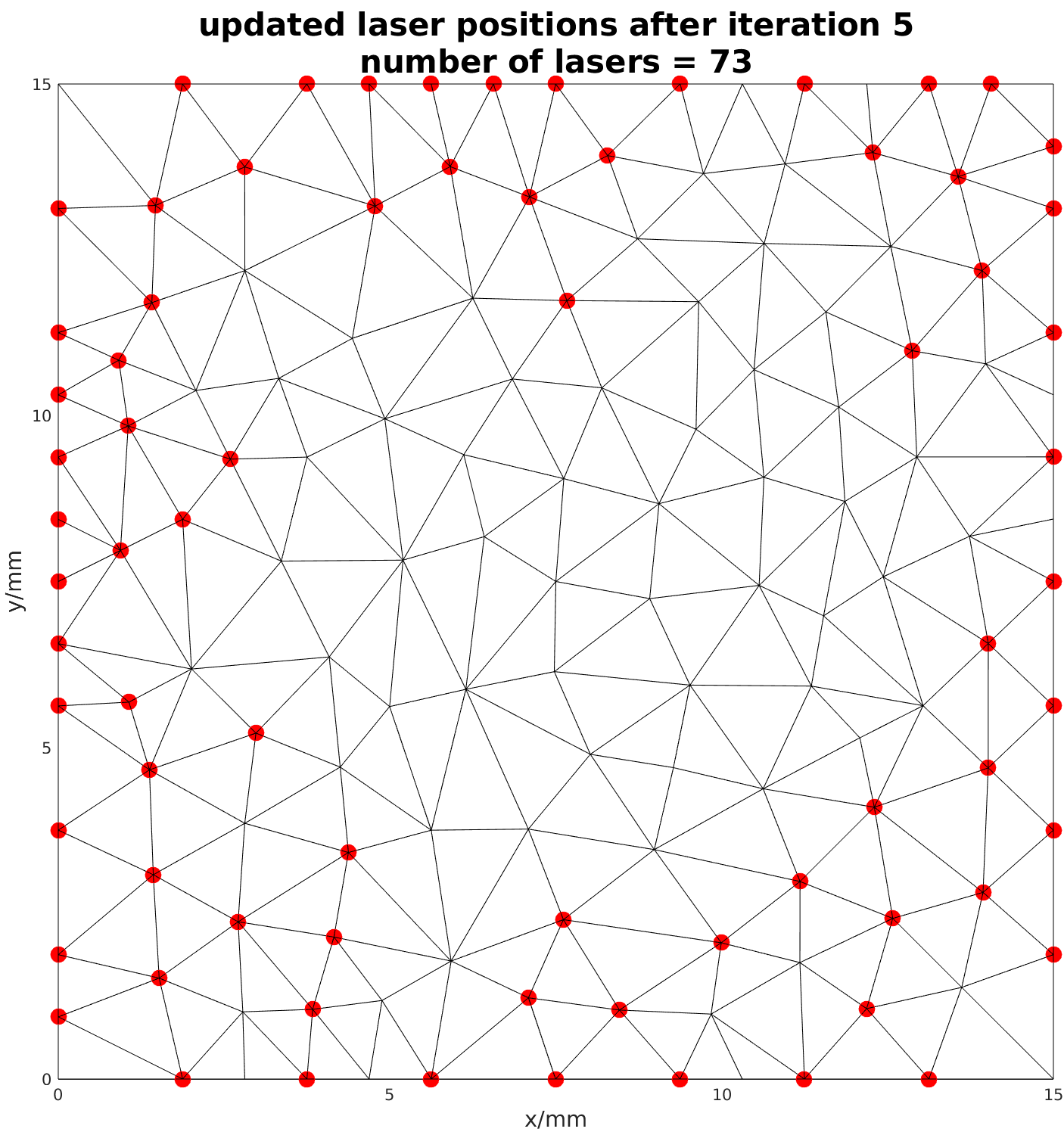}&
   \includegraphics[width=0.4\textwidth]{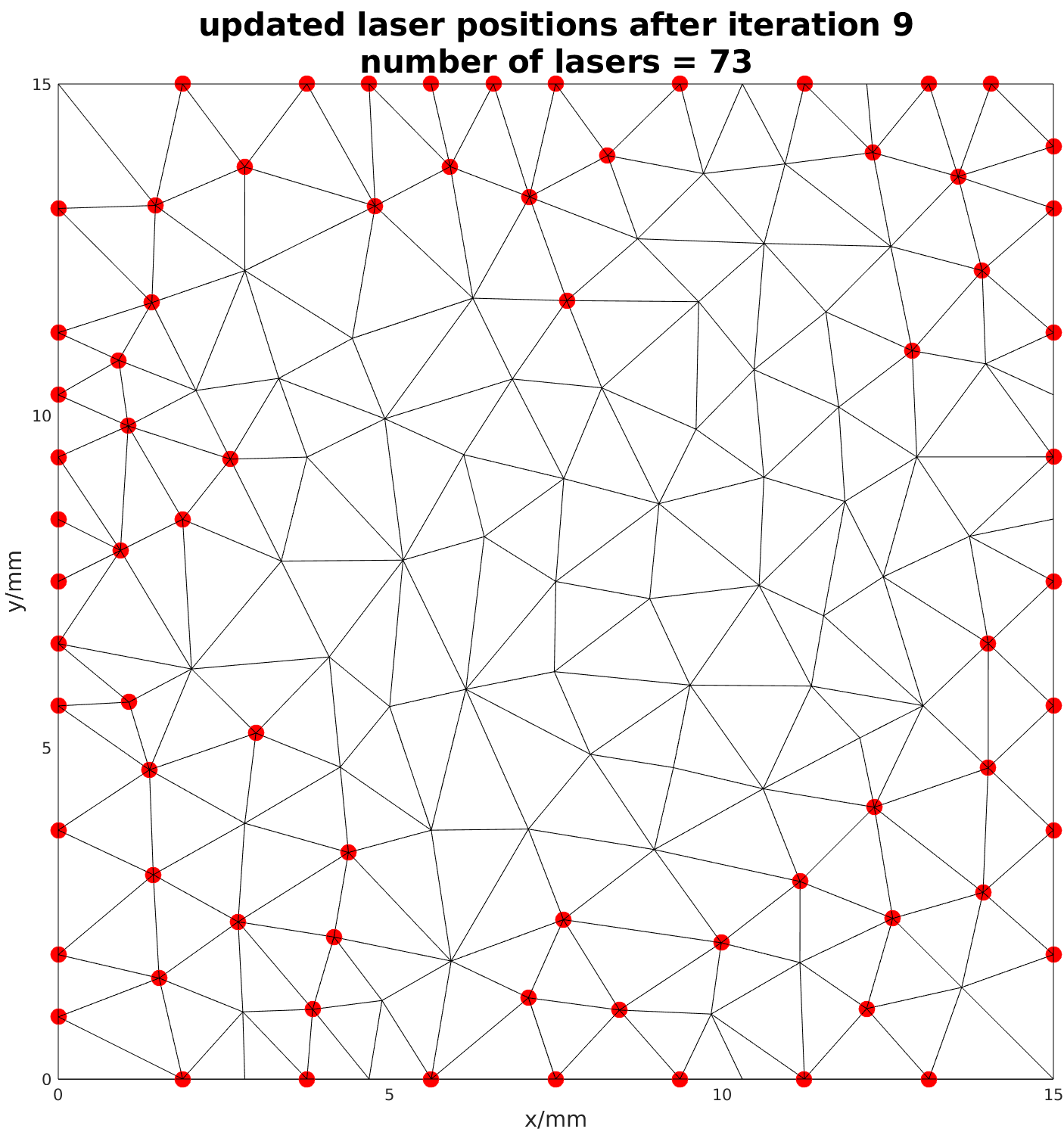}
 \end{tabular}
 \end{center}
\caption{Updated spatial layout of illumination points after round 1, 2, 3, 4, 5 and 9 of iteration starting from the new initial illumination pattern in Table \ref{cornerfmt}. Red points represent the lasers. The number of lasers (illumination points) are indicated above each figure.}
\label{new_illu_pattern}
\end{table}
The updated illumination patterns behave almost the same as the ones in the previous experiment: quickly shaping into the optimal illumination pattern as before and the updated patterns become very stable as we iterate up to 10 rounds.  

\begin{table}[h!]
 \setlength\tabcolsep{2pt}
  \begin{center}
   \begin{tabular}{cc}
    \includegraphics[width=0.5\textwidth]{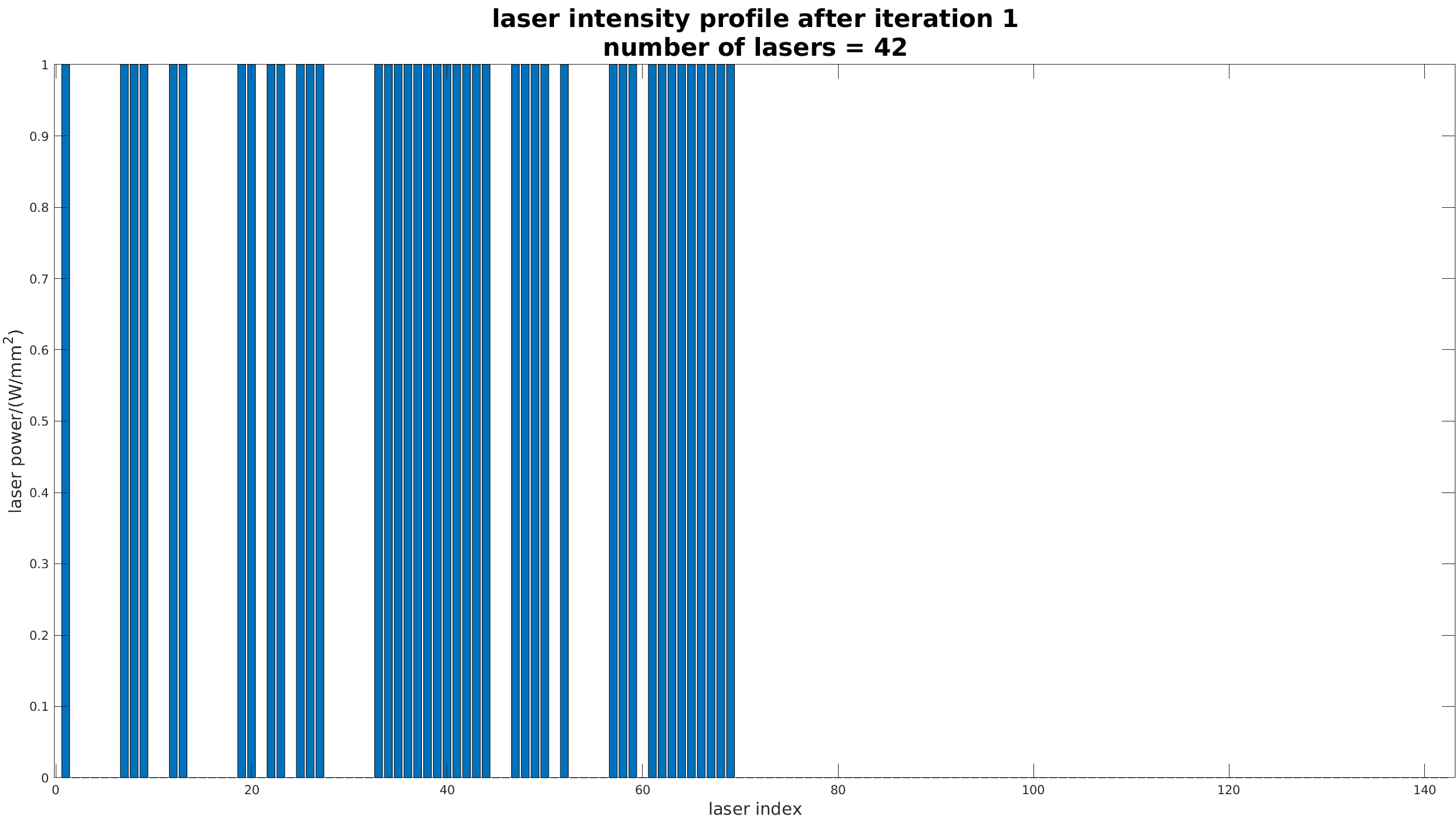}&
    \includegraphics[width=0.5\textwidth]{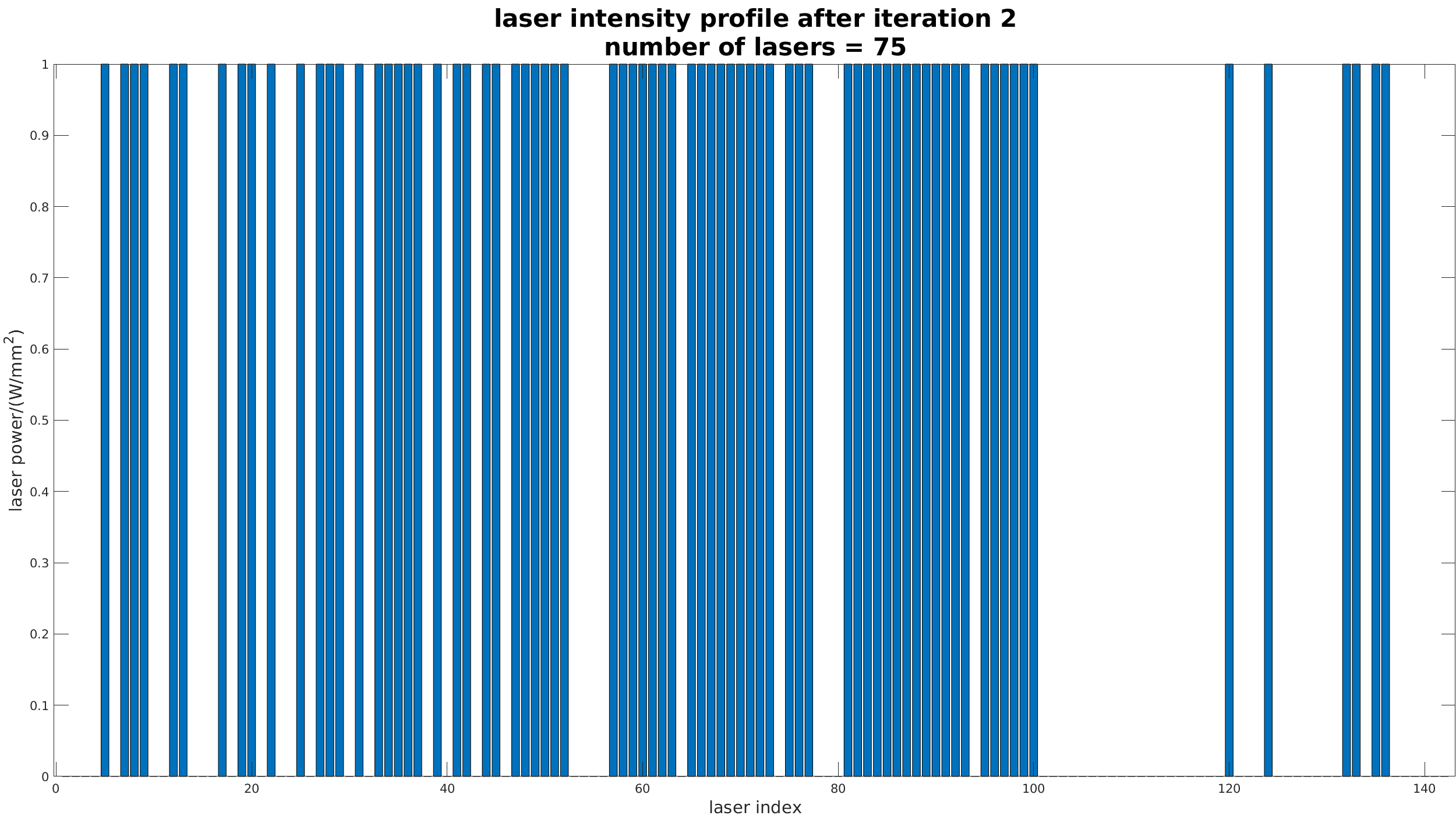}\\
    \includegraphics[width=0.5\textwidth]{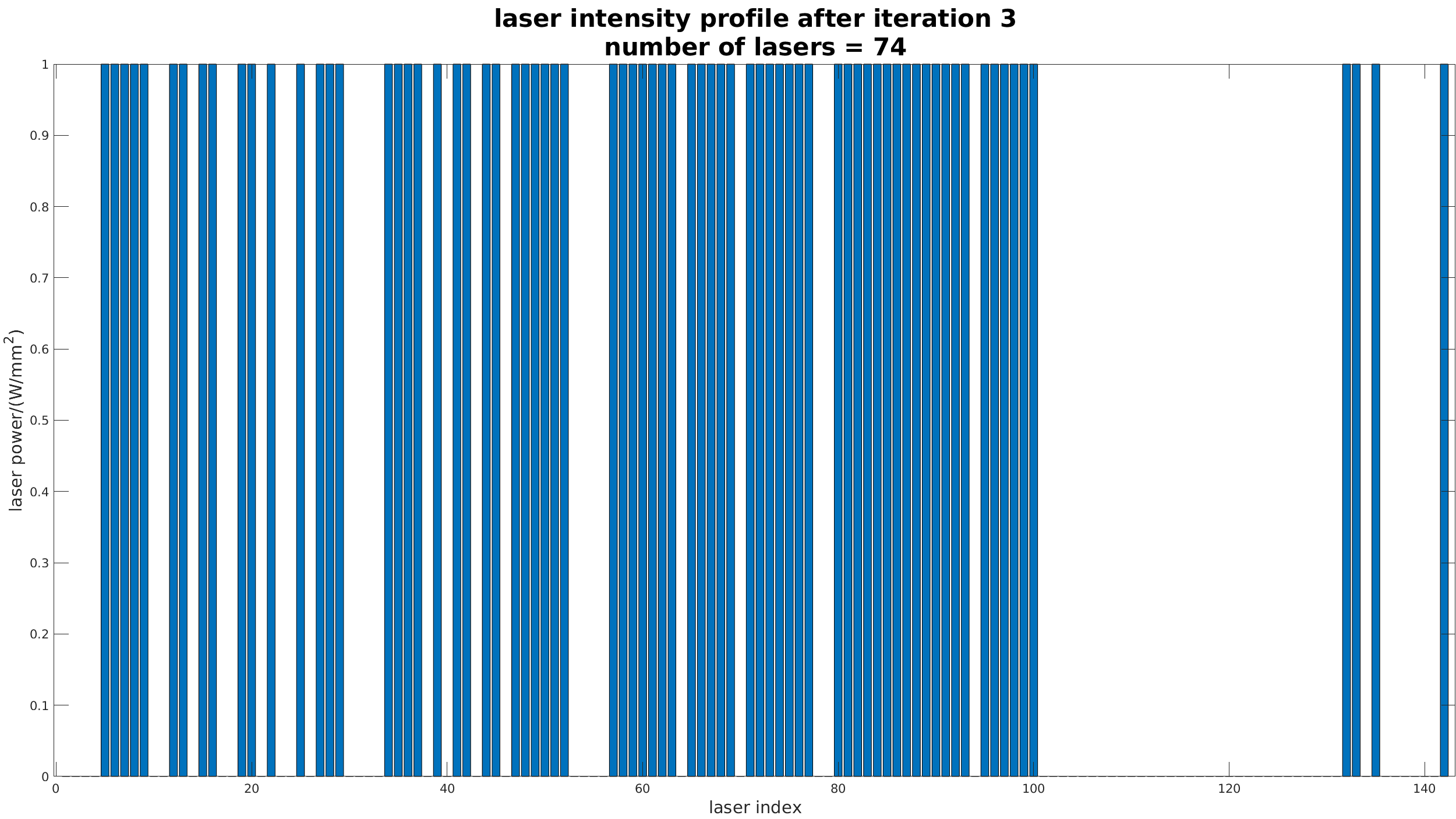}&
    \includegraphics[width=0.5\textwidth]{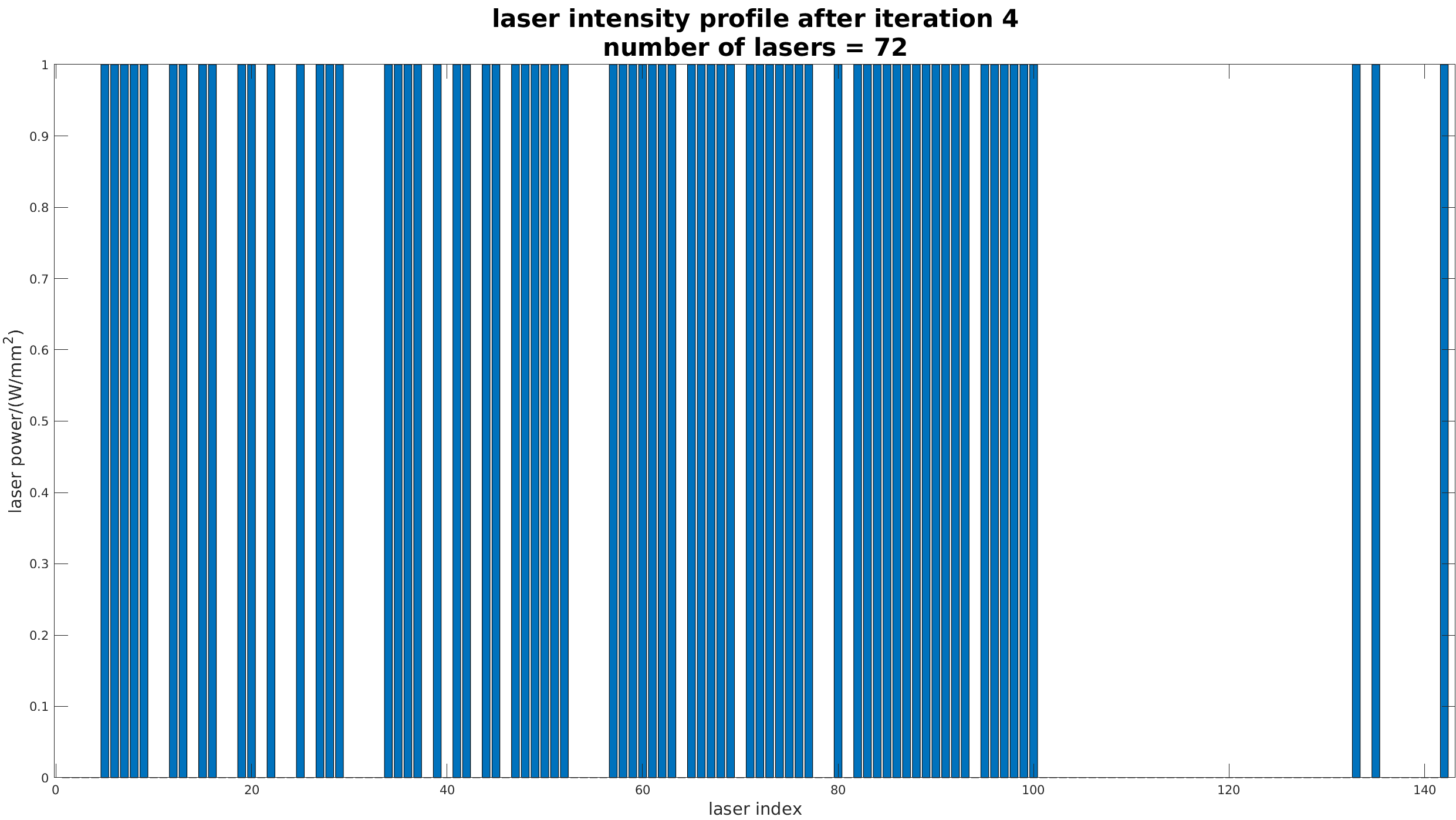}\\
    \includegraphics[width=0.5\textwidth]{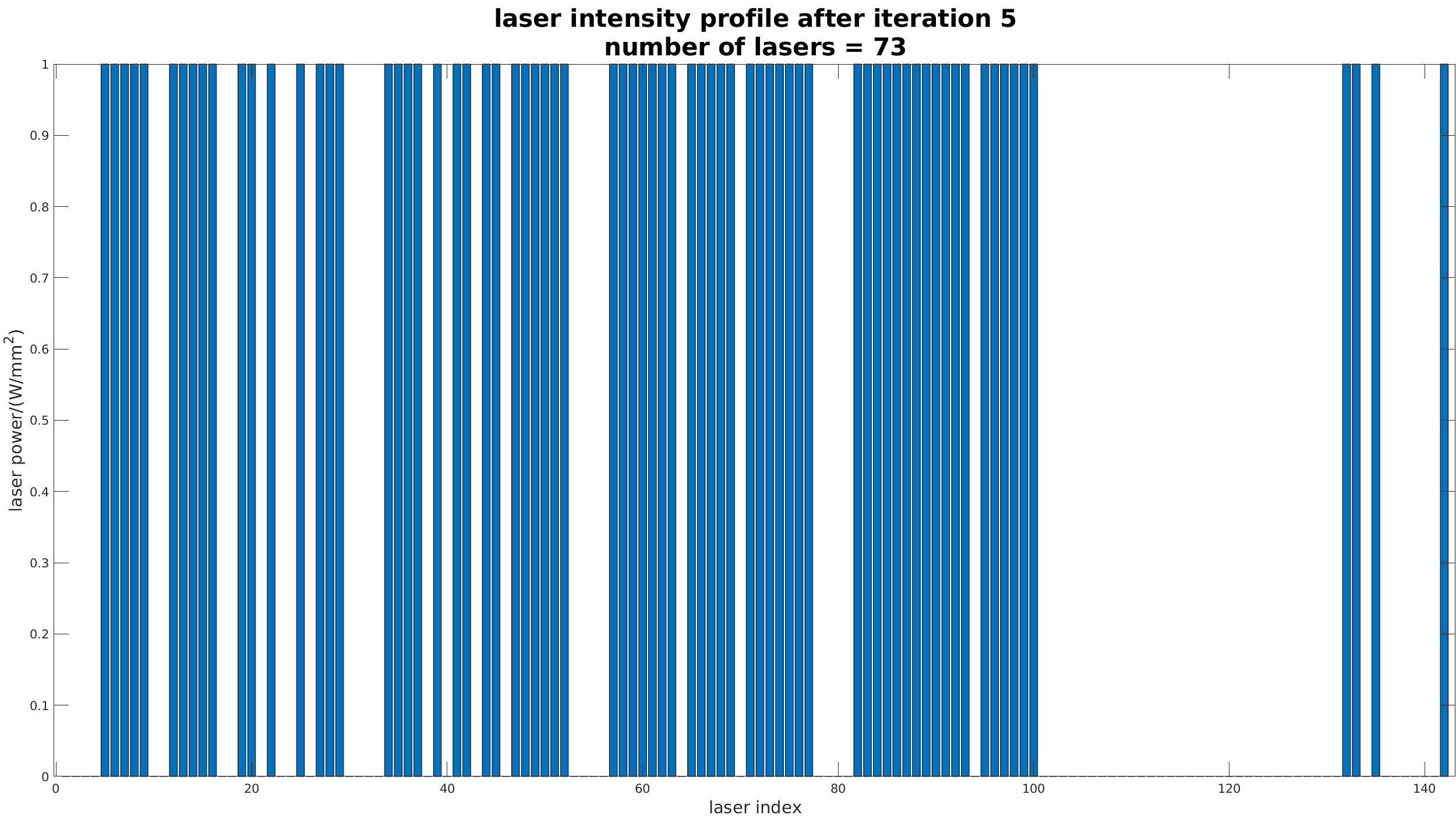}&
    \includegraphics[width=0.5\textwidth]{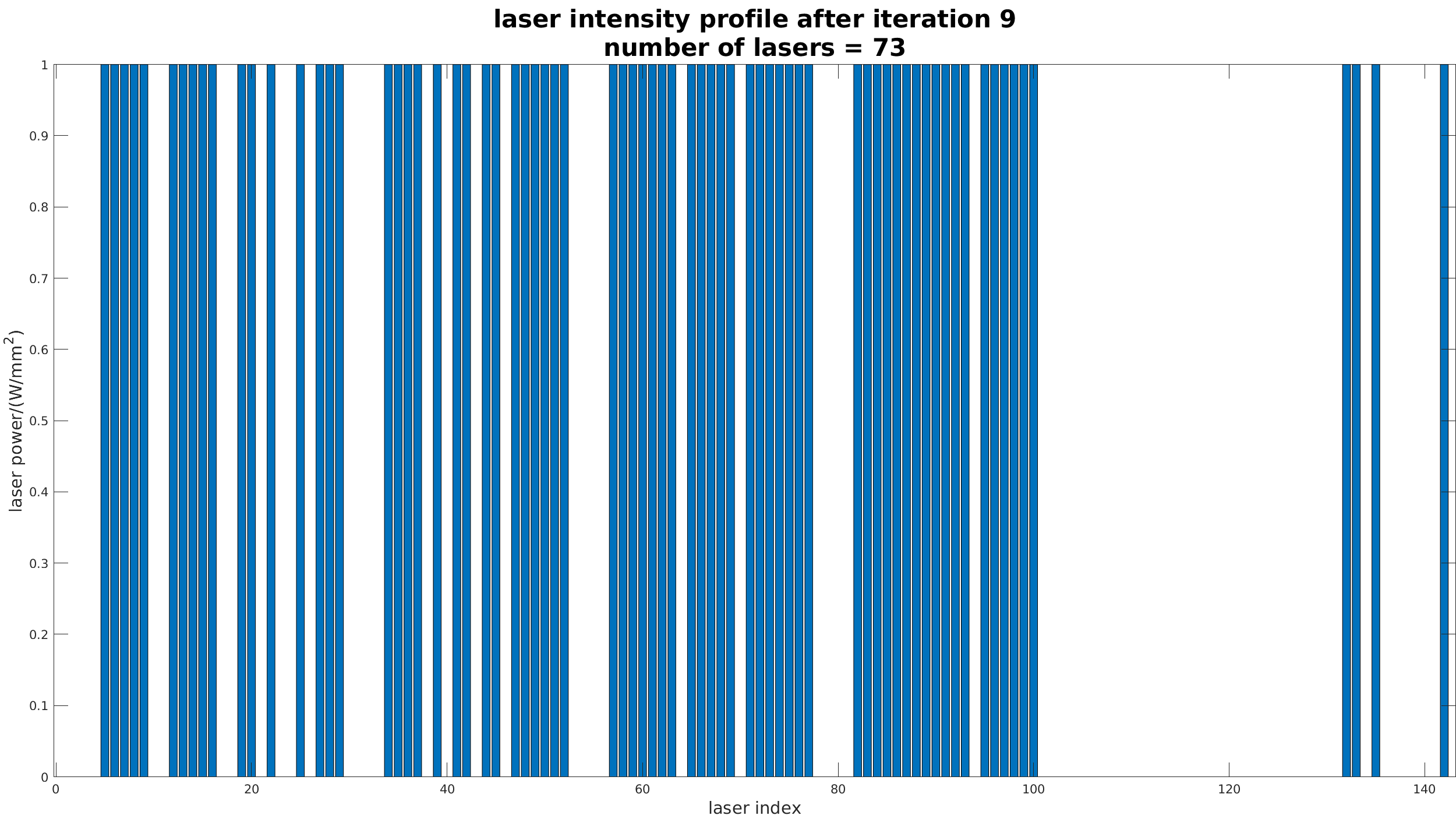}
   \end{tabular}
  \end{center}
\caption{Laser intensity profiles of each updated illumination pattern starting from the new initial laser setting. First column from top to bottom: laser profile for 1st, 3rd, and 5th updated illumination patterns. Second column from top to bottom: laser profile for 2nd, 4th, and 9th updated illumination pattern.}
 \label{newlaserprofile}
 \end{table}
The laser intensity profiles of every round tend to resemble those from the previous experiment as the number of iteration rounds increases. Slightly different from Table \ref{laserprofile}, the intensity of each laser in all rounds of iteration is identical and reaches the maximal intensity 1$W/mm^2$.  

\begin{table}[h!]
\begin{center}
 \begin{tabular}{cccc}
  Illumination 0&\includegraphics[width=0.2\textwidth]{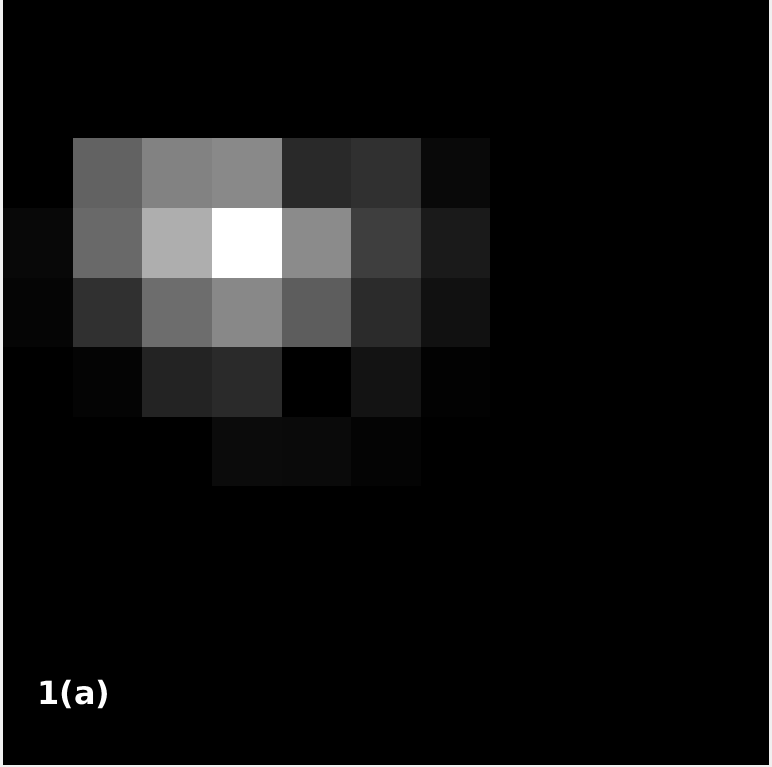}&
  \includegraphics[width=0.2\textwidth]{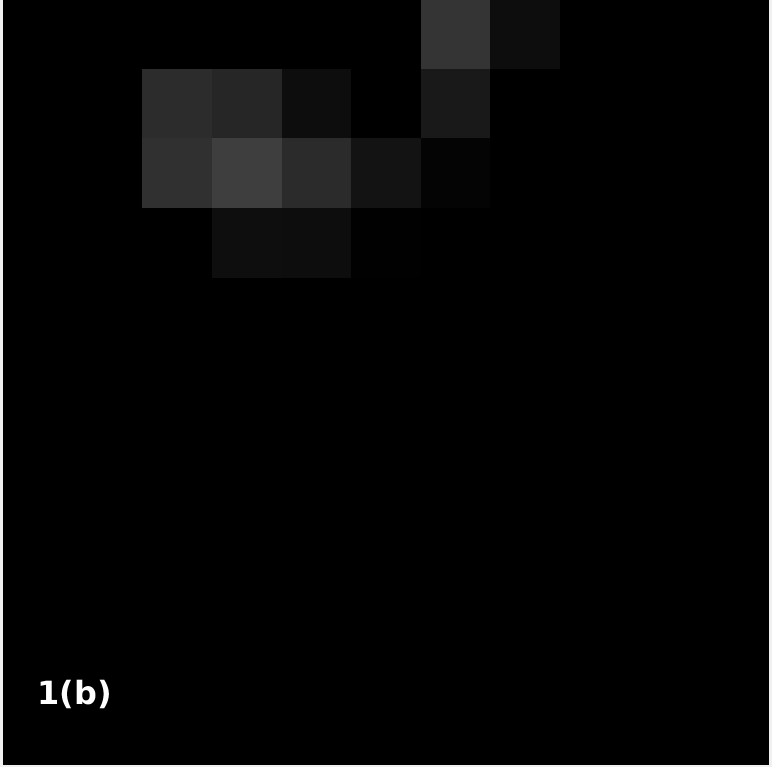}&
  \includegraphics[width=0.2\textwidth]{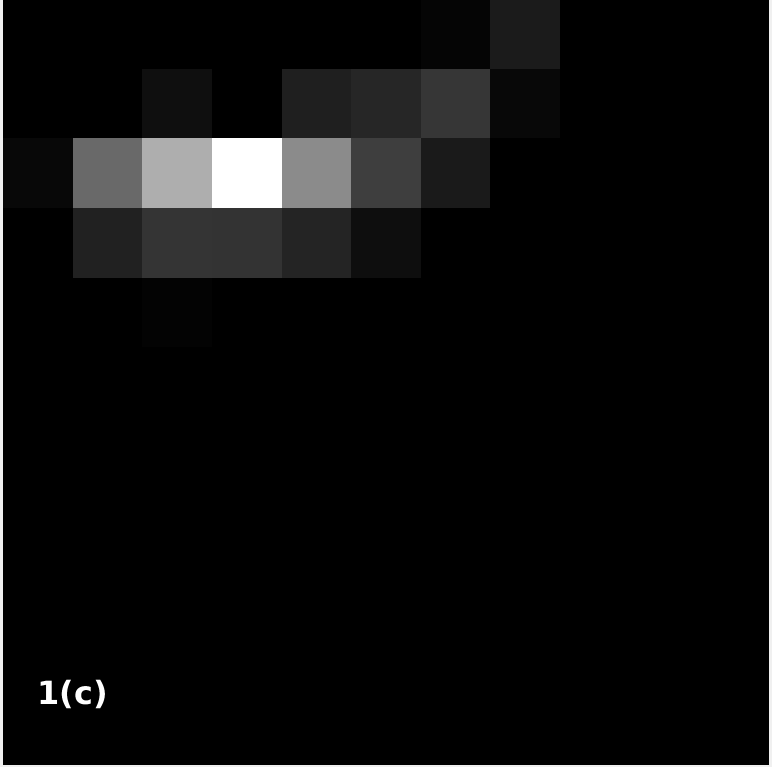}\\
  Illumination 1&\includegraphics[width=0.2\textwidth]{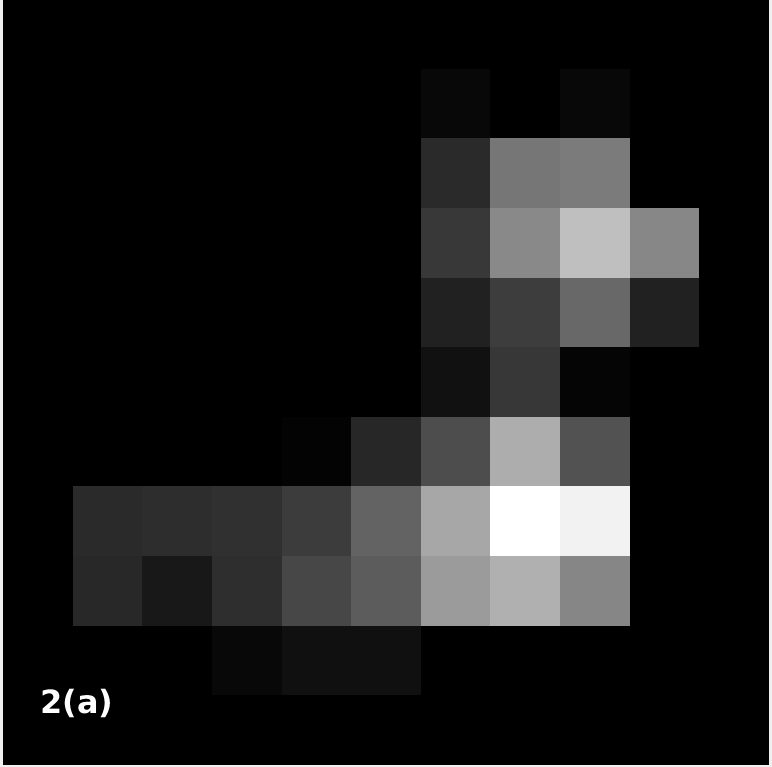}&
  \includegraphics[width=0.2\textwidth]{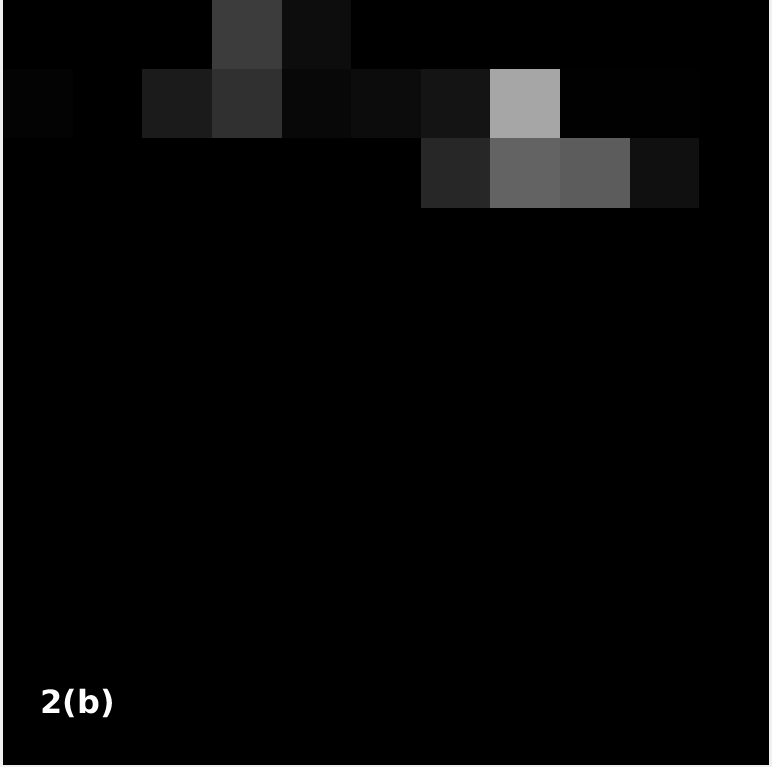}&
  \includegraphics[width=0.2\textwidth]{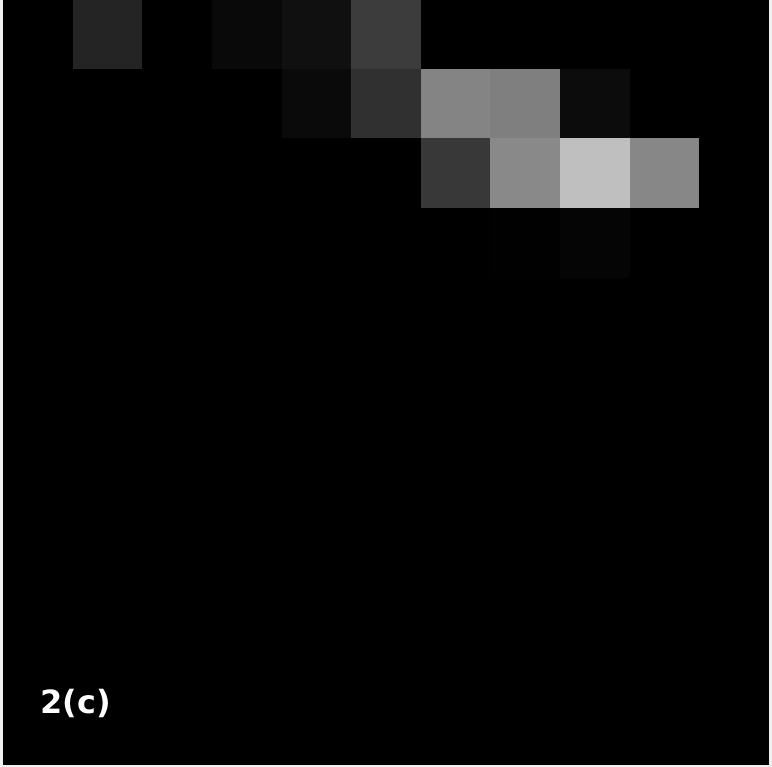}\\
  Illumination 2&\includegraphics[width=0.2\textwidth]{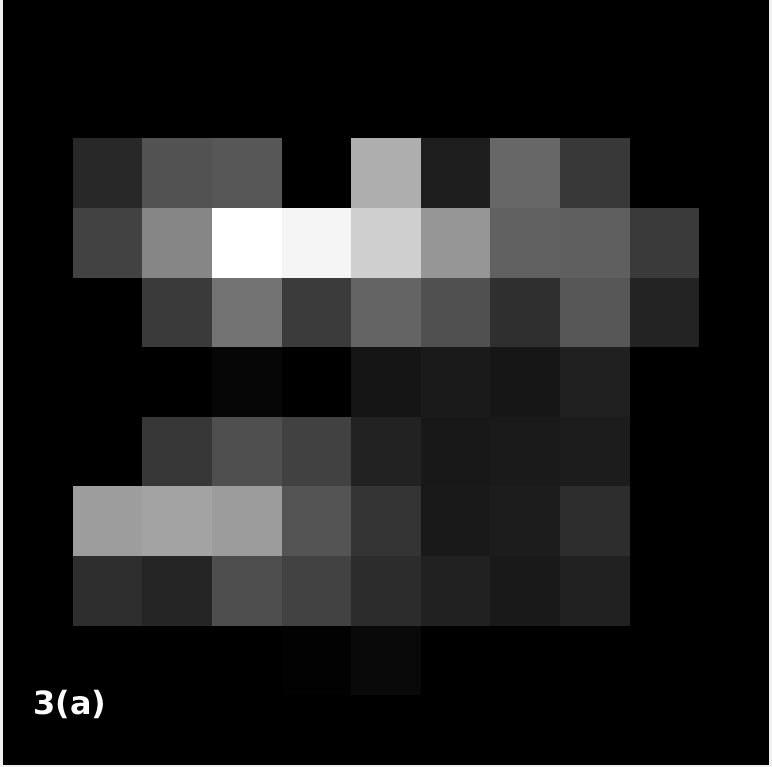}&
  \includegraphics[width=0.2\textwidth]{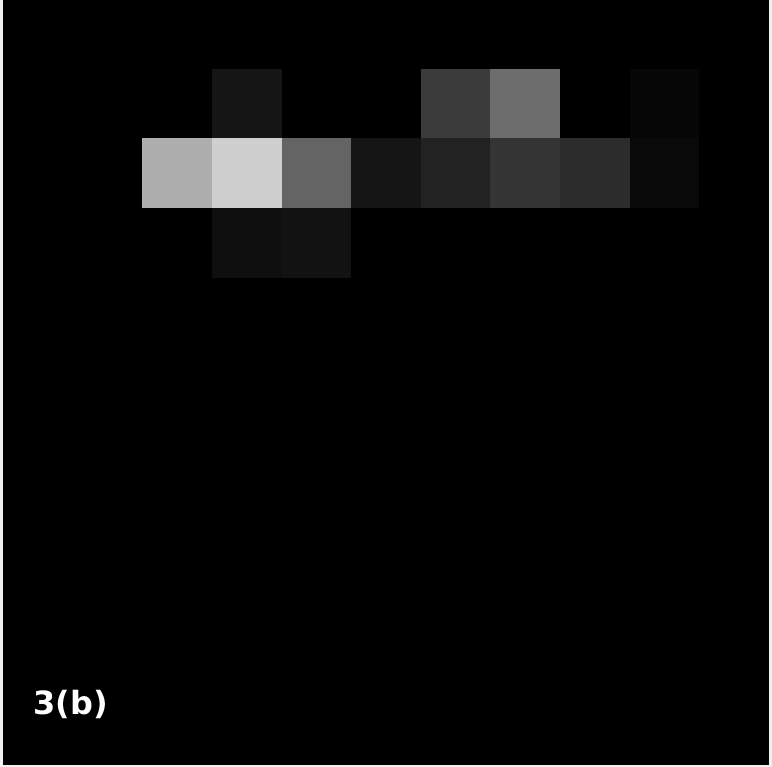}&
  \includegraphics[width=0.2\textwidth]{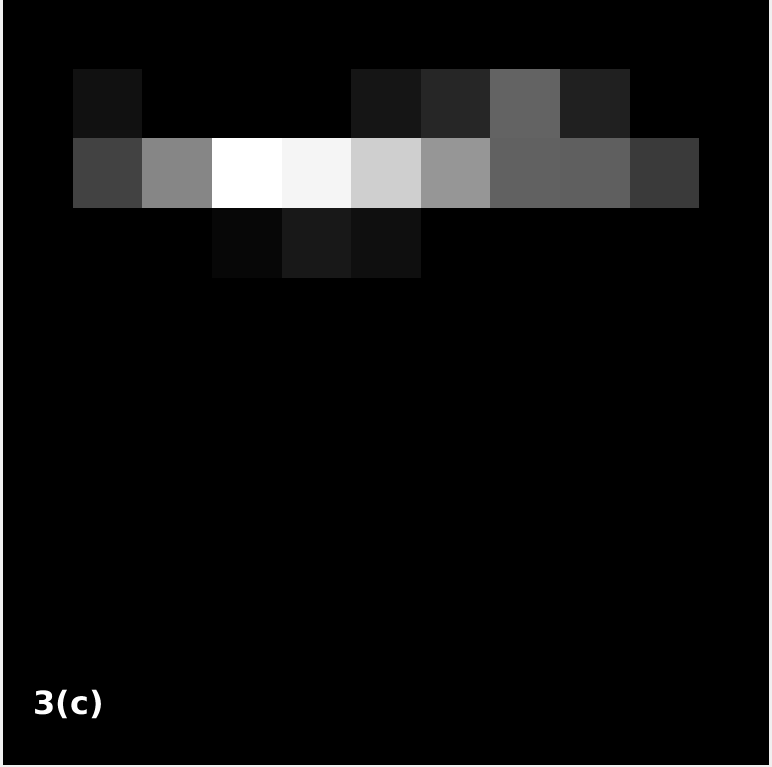}\\
  Illumination 3&\includegraphics[width=0.2\textwidth]{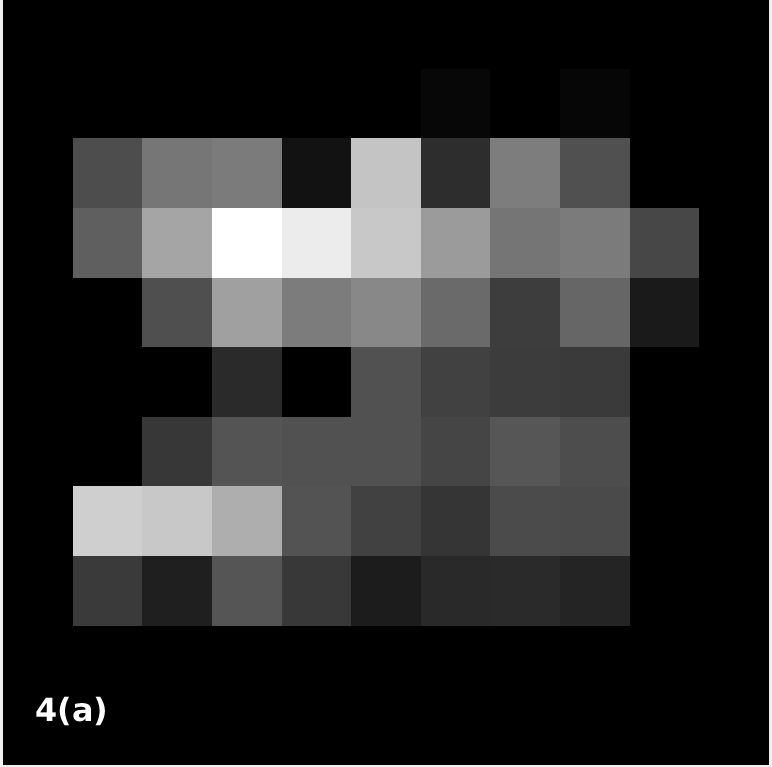}&
  \includegraphics[width=0.2\textwidth]{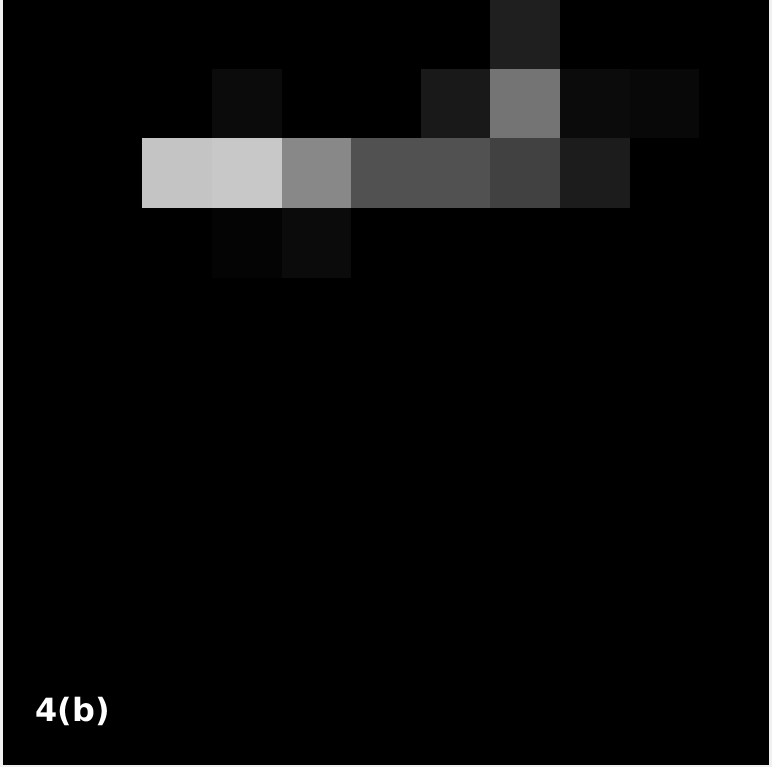}&
  \includegraphics[width=0.2\textwidth]{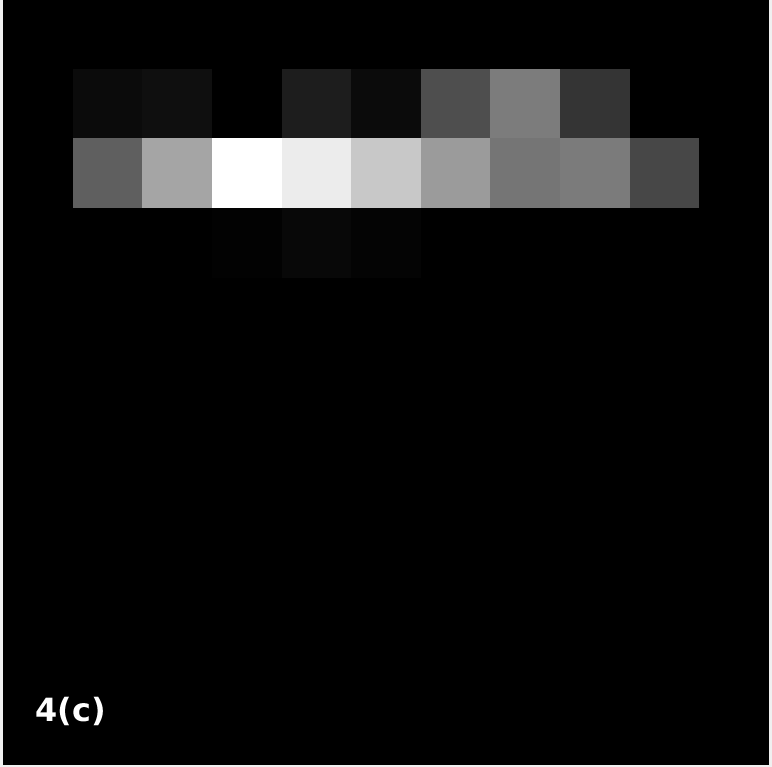}\\
  Illumination 4&\includegraphics[width=0.2\textwidth]{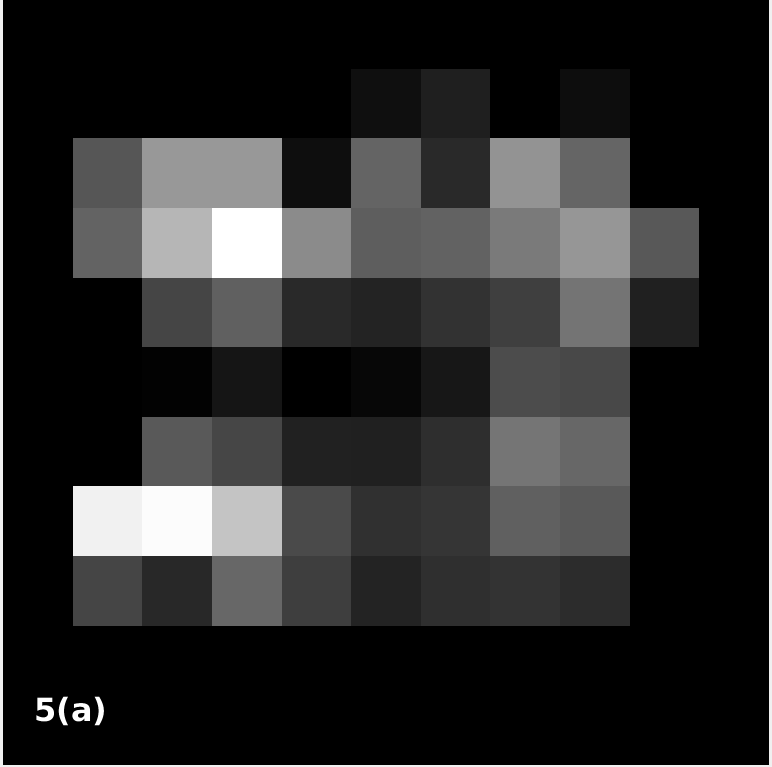}&
  \includegraphics[width=0.2\textwidth]{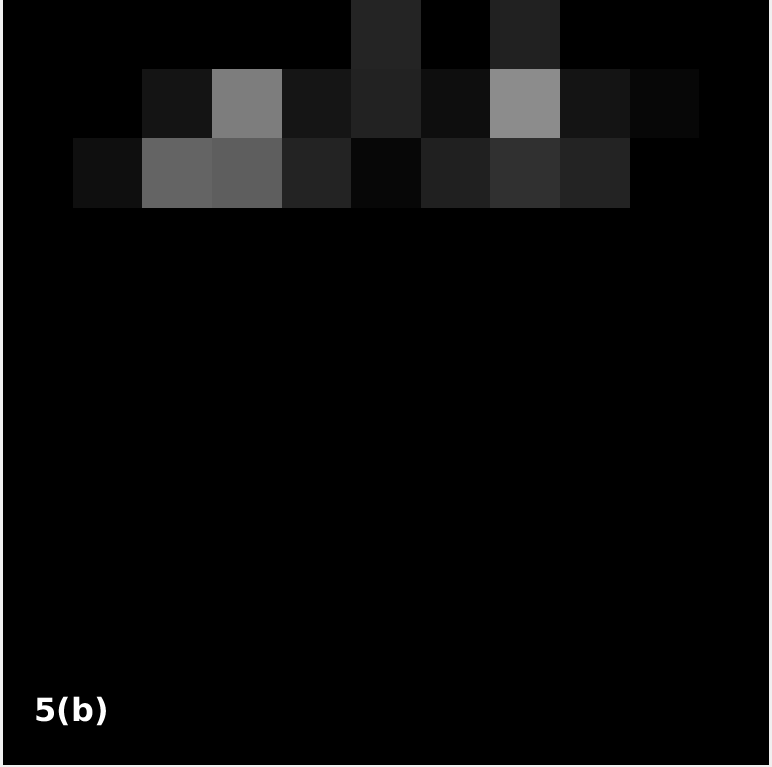}&
  \includegraphics[width=0.2\textwidth]{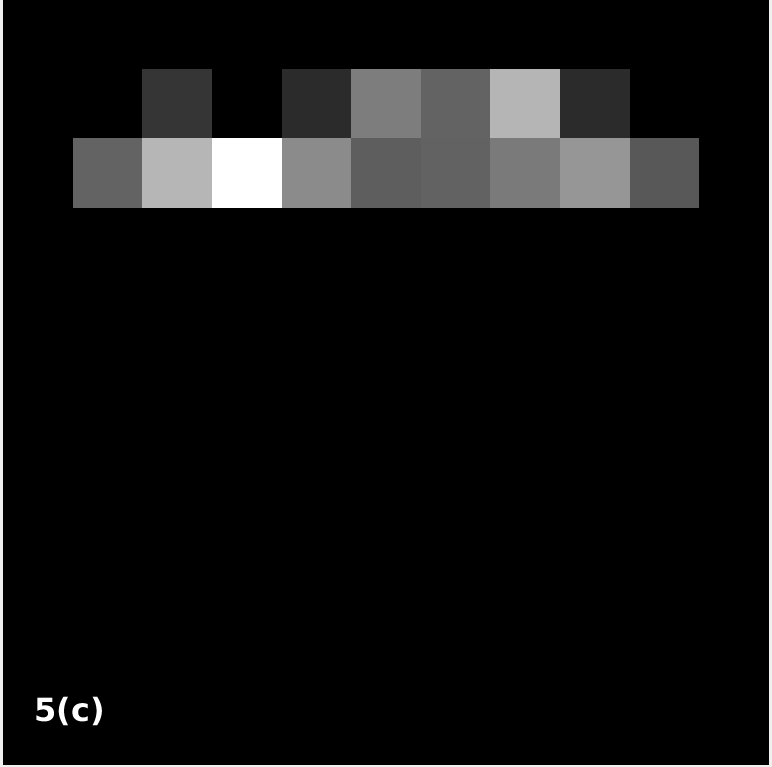}\\
  Illumination 5&\includegraphics[width=0.2\textwidth]{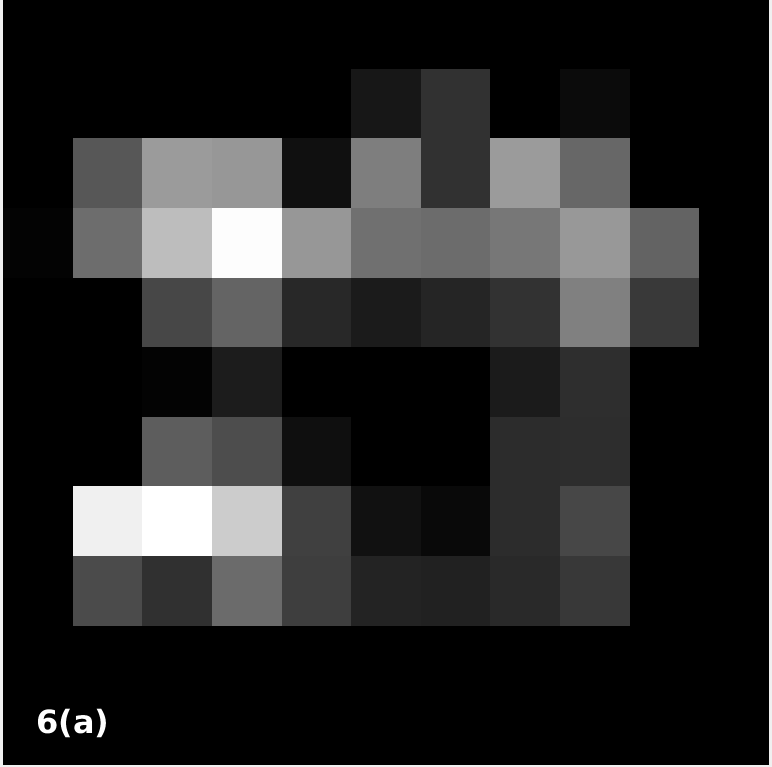}&
  \includegraphics[width=0.2\textwidth]{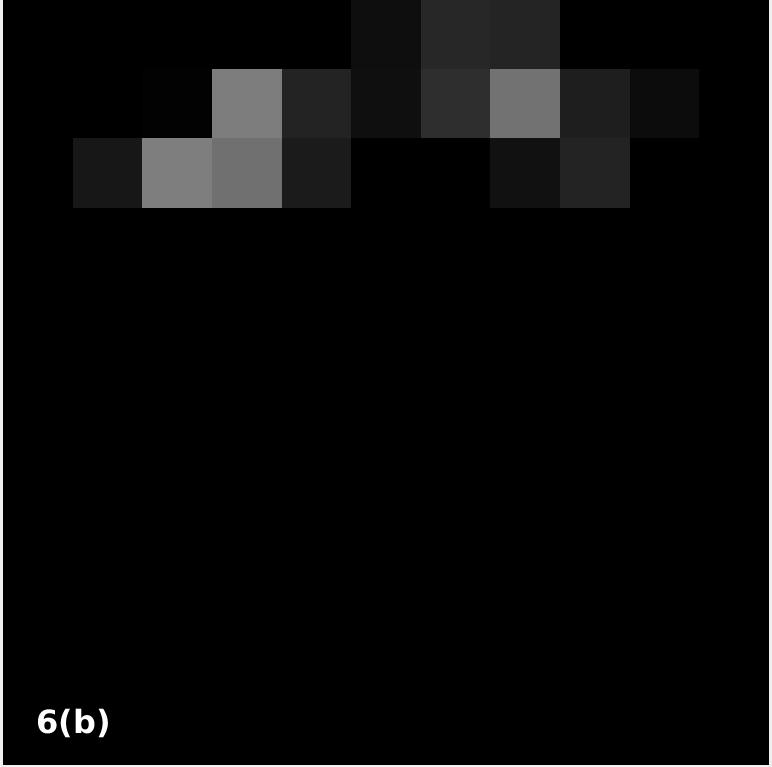}&
  \includegraphics[width=0.2\textwidth]{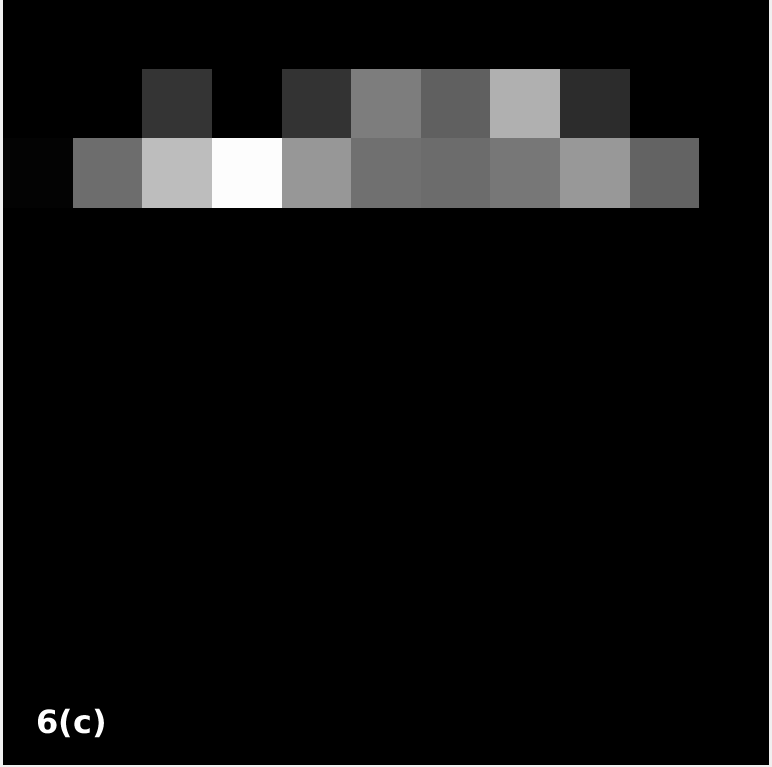}\\
  Illumination 9&\includegraphics[width=0.2\textwidth]{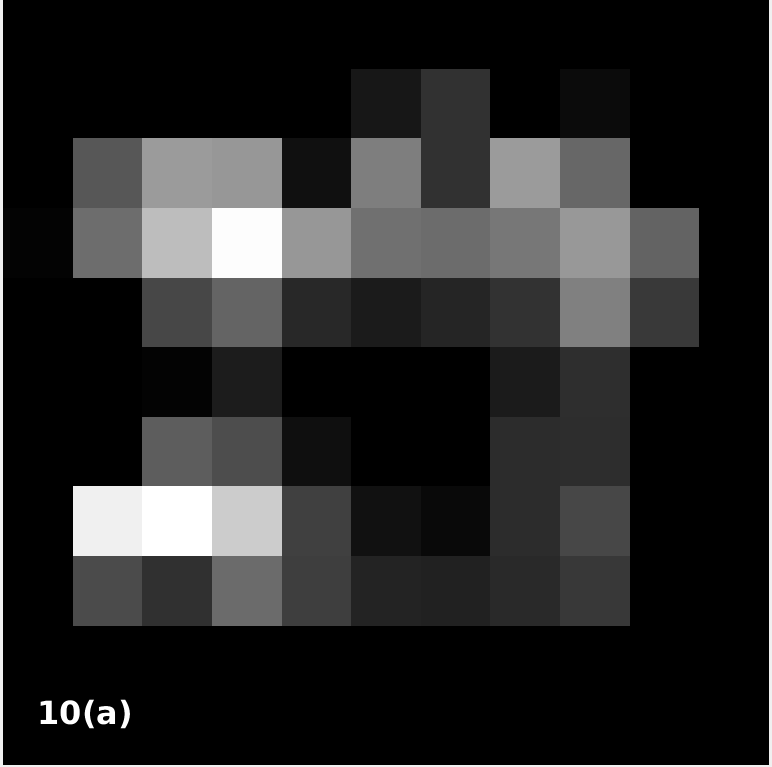}&
  \includegraphics[width=0.2\textwidth]{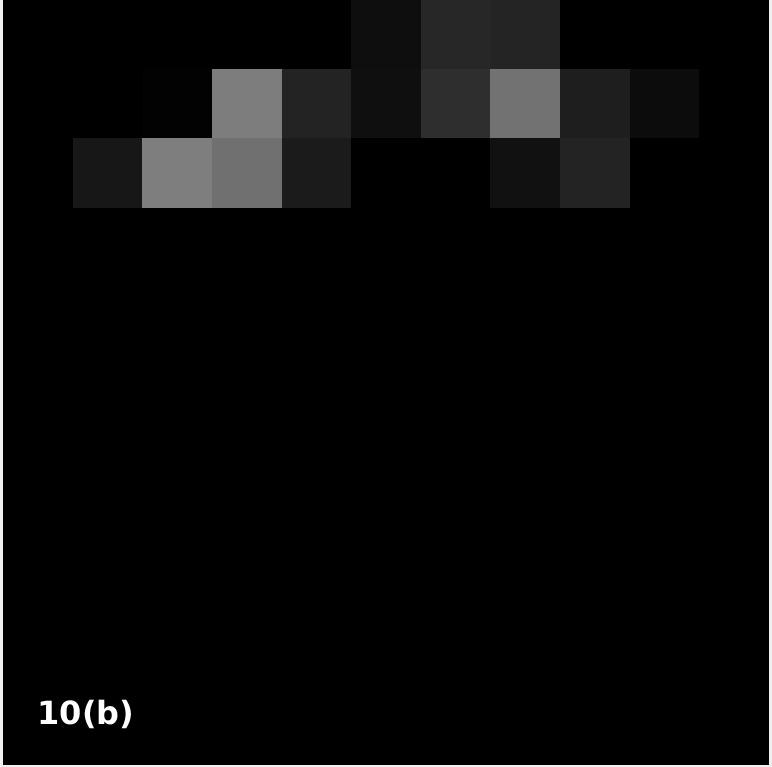}&
  \includegraphics[width=0.2\textwidth]{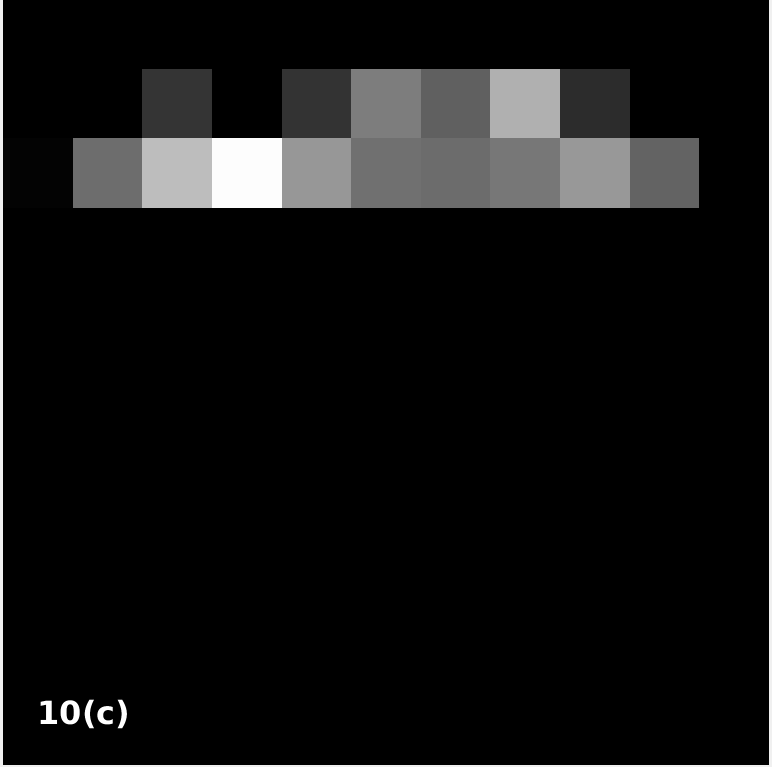}\\
 \end{tabular}
 \end{center}
\caption{Reconstruction results obtained from different rounds of iteration under the new initial laser setting. Reconstruction results are sliced at the same reference point $(13,9,6)$ as the ground truth in Table \ref{fluo_gt}. The first row (Illumination 0) contains reconstruction results of the initial laser setting viewed from the top (a), left (b) and front (c) of the phantom. Row 2 to row 7 are reconstruction results from the 1st, 2nd, 3rd, 4th, 5th and 9th updated illumination pattern.}
\label{new_recon}
\end{table}

The reconstruction results in the new experiment are displayed in Table \ref{new_recon}. The first reconstruction result that is based on the new initial illumination pattern does not capture the complete information of the fluorescence distribution, especially in 1(a). The result after the first updated illumination pattern has more information obtained as is seen in 2(a). Interestingly, the information content in 2(a) is not complete either, and the lost part seems to be that in 1(a). The reason for this might be explained by the comparing the initial illumination pattern and the first updated illumination pattern in Table \ref{new_illu_pattern}. It can be observed that all the lasers are inclined to run away from the the left bottom corner where all lasers were in the initial setting, so the information from the left bottom corner is consequently missing. However, as more iterations are performed, the lasers quickly find their optimal positions on the top surface, hence the corresponding reconstruction results are improved. In the end, the results remain unchanged in Illumination 4, 5 and 9 because the updated illumination patterns in these rounds are the same as well. Even though the new initial illumination pattern produces worse first reconstruction results 1(a), 1(b), and 1(c) in Table \ref{new_recon}, the updated illumination patterns and the reconstruction results still converge to the optimal one very quickly.

\begin{table}[h!]
  \begin{center}
   \begin{tabular}{cc}
    \includegraphics[width=0.47\textwidth]{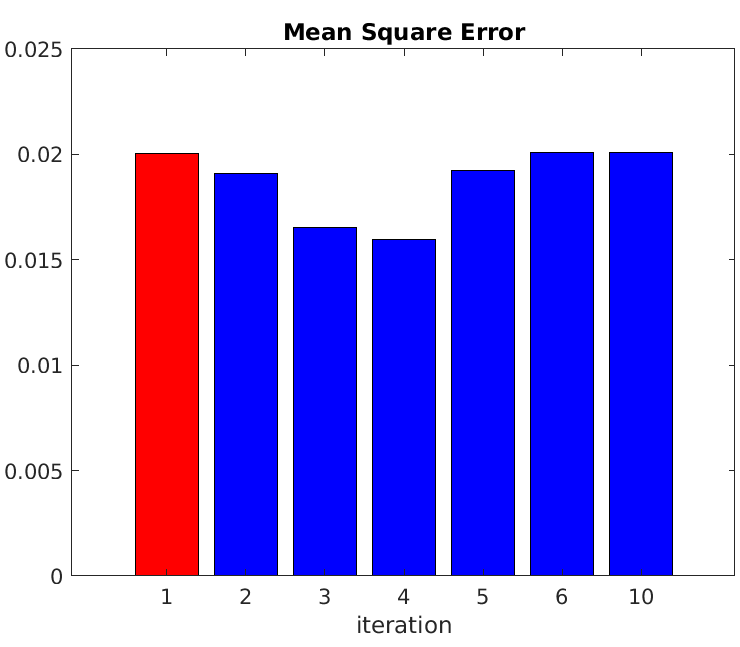}&
    \includegraphics[width=0.45\textwidth]{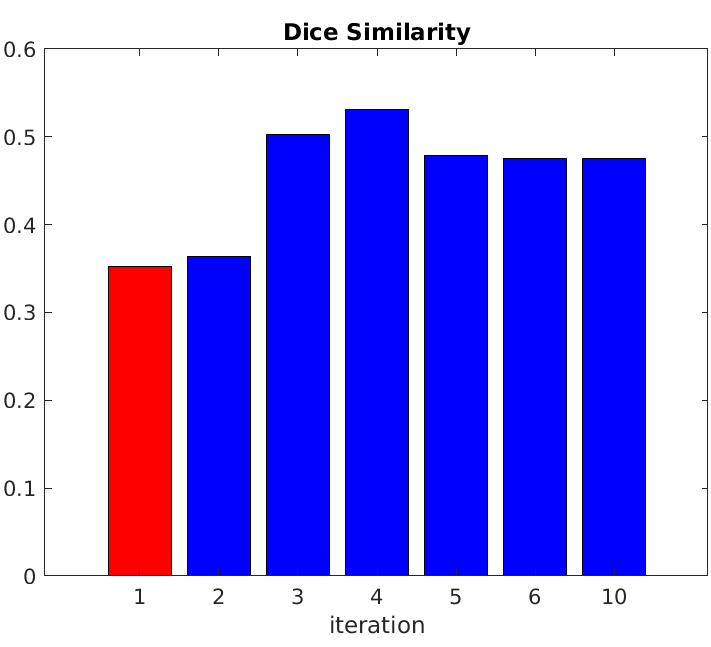}\\
    \includegraphics[width=0.45\textwidth]{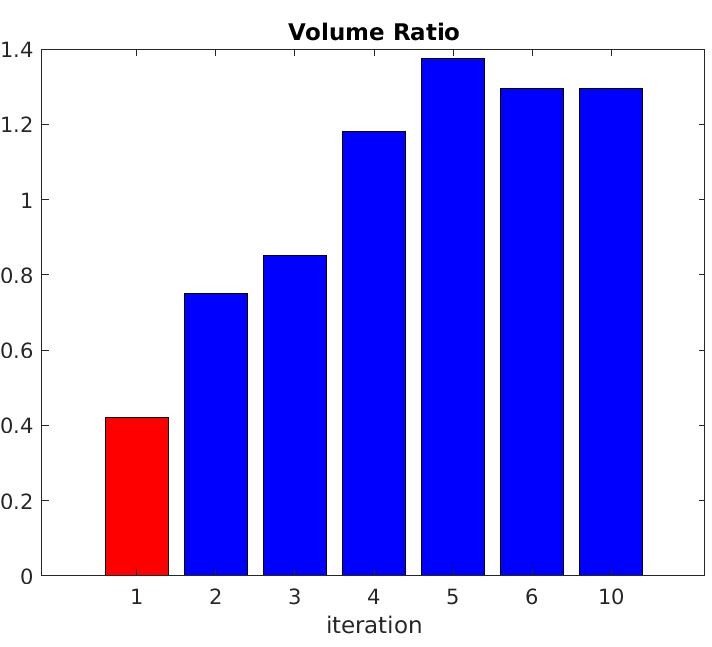}&
    \includegraphics[width=0.45\textwidth]{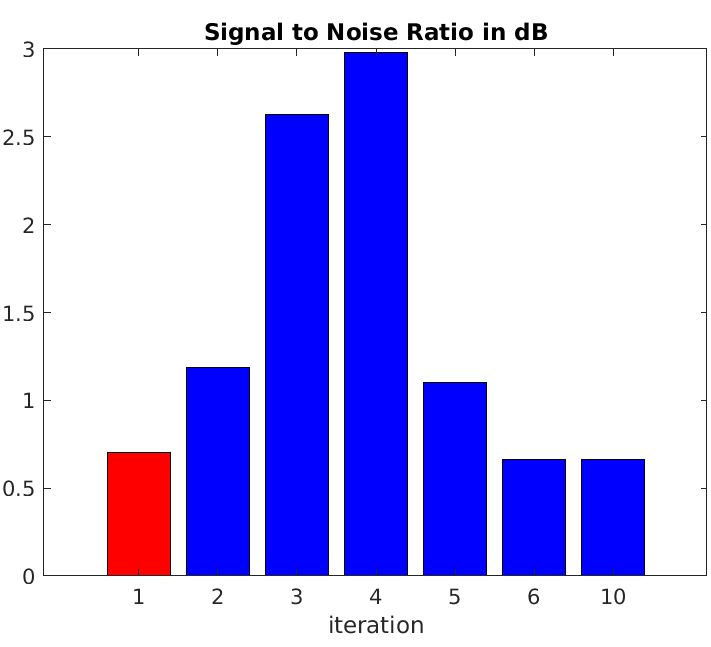}
   \end{tabular}
   \end{center}
   \caption{Mean square error, Dice similarity, volume ratio and signal to noise ratio of the reconstruction results at round 1-6 and round 10. The error of the reconstruction result based on the initial laser setting, i.e. the first round of iteration is indicated by the red bars. The blue bars represent the error of following rounds of iterations that are based on updated illumination patterns.}
  \label{new_error}
  \end{table}

  \begin{table}[h!]
   \begin{center}
    \begin{tabular}{ccccc}
    \toprule[1.5pt]
    &MSE&Dice similarity&VR&SNR in dB\\
    \midrule
    $1^{st}$ experiment initial&0.018436&0.47761&0.52273&1.5303\\
    \midrule
    $1^{st}$ experiment $3^{rd}$ round&0.01489&0.53631&1.0341&3.6664\\
    \midrule
    $2^{nd}$ experiment initial&0.020026&0.35200&0.42045&0.70309\\
    \midrule
    $2^{nd}$ experiment $4^{th}$ round&0.015951&0.53125&1.1818&2.9783\\
  \bottomrule
    \end{tabular}
   \end{center}
\caption{Comparison of quantitative error under the four metrics between the first reconstruction
 result based on the initial laser setting and the best reconstruction result of the first and the second experiments.}
  \label{Comparison_error}
  \end{table}

Error analysis in Table \ref{new_error} shows that the 4th round produces the best reconstruction result. Furthermore, by comparing the quantitative error using the four metrics defined before between the best reconstruction result in the previous experiment (appears in the 3rd round of iteration) and this experiment (appears in the 4th round of iteration) as is shown in Table \ref{Comparison_error}, the quality of the best result in both experiments are nearly the same. Besides, the reconstruction quality is greatly improved compared the that of the initial setting in both cases.  Summing up the conclusions before, it is safe to say that the optimal illumination pattern is independent of the initial laser setting.

\section{Discussion and conclusion}

\subsection{Discussion}
 When actually implemented in the numerical experiments section, both FISTA algorithm and the Cyclic Coordinate Descent prove to be quite fast.
Notice that Cyclic Coordinate Descent is very similar to stochastic gradient descent (SGD) which randomly chooses one coordinate to calculate the gradient at each step. In our case, Cyclic Coordinate Descent has an excellent performance in terms of computational speed, so SGD is not considered. In practice, if the system becomes larger and Cyclic Coordinate Descent becomes costly, we can simply transform it into SGD to save computational time.

 It is also worth pointing out that $\mu$ needs to have different values for design matrix $V$ and matrix $S_p$ if we wish to restrict lasers on a certain region of the phantom surface. If we regard $\mu$ as a first-level parameter, after determining appropriate $\mu$, a second-level parameter that restricts the lasers to the desired region of the object surface needs to be determined as well. During phantom experiments, more penalty was enforced on the matrix $S_p$ so that the nonzero entries of the solution are confined solely on the top surface of the phantom. Overall, the proper approach for the choice of the value of the regularization weight should be that we first adjust $\mu$ to make sure the solution is stable and meaningful, in other words, a convergence-type analysis for $\mu$ is needed; then slightly adjust the finer penalty parameter placed on $S_p$ to attain the desirable number and location of lasers according to our need. However, choosing the proper value of $\mu$ is non-trivial because $\mu $ is a highly sensitive parameter. Suitable values of $\mu$ produce desired number of lasers whereas tiniest changes in $\mu$ can lead to the solution oscillating between no lasers at all and over-numbered lasers. Currently, there exists no systematic way of determining the regularization weight $\mu$. One possibility might be to tune the value of $\mu$ using machine learning techniques. Further investigation of the factors that influence the value of $\mu$ will also shed light on the choice of $\mu$ and thus enable our method to adapt to various experiment setups in practice.

 In addition, the imaging of time-varying targets is becoming a trendy topic, too. Dynamical imaging methods have been developed in fields such as ultrafast ultrasound imaging \cite{dynamic_imaging} in which the reconstruction is combined with some tracking technique, and electrical impedance tomography \cite{dynamic_eit} where the determination of optimal current pattern in cases of imaging time-variant targets is made possible by using statistical tools. These dynamical methods provide good inspiration for extending the static methods developed in this work to dynamical FMT in the future. 

\subsection{Conclusion}
We start with a complete theoretical analysis of the diffusion equation that describes the physical processes of FMT and also the discrete formulation of the equation systems. Together with the free-space light propagation model due to the noncontact configuration, a comprehensive forward model is derived. Based on the forward model, we introduced sparse regularization techniques for the inverse problem of recovering the fluorescence distribution. Sparse regularizers have an advantage in promoting the sparsity of the solution to the linear system thus producing reconstruction results with higher spatial resolution.

Biologists have noticed that designing optimized illumination patterns plays a very key role in improving the reconstruction quality of FMT. In order to design the optimal illumination pattern, we first introduce a proper mathematical definition of an illumination pattern $\Sigma$ making use of the discrete formulation. With such a definition, we regard the restrictions on the laser properties including intensity, location and total number as restrictions on different norms of $\Sigma$ and hence define the admissible set $\Delta$ of illumination patterns. Designing the optimal illumination pattern thus becomes solving another linear system with the illumination pattern vector $\Sigma$ as the unknown subject to all the constraints in the admissible set. We also performed a detailed analysis on the inverse problem of designing $\Sigma$ and proposed a fast iterative algorithm, the Cyclic Coordinate Descent algorithm, to solve the inverse problem. 

At the end, we combine the sparse regularization techniques and designing the optimal illumination pattern as a two-step approach to further improve the reconstruction quality. Numerical experiments on a small cubic phantom are performed on different initial laser settings. It turns out that with suitable choice of $\mu$ the updated illumination pattern after each round converges quickly and remains very stable after merely four or five rounds. In addition, the optimal illumination pattern does not depend on the initial laser setting. At the same time, the reconstruction result corresponding to the optimal illumination pattern proves to be better than the reconstruction result based on the initial unoptimized illumination pattern in all metrics. The spatial resolution is increased with a higher Dice similarity and lower mean square error. The signal to noise ratio is greatly increased as well. Numerical experiments have demonstrated that our two-step approach outperforms the existing nonsparse reconstruction methods without optimizing the illumination pattern.

 \section*{Acknowledgements} 
 We would like to thank Prof. Markus Rudin (ETH Z\"urich) and Prof. Martin Wolf (University Hospital Z\"urich) for their very constructive suggestions. 

\bibliographystyle{abbrvnat}
\bibliography{ref_thesis} 
\addcontentsline{toc}{section}{References}

\end{document}